\documentclass{aa}  

\usepackage{graphicx, color}
\usepackage{txfonts}
\usepackage{comment}
\usepackage{siunitx}  % Used by Dag
\usepackage{booktabs} % Used by Dag
\usepackage[normalem]{ulem} % Paul added this one to cross something out
\usepackage{orcidlink}
\begin{document}

   \title{A magnetic field study of two fast-rotating, radio bright M dwarfs}
\subtitle{StKM~1-1262 and V374~Peg}
   \titlerunning{A magnetic field study of two radio bright M dwarfs}

   \author{S. Bellotti \inst{1,2}\orcidlink{0000-0002-2558-6920}
          \and 
          P. I. Cristofari \inst{1,3}\orcidlink{0000-0003-4019-0630}
          \and
          J. R. Callingham\inst{4,5}\orcidlink{0000-0002-7167-1819}
          \and
          J. Morin \inst{6}\orcidlink{0000-0002-4996-6901}
          \and
          P. Petit \inst{2}\orcidlink{0000-0001-7624-9222}
          \and 
          A. A. Vidotto \inst{1}\orcidlink{0000-0001-5371-2675}
          \and
          M. Jardine\inst{7}\orcidlink{0000-0002-1466-5236}
          \and
          L. Arnold\inst{8}\orcidlink{0000-0002-0111-1234}
          \and
          R. D. Kavanagh\inst{5,4}\orcidlink{0000-0002-1486-7188}
          \and
          J. Llama\inst{9}\orcidlink{0000-0003-4450-0368}
          \and
          H. Vedantham\inst{4,10}\orcidlink{0000-0002-0872-181X}   
          }

   \authorrunning{Bellotti et al.}
    
   \institute{
            Leiden Observatory, Leiden University,
            PO Box 9513, 2300 RA Leiden, The Netherlands\\
            \email{bellotti@strw.leidenuniv.nl}
        \and
            Institut de Recherche en Astrophysique et Plan\'etologie,
            Universit\'e de Toulouse, CNRS, IRAP/UMR 5277,
            14 avenue Edouard Belin, F-31400, Toulouse, France  
        \and
            Center for Astrophysics | Harvard \& Smithsonian,
            60 Garden Street,
            Cambridge, MA 02138, United States
        \and 
            ASTRON, Netherlands Institute for Radio Astronomy, Oude Hoogeveensedijk 4, Dwingeloo, 7991 PD, The Netherlands.
        \and
            Anton Pannekoek Institute for Astronomy, University of Amsterdam, Science Park 904, 1098 XH, Amsterdam, The Netherlands        
        \and
             Laboratoire Univers et Particules de Montpellier,
             Universit\'e de Montpellier, CNRS,
             F-34095, Montpellier, France
        \and
            School of Physics and Astronomy, University of St Andrews, North Haugh, St Andrews, Fife, Scotland, KY16 9SS
        \and
            Canada-France-Hawaii Telescope, 65-1238 Mamalahoa Hwy, Kamuela, HI 96743, USA
        \and 
            Lowell Observatory, 1400 W. Mars Hill Rd. Flagstaff, AZ. 86001. USA
        \and
            Kapteyn Astronomical Institute, University of Groningen, PO Box 800, 9700 AV Groningen, The Netherlands
             }
   \date{Received ; accepted }

% \abstract{}{}{}{}{} 
% 5 {} token are mandatory
 
  \abstract
   {Radio observations at low frequencies are sensitive to the magnetic activity of stars and the plasma environment surrounding them, so one can scrutinize the conditions under which stellar space weather develops and impacts exoplanets. The accurate interpretation of the processes underlying the radio signatures requires a detailed characterisation of the stellar magnetism.}   
   {We study two M~dwarfs, namely StKM~1-1262 (M0 type, P$_\mathrm{rot}=1.24$\,d) and V374~Peg (M4 type, P$_\mathrm{rot}=0.4455$\,d), which were detected recently with the LOw Frequency ARray (LOFAR). StKM~1-1262 exhibited the typical signature of a type-II radio burst, potentially resulting from a coronal mass ejection event. V374~Peg manifested low-frequency radio emission with high brightness temperature and low degree of polarisation, suggesting an electron-cyclotron maser instability emission mechanism. In this work, we provide recent observational constraints on the magnetic field of both stars.}
   {We analysed spectropolarimetric observations of these M~dwarfs collected with the SpectroPolarim\`etre InfraRouge (SPIRou). Firstly, we refined the stellar parameters such as effective temperature, surface gravity, and metallicity, and measured the average surface magnetic flux via modelling of Zeeman broadening in unpolarised spectra. We then applied Zeeman-Doppler imaging to least-squares deconvolution line profiles in circular polarisation to reconstruct their large-scale magnetic fields. We also reconstructed a brightness map for the two stars by means of Doppler imaging.}
   {StKM~1-1262 has a total, unsigned magnetic field of $3.53\pm0.06$\,kG on average and the large-scale magnetic field topology is predominantly poloidal, dipolar and moderately axisymmetric, with an average strength of 300\,G. V374~Peg has an unsigned magnetic field of $5.46\pm0.09$\,kG and the large-scale field is poloidal, dipolar and axisymmetric, with an average strength of 800\,G. For StKM~1-1262, we found a strong (Pearson $\rho=-0.96$) anti-correlation between the total magnetic field and the effective temperature which is reminiscent of the tight link between small-scale magnetic fields and surface inhomogeneities. For V374~Peg, we found a moderate ($\rho=-0.43$) anti-correlation, possibly due to a more even distribution of surface features.
   }
   {The large-scale magnetic field topology of StKM~1-1262 is similar to other stars with similar fundamental parameters like mass and rotation period, and the brightness map features one dark spot which is responsible for the rotational modulation of the total magnetic field and the retrieved effective temperature. For V374~Peg, the magnetic topology and the brightness map are similar to previous reconstructions indicating a temporal stability of approximately 14-yr.} 


   \maketitle

%
%-------------------------------------------------------------------

\section{Introduction}

A central goal of exoplanetology is to find Earth-like planets that potentially host life. According to the canonical definition, assessing the potential habitability of a planet requires knowledge about the distance (or insolation) of the planet from the star \citep{Kasting1993,Kopparapu2013}. However, this is a first-order assessment, and additional details such as the stellar magnetic field \citep{Driscoll2013}, the planetary volcanic and tectonic activity \citep{Lenardic2016,Seales2021}, the distribution of oceans \citep{Cockell2016}, the asteroid impacts \citep{Childs2022}, and the binarity of the host star \citep{Jaime2014,Cuntz2015,Cuntz2020} are needed to obtain a comprehensive definition of habitability. The space weather and radiation environment in which exoplanets orbit is also crucial in this context \citep{Vidotto2013,Airapetian2017,Airapetian2020}. For instance, X-ray and UV radiation can drive atmospheric escape and alter its chemical constituents, ultimately shaping the evolution of planetary atmospheres \citep{Lammer2003,Penz2008,Owen2012,Rugheimer2015,Carolan2019,vanLooveren2024}. 

Coronal mass ejections (CMEs) are expulsions of hot coronal plasma and magnetic field into the heliosphere with speeds ranging from 100 to 2000~km~s$^{-1}$ \citep{Nindos2008,Webb2012}. They represent a major constituent of the space weather in the Solar System and a continuous exposure is expected to lead to erosion of planetary atmospheres over time \citep{Lammer2007,Kodachenko2007}. Similar effects are anticipated on other stars \citep{Lammer2013,Cherenkov2017}, but there has yet to be an unambiguous detection of an extrasolar CME, despite substantial observational effort \citep[e.g.][]{Crosley2016,Crosley2018}. The incidence and energetics of these phenomena are thus not constrained empirically, which prevents us from assessing the habitability of other systems in a comprehensive manner. 

Radio observations of stellar systems allow one to investigate stellar magnetic activity and the space weather conditions of extrasolar planets \citep[see][and references therein]{Callingham2024}. Observing a type~II radio burst represents a detection of plasma radial motion away from the star, because electrons are accelerated by shocks at the leading edges of outward-moving CMEs which generate the typical frequency sweep \citep{Cane2002,Gopalswamy2008,Osten2017,Majumdar2021}. A type~II radio burst is thus a `smoking gun' signature of CMEs, but it has been elusive for stars other than the Sun. 

Another puzzle coming from radio observations of stellar systems concerns the low-frequency radio emission detected from both active and quiescent M~dwarfs within the LOFAR Two-meter Sky Survey \citep[LoTSS][]{Shimwell2017,Shimwell2022}. Specifically, these observations unravelled a population of M1-M6~dwarfs manifesting highly circularly polarised ($\geq$~60\%), high brightness temperature ($>$~10$^{12}$~K), coherent, low-frequency (on the order of MHz) radio emission \citep{Callingham2021}. To explain these findings, there are two main categories of processes: plasma emission or electron-cyclotron maser instability (ECMI) emission. The former occurs when flares and CMEs inject hot plasma into colder plasma \citep{Dulk1985,Matthews2019} and is correlated with stellar magnetic activity \citep{Mclean2012,Villadsen2019,Callingham2021}. The M~dwarfs detected within the LoTSS survey span various activity levels, suggesting that the radio emission may not be only activity-driven which finds support from analyses of flare activity conducted with the Transiting Exoplanet Survey Satellite \citep[TESS;][]{Ricker2014} as shown by \citet{Pope2021}.

For ECMI-driven emission \citep{Wu1979,Treumann2006}, the scenario sees an acceleration of electrons towards the star, powering auroral processes similar to Jupiter \citep{Zarka1998}. A first possibility is that, at a certain distance from the star, the stellar magnetic field cannot impose co-rotation of the surrounding plasma. This breakdown of co-rotation generates a lag between the plasma disk and the star's magnetosphere, and ultimately a current system that accelerates electrons towards the star \citep{Nichols2012,Pineda2017,Marques2017}. A second possibility translates the Jupiter interaction with its moon Io into a star-planet one, for which a planet orbits within the Alfv\'en surface of the host star and perturbs or reconnects with the stellar magnetic field, driving electron acceleration \citep{Lazio2004,Zarka2007,Turnpenney2018,Vidotto2019,Vedantham2020}. Although there are candidates for star-planet interactions, no exoplanet has been conclusively detected via low-frequency radio emission yet \citep{Lynch2018,Vedantham2020,Pope2021,Turner2021,PerezTorres2021}.

Considering the sensitivity of radio observations to the plasma environment around a star, interpreting such observations requires an accurate model of the stellar magnetic environment and wind \citep[e.g.][]{Vidotto2014,Kavanagh2019,Kavanagh2021,Kavanagh2022,Alvarado-Gomez2022,Elekes2023,PenaMonino2024}. In turn, such model is dictated by our knowledge on the magnetic field of the star, whose large-scale configuration can be reconstructed from spectropolarimetric time series via Zeeman-Doppler imaging \citep[ZDI][]{Semel1989,DonatiBrown1997}. Over the years, ZDI has been applied extensively, revealing a variety of field geometries for low-mass stars as well as their temporal evolution \citep[e.g.][]{Petit2005,Donati2008,Morin2008,Morin2010,Hebrard2016,Kochukhov2017,Lavail2018,Folsom2018,Bellotti2023b,Bellotti2024a}.

Here, we analyse recent spectropolarimetric observations of two stars, StKM~1-1262 and V374~Peg, and characterise their large-scale magnetic field. These two stars were detected within the LoTSS survey \citep{Tasse2021,Yiu2024}. StKM~1-1262 is an early M dwarf for which a candidate type-II radio burst was detected as an isolated event, with the characteristic frequency sweep of a typical solar type-II burst but more than four orders of magnitude radio luminous \citep{Callingham2025}. The constraint of the large-scale magnetic field geometry in the current work is used to determine whether a CME could have sufficient kinetic energy to escape the large-scale field and could drive a super-Alfv\'{e}nic shock \citep{Callingham2025}. V374~Peg is an active, fully convective mid-M~dwarf known to exhibit frequent flares \citep{Korhonen2010,Vida2016}, a strong ($\sim1$\,kG), poloidal, and axisymmetric large-scale magnetic field \citep{Morin2008a}, and a powerful wind \citep{Vidotto2011}. Radio emission from V374~Peg was detected with the Very Large Array in a frequency band of 4-8\,GHz \citep{Hallinan2009}, and the modulation of the radio light curve due to stellar rotation was fit by \citet{Llama2018} using an ECMI-driven model. Additional radio observations were obtained within the LoTSS survey (120–168 MHz) in 2018 and more recently in 2024, bracketing our spectropolarimetric observations. The reconstructed large-scale magnetic field in this work will then be used as boundary condition to simulate the stellar environment of V374~Peg and better understand which ECMI mechanism is powering the observed low-frequency radio emission in a follow-up study.

This paper is structured as follows. In Sect.~\ref{sec:obs} we describe the spectropolarimetric data set. In Sect.~\ref{sec:ZBro} we outline the modelling of Zeeman broadening and intensification of unpolarised spectra to measure the total, unsigned magnetic field intensity. In Sect.~\ref{sec:Blon} and Sect.~\ref{sec:ZDI}, we describe the analysis of circular polarisation spectra to measure the longitudinal magnetic field and to reconstruct the large-scale magnetic field geometry by means of ZDI, respectively. We then show the reconstruction of the brightness maps of StKM~1-1262 and V374~Peg with Doppler imaging in Sect.~\ref{sec:DI} and finally draw our conclusions in Sect.~\ref{sec:conclusions} .

\section{Observations}\label{sec:obs}

The observations analysed in this work were obtained in circular polarisation mode with the SpectroPolarim\`etre InfraRouge \citep[SPIRou][]{Donati2020}, the stabilised high-resolution near-infrared spectropolarimeter mounted on the 3.6\,m Canada–France–Hawaii Telescope (CFHT) atop Maunakea, Hawaii. SPIRou operates between 0.96 and 2.5~$\mu$m ($YJHK$ bands) at a spectral resolving power of $R \sim 70\,000 $. Optimal extraction of spectra in unpolarised (Stokes~$I$) and circularly polarised (Stokes~$V$) light was carried out with {\it A PipelinE to Reduce Observations} (\texttt{APERO}), a fully automatic reduction package installed at CFHT \citep{Cook2022}. The observations of both stars were reduced with APERO v0.7.288. The journal of the observations is given in Table~\ref{tab:log} and Table~\ref{tab:log2}.

StKM~1-1262 was observed for 17 nights between January and March 2024, spanning a period of 41 days. We recorded a S/N at 1650~nm per spectral element between 210 and 278, with a mean of 246. Two observations (23 and 28 February) did not reach enough S/N to produce meaningful Stokes~$V$ LSD profiles (see Sect.~\ref{sec:Blon}), so they were discarded. V374~Peg was observed 55 times between August 14th and 23th in 2021. The recorded S/N oscillated between 86 and 189, with an average of 162. In Table~\ref{tab:stars_properties}, we provide the fundamenteal stellar aparameters derived and used in this work for both stars.

The observations of the StKM~1-1262 and V374~Peg are phased according to the ephemeris
\begin{align}
    \mathrm{HJD}= \mathrm{HDJ}_0 + \mathrm{P}_\mathrm{rot}\cdot n_\mathrm{cyc}.
    \label{eq:ephemeris}
\end{align}
In the formula, HJD$_0$ is the reference heliocentric Julian date (2460330.1563 for StKM~1-1262, and 2459440.7902 for V374~Peg), P$_\mathrm{rot}$ is the stellar rotation period (see Table~\ref{tab:stars_properties}), and $n_\mathrm{cyc}$ is the number of rotation cycles.

\begin{table}[!t]
\caption{Properties of StKM~1-1262 and V374~Peg.} 
\label{tab:stars_properties}     
\centering                       
\begin{tabular}{l r r}    
\toprule
Name & StKM~1-1262 & V374~Peg\\
\midrule
Spectral Type & M0V & M4V\\
$H$  [mag] & 8.15$^a$ & 7.04$^a$\\
Distance [pc] & 40.9$^b$ & 9.1$^b$\\ 
T$_\mathrm{eff}$ [K] & $3933\pm31^\dagger$ & $3228\pm30^\dagger$\\
$\log g$ [cgs] & $4.65\pm0.05^\dagger$ & $4.72\pm0.05^\dagger$\\
M/H [dex] & $-0.07\pm0.10^\dagger$ & $0.07\pm0.10^\dagger$\\
Mass [M$_\odot$] & 0.64$^c$ & 0.30$^c$\\
Radius [R$_\odot$] & 0.63$^c$ & 0.31$^c$\\
P$_\mathrm{rot}$ [d] & 1.24$^d$ & 0.4455$^e$ \\
$v_\mathrm{eq}\sin i$ [km s$^{-1}$] & $25.04\pm0.10^\dagger$ & $36.73\pm0.17^\dagger$\\
$i$ [$^\circ$] & $>70^\dagger$ & $70^e$\\
\bottomrule 
\end{tabular}
\tablefoot{The listed properties are: identifier, spectral type, $H$ band magnitude, distance, effective temperature, surface gravity, metallicity, stellar mass, stellar radius, rotation period, equatorial projected velocity, and inclination used in ZDI. The parameters marked with $\dagger$ are derived in this work. The references are: $a)$ \citet{Cutri2003}, $b)$ \citet{GaiaCollaboration2020}, $c)$ \citet{Kervella2022}, $d)$ \citet{Colman2024}, and $e)$ \citet{Morin2008a}.}
\end{table}

\section{Total magnetic field and atmospheric characterisation}\label{sec:ZBro}

We estimated the average small-scale magnetic field strength at the stellar surface of StKM~1-1262 and V374~Peg following the approach described in~\citet{Cristofari2023a, Cristofari2023b}. Our process relies on synthetic spectra computed from MARCS model atmospheres~\citep{Gustafsson2008} with \texttt{ZeeTurbo}~\citep{Cristofari2023a}, a tool built from the \texttt{Turbospectrum}~\citep{Plez2012} and \texttt{Zeeman}~\citep{Landstreet1988, Wade2001, Folsom2016} codes. Specifically, we computed a grid of synthetic spectra for different set of atmospheric parameters, with $T_{\rm eff}$ ranging from 3200 to 4400~K in steps of 100~K, $\log{g}$ ranging from $4.0$ to $5.0$~dex in steps of $0.5$~dex, and $\rm [M/H]$ ranging from $-0.75$ to $0.75$~dex in steps of $0.25$~dex. The chosen ranges are wide enough to encompass typical parameters for M~dwarfs and act as uniform priors. For each set of atmospheric parameters, we computed spectra with different magnetic field intensities, ranging from 0 to 10\,kG in steps of 2\,kG, assuming that the magnetic field is radial in all points of the photosphere.

A model of the observed spectrum is obtained with a linear combination of the spectra computed for the different magnetic fields intensities so that $S=\sum_i f_iS_i$, with $S$ the model spectrum, $S_i$ the spectrum computed for a $i$\,kG field, and $f_i$ the associated filling factor. We rely on an MCMC process to derive the set of atmospheric parameters, filling factors, and projected rotational velocity ($v_\mathrm{eq}\sin i$) leading the best fit to the data and estimate formal error bars~\citep[see also][for more details]{Cristofari2023a, Cristofari2023b}. Our method is inherently sensitive to systematics which are not fully reflected by our formal error bars. Following~\citet{Cristofari2022a}, we inflate the error bars by quadratically adding 30\,K, 0.05\,dex, and 0.10\,dex to $T_{\rm eff}$, $\log{g}$, and $\rm [M/H]$, respectively. 

For both stars, we begin by applying our process to template spectra obtained by taking the median of all available spectra.
For StKM~1-1262, our process yields $T_{\rm eff}=3933\pm31$\,K, $\log{g}=4.65\pm0.05$\,dex, $\rm [M/H]=-0.07\pm0.10$\,dex and $v_\mathrm{eq}\sin i=25.04\pm0.10$\,$\rm km\,s^{-1}$. For V374 Peg, our analysis yields $T_{\rm eff}=3228\pm30$\,K, $\log{g}=4.72\pm0.05$\,dex, $\rm [M/H]=0.07\pm0.10$\,dex and $v_\mathrm{eq}\sin i=36.73\pm0.17$\,$\rm km\,s^{-1}$. Our atmospheric estimates are reported in Table~\ref{tab:stars_properties}. 

\begin{figure}
    \centering
    \includegraphics[width=\linewidth]{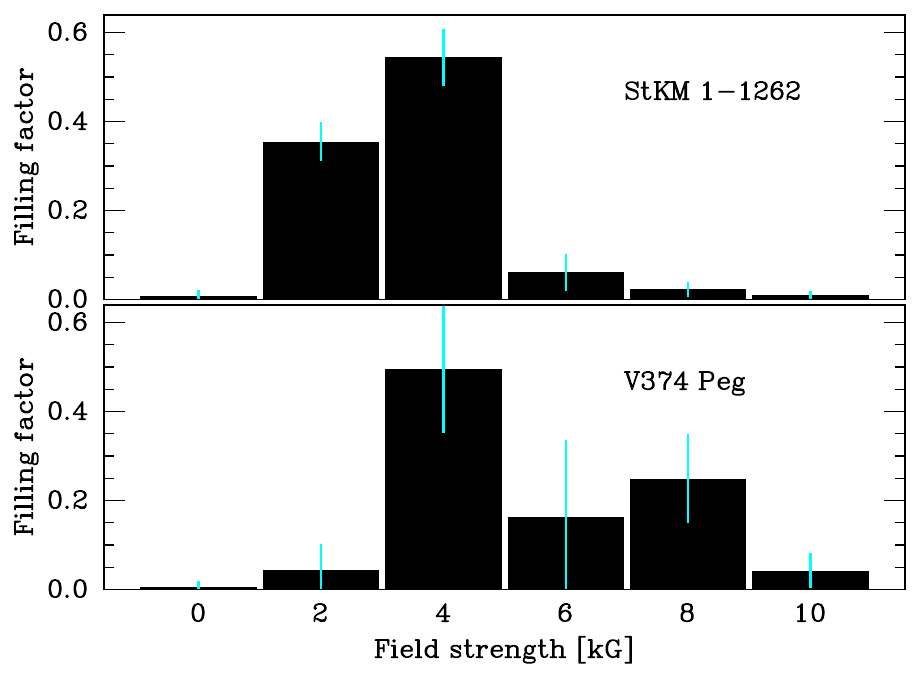}
    \caption{Distribution of the magnetic flux on the different magnetic components. The top panel refers to StKM~1-1262 and the bottom panel to V374~Peg.}
    \label{fig:b_distrib}
\end{figure}

For StKM~1-1262, our best fit on the template spectrum yields an average small-scale magnetic field of $\langle B_I\rangle=3.53\pm0.06$\,kG, with most of the magnetic intensity distributed on the 2 and 4\,kG components ($\sim 87$\,\% of the total magnetic flux, see Fig.~\ref{fig:b_distrib}). The value $\langle B_I\rangle=3.53$\,kG is consistent with that obtained for fast-rotating, early M~dwarfs like OT~Ser \citep[$\langle B_I\rangle=3.2$\,kG;][]{Shulyak2019} and YY~Gem \citep[$\langle B_I\rangle=3.2-3.4$\,kG;][]{Kochukhov2019}. For V374 Peg, our process yields an average small-scale magnetic field of $\langle B_I\rangle=5.46\pm0.09$\,kG. The distribution of filling factors shows more contribution from the higher components, with the 4, 6 and 8\,kG components accounting for 90\,\% of the total magnetic flux. In contrast to StKM~1-1262, the 2\,kG component only amount to 4\,\% of the magnetic flux of V374~Peg (see Fig.~\ref{fig:b_distrib}).

\begin{figure}[t]
    \includegraphics[width=\columnwidth]{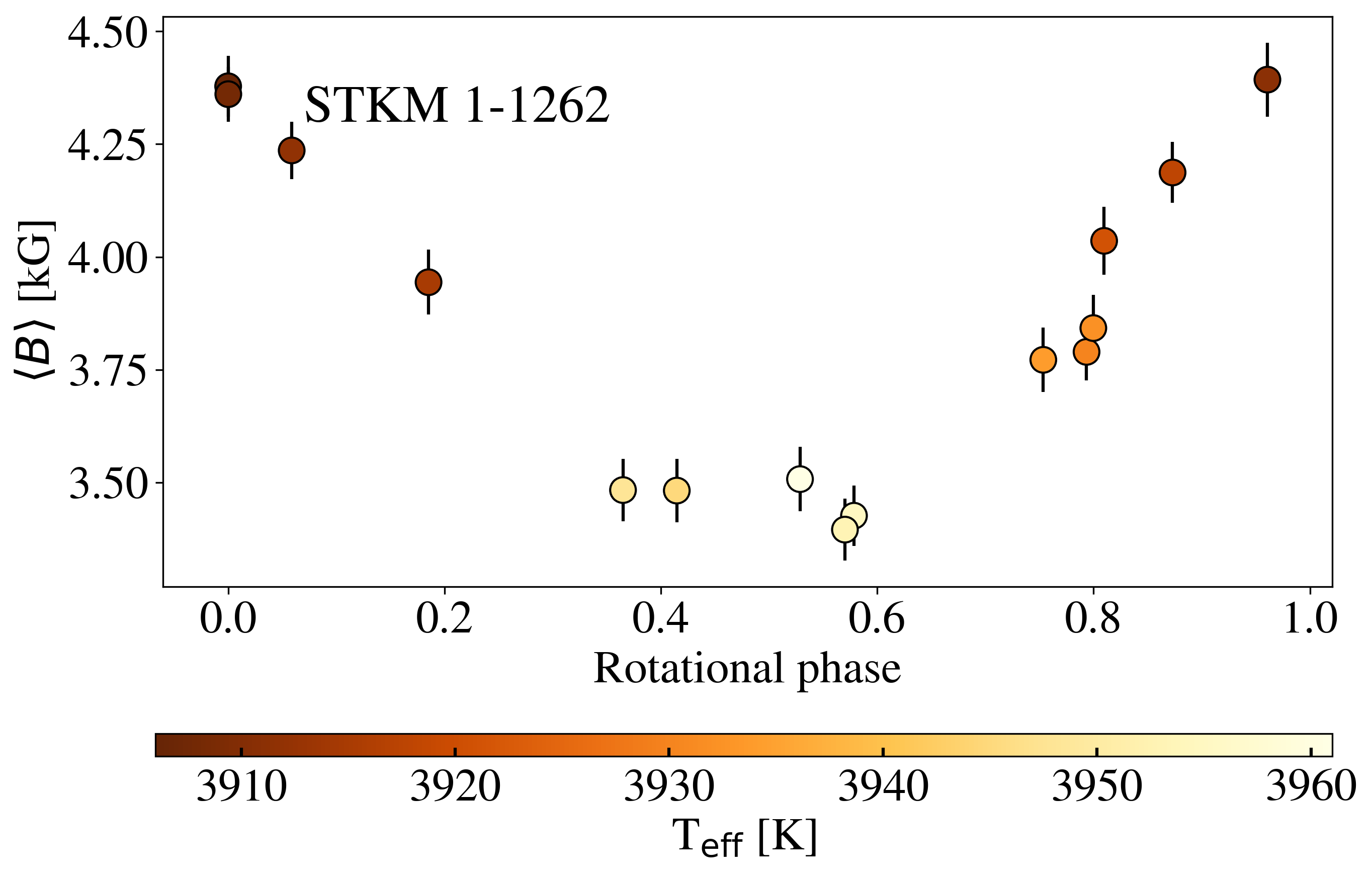}
    \includegraphics[width=\columnwidth]{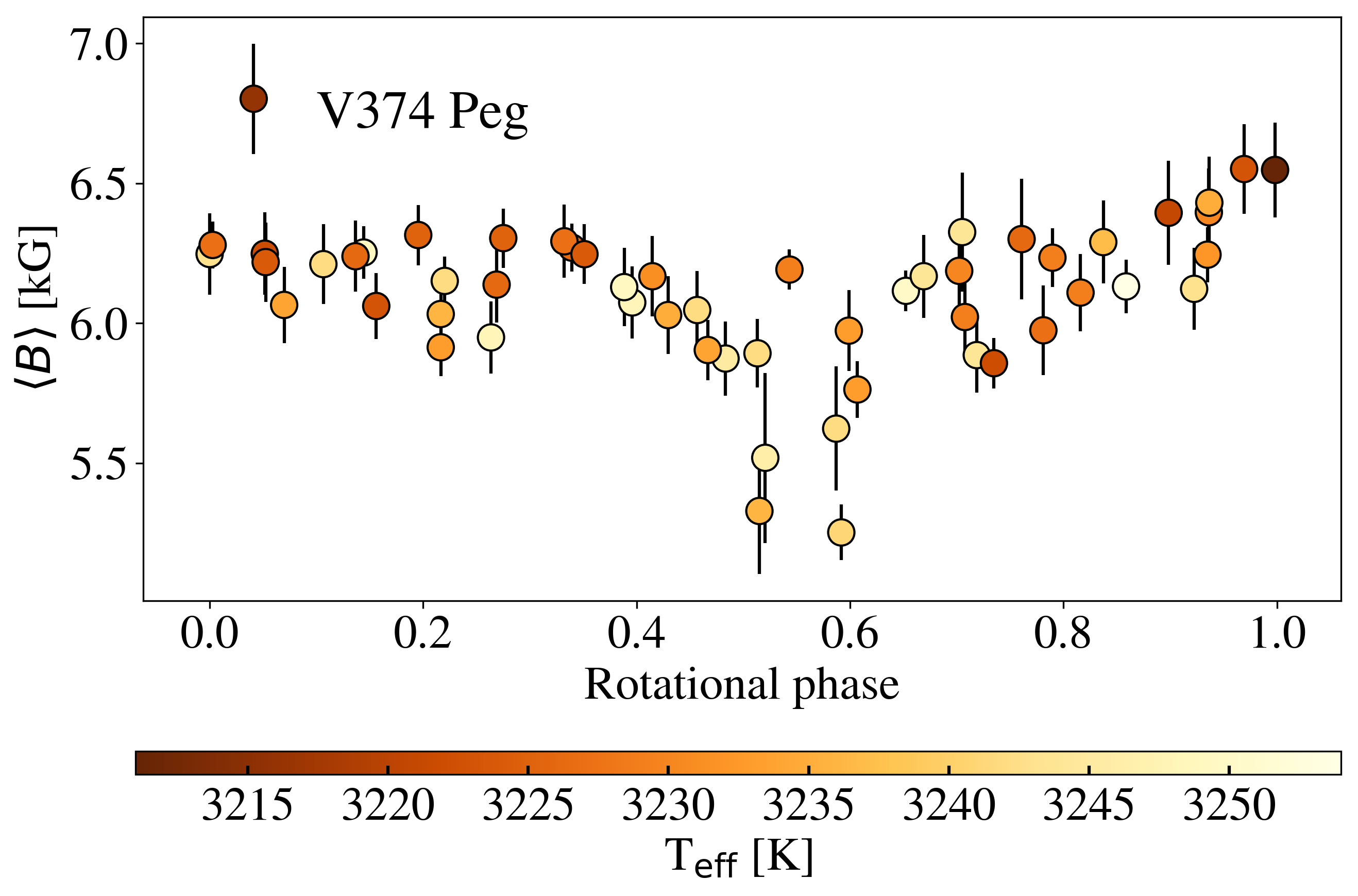}
    \caption{Total magnetic field measurements of StKM~1-1262 and V374~Peg as a function of rotational phase. The data points are colour-coded with the effective temperature obtained from spectral modelling (see Sect.~\ref{sec:ZBro}). The rotational phases are computed according to Eq.~\ref{eq:ephemeris}, using the stellar rotation period listed in Table~\ref{tab:stars_properties}.}
    \label{fig:BI}
\end{figure}

For each star, we repeat our process, this time analysing each individual spectrum, and fixing the values of $\log{g}$ and $\rm [M/H]$ to those listed in Table~\ref{tab:stars_properties}. This approach allows us to simultaneously explore the temporal variations of $\langle B_I\rangle$ and temperature. Figure~\ref{fig:BI} shows $\langle B_I\rangle$ computed for each observation of StKM~1-1262 and V374~Peg, colour-coded with the associated value of T$_\mathrm{eff}$.

For StKM~1-1262, the $\langle B_I\rangle$ range from $\sim$4.50 to $\sim$3.25\,kG, and exhibit an evident modulation at the stellar rotation period of 1.24\,d. We note a clear anti-correlation between $\langle B_I\rangle$ and T$_\mathrm{eff}$ (Pearson correlation coefficient $\rho=-0.96$). Specifically, we observe $\langle B_I\rangle$ to be highest around phase 0.0, when the retrieved T$_\mathrm{eff}$ is lowest. Such behaviour was already captured for the active M~dwarfs AU~Mic \citep{Artigau2024}, EV~Lac, and DS~Leo \citep{Cristofari2025}. As we will describe from Doppler imaging in Sect.~\ref{sec:DI}, such behaviour tracks the presence of a starspot or a group of starspots, which are cooler than the quiet photosphere.

We further use the anti-correlation between $\langle B_I\rangle$ and temperature to estimate the spot coverage of StKM~1-1262. We fit a line $\langle B_I\rangle$ as a function of temperature, yielding a slope of $-54\pm5$\,$\rm K\,kG^{-1}$ and an intercept of $4139\pm18$\,K (see also Fig.~\ref{fig:correlations}). We attribute the intercept of this fit to the temperature of the quiet photosphere ($T_{\rm max}$). Following the relation of~\citet{berdyugina-2005}, we derive a spot temperature for StKM~1-1262 of $T_{\rm spot}=3303\pm14$\,K. Given a temperature $T=3933$~K (see also Table~\ref{tab:stars_properties}), we then compute the fraction of stellar surface occupied by spots, $f$, such that
\begin{equation}
    f=\frac{T_{\rm max}-T}{T_{\rm max}-T_{\rm spot}}.
    \label{eq:spot_coverage}
\end{equation}
Injecting the effective temperature obtained for StKM~1-1262 into equation~\ref{eq:spot_coverage}, we derived a spot coverage of about $f=25\%$.

For V374~Peg, no clear rotational modulation of $\langle B_I\rangle$ is observed, which can be attributed to lower S/N of individual spectra rendering the fitting procedure more challenging. We find a moderate anti-correlation (Pearson correlation coefficient $\rho=-0.43$) between our temperature and $\langle B_I\rangle$ measurements, with a linear fit yielding a slope of $-37\pm7$\,$\rm K\,kG^{-1}$ and an intercept of $3461\pm43$\,K. Relying on the relation of~\citet{berdyugina-2005}, we derive a spot temperature $T_{\rm spot}=2979\pm15$, and deduce a spot coverage of about 48\%.

\section{Longitudinal magnetic field}\label{sec:Blon}

We computed mean line profiles using least-squares deconvolution \citep[LSD;][]{Donati1997,Kochukhov2010a} and in particular the Python \texttt{LSDpy} package \citep{Folsom2025}.\footnote{The code is available at \href{https://github.com/folsomcp/LSDpy}{https://github.com/folsomcp/LSDpy}} Both the Stokes~$I$ and $V$ spectra are deconvolved with a line list, to obtain individual, high-signal-to-noise ratio (S/N) profiles summarising the properties of hundreds of spectral lines. The line list contains absorption lines in the stellar spectrum and with the associated line features such as depth, and sensitivity to Zeeman effect (commonly known as the Land\'e factor $g_\mathrm{eff}$). For our stars, we used a synthetic line list corresponding to a local thermodynamic equilibrium model \citep{Gustafsson2008} characterised by $\log g=$ 5.0\,[cm s$^{-2}$], $v_{\mathrm{micro}}=$ 1\,km s$^{-1}$, and $T_{\mathrm{eff}}=3500$\,K. The adopted line list contains 1400 atomic photospheric lines between 950--2600\,nm with depth larger than 3\,\% the continuum level. The masks were synthesised using the Vienna Atomic Line Database\footnote{\url{http://vald.astro.uu.se/} using the Montpellier mirror to request locally MARCS model atmospheres.} \citep[VALD,][]{Ryabchikova2015}. We set the LSD normalisation wavelength and $g_\mathrm{eff}$ to 1700~nm and 1.2, respectively.

We computed the disk-integrated, line-of-sight projected component of the large-scale magnetic field following \citet{Donati1997}. We used the general formula as in \citet{Cotton2019}, 
\begin{equation}
\mathrm{B}_\ell = \frac{h}{\mu_B \lambda_0 \mathrm{g}_{\mathrm{eff}}}\frac{\int vV(v)dv}{\int(I_c-I(v))dv} \,,
\label{eq:Bl}
\end{equation}
where $\lambda_0$ and $\mathrm{g}_\mathrm{eff}$ are the normalisation wavelength and Land\'e factor of the LSD profiles, $V$ is the LSD Stokes profile in circular polarisation, $I$ is the LSD Stokes profile in unpolarised light, $I_c$ is the continuum level, $v$ is the radial velocity associated to a point in the spectral line profile in the star's rest frame, $h$ is the Planck's constant and $\mu_B$ is the Bohr magneton. To express it as in \citet{Rees1979} and \citet{Donati1997}, one can use $hc/\mu_B=0.0214$\,Tm, where $c$ is the speed of light in m\,s$^{-1}$.

\begin{figure}[!t]
    \includegraphics[width=\columnwidth]{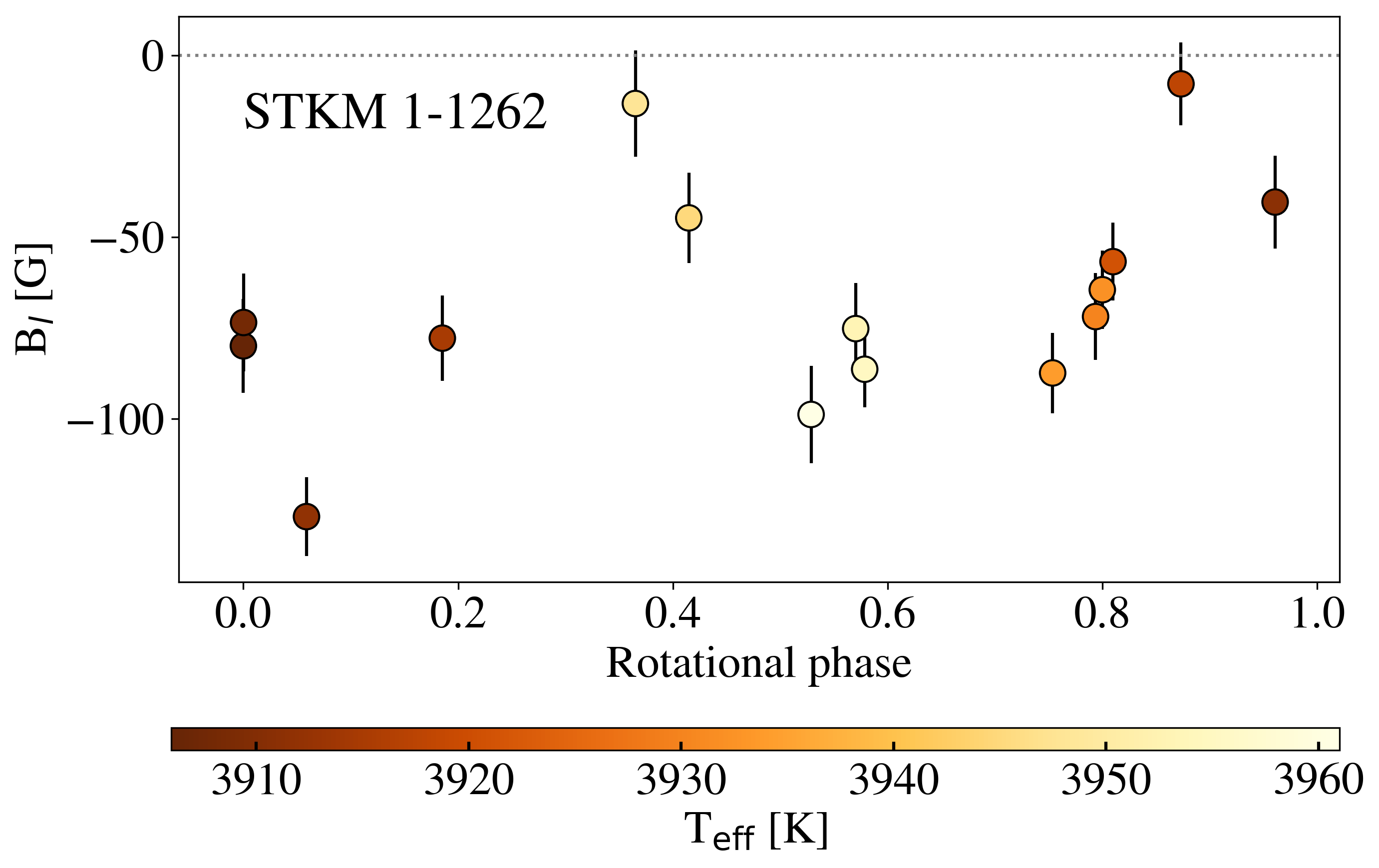}
    \includegraphics[width=\columnwidth]{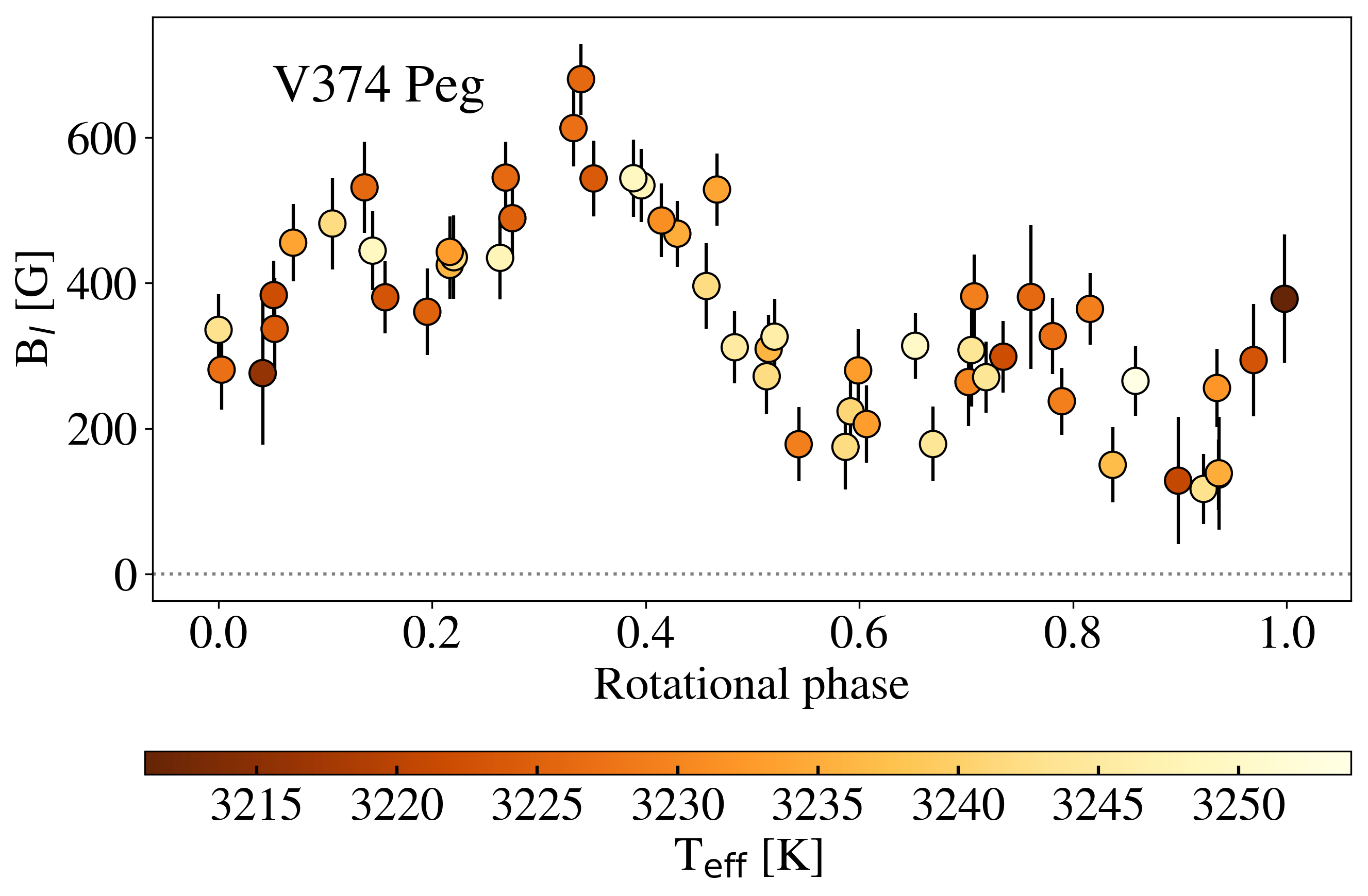}
    \caption{Longitudinal magnetic field measurements of StKM~1-1262 and V374~Peg. The panels show B$_\ell$ as a function of rotational phase for StKM~1-1262 and V374~Peg. The rotational phases are computed according to Eq.~\ref{eq:ephemeris}, using the rotational period listed in Table~\ref{tab:stars_properties}.}
    \label{fig:Bl}
\end{figure}

For StKM~1-1262, the computation of B$_l$ was carried out within $\pm55$\,km\,s$^{-1}$ from line centre at $-3.9$\,km\,s$^{-1}$ for both Stokes~$I$ and $V$ LSD profiles. As shown in Fig.~\ref{fig:Bl}, the values range between $-127$ and $-8$\,G, with a median value of $-74$\,G and a median error bar of $12$\,G. The error bars are estimated from formal propagation of Eq.~\ref{eq:Bl}. The phase-folded B$_\ell$ time series does not feature an evident modulation at the stellar rotation period, but rather higher frequency variations around the first harmonic of P$_\mathrm{rot}$. When the B$_\ell$ data points are colour-coded based on T$_\mathrm{eff}$ in a similar manner as in Fig.~\ref{fig:BI}, we do not observe a correlation (Pearson $\rho=0.01$, see also Fig.~\ref{fig:correlations}).

For V374~Peg, we set an integration range of $\pm80$\,km\,s$^{-1}$ from line centre at $-3.0$\,km\,s$^{-1}$, and we found values between 117 and 680\,G, with a median of 330\,G and a median error bar of 50\,G (see Fig.~\ref{fig:Bl}). The modulation of the B$_\ell$ time series sees an increase around phase 0.3 and a decrease around phase 0.8, but the values are always positive. This can indicate a tilted dipolar configuration of the large-scale field, as it will be explained in Sect.~\ref{sec:ZDI}. When colour-coded by T$_\mathrm{eff}$, the B$_\ell$ data points do not reveal any evident correlation (Pearson $\rho=-0.08$), as also shown in Fig.~\ref{fig:correlations}.

\begin{figure}[!t]
    \includegraphics[width=\columnwidth]{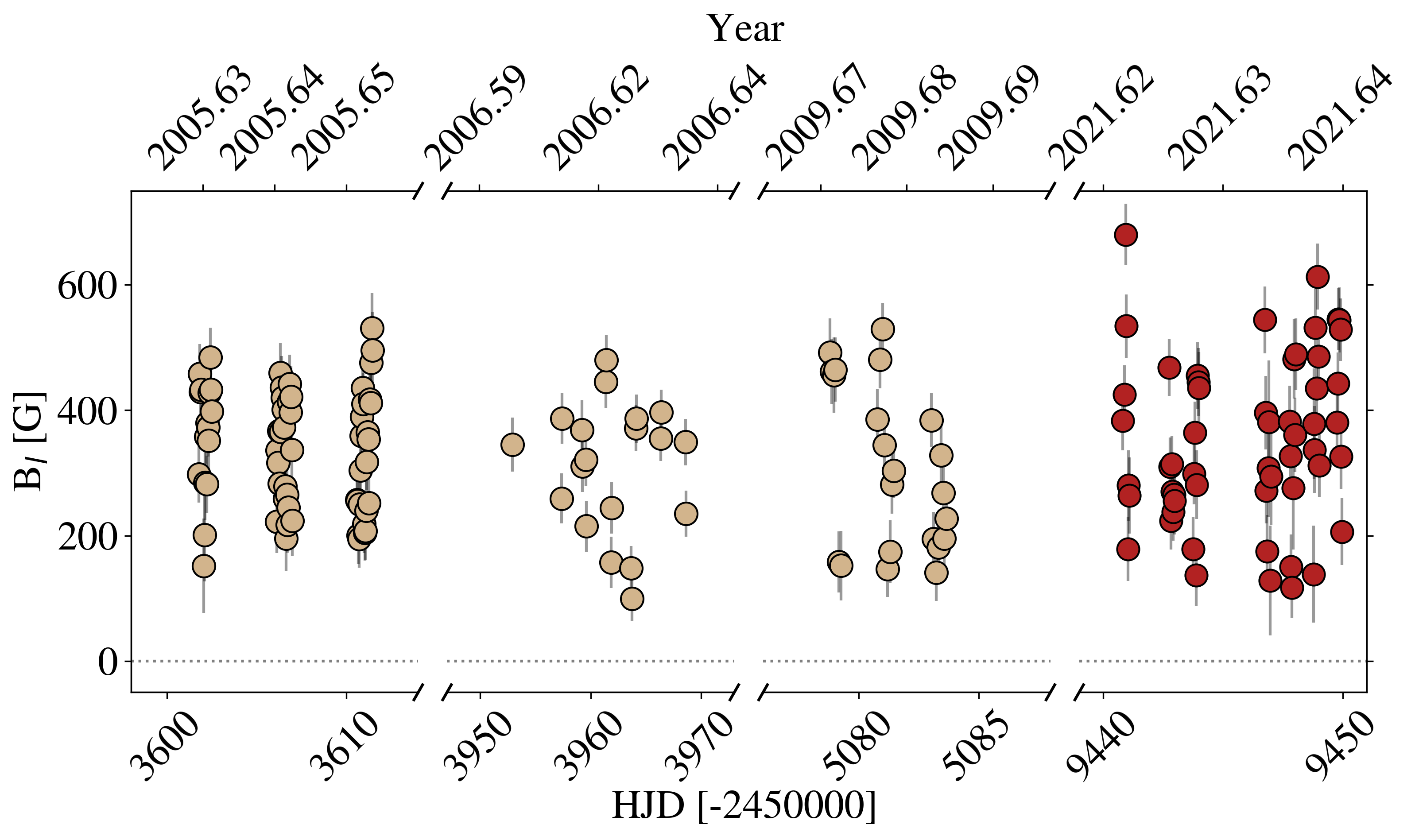}
    \caption{Long-term measurements of V374~Peg's longitudinal magnetic field. The leftmost epochs (couloured with light brown) represent the August 2005, August 2006, and September 2009 time series analysed by \citet{Morin2008a}. The rightmost (dark brown) time series corresponds to the data analysed in this work.}
    \label{fig:Bl_longterm}
\end{figure}

Our range of B$_\ell$ measurements are consistent with previous ESPaDOnS observations performed by \citet{Morin2008a}, as shown in Fig.~\ref{fig:Bl_longterm}. Their observations in August 2005, August 2006, and September 2009 resulted in average B$_\ell$ values of 338\,G, 309\,G, and 307\,G. Combined with our observations, we do not see significant variations in the mean and range of B$_\ell$ values, potentially suggesting that the large-scale magnetic field has remained stable. 

\section{Magnetic field topology}\label{sec:ZDI}

The characterisation of the large-scale magnetic field of StKM~1-1262 and V374~Peg was performed with Zeeman-Doppler imaging \citep[][]{Semel1989,DonatiBrown1997}. The algorithm compares observed and model Stokes~$V$ LSD profiles iteratively, fitting the spherical harmonics coefficients $\alpha_{\ell,m}$, $\beta_{\ell,m}$, and $\gamma_{\ell,m}$ (with $\ell$ and $m$ the degree and order of the mode, respectively), until they match within a target reduced $\chi^2$. Because the inversion problem is ill-posed, a maximum-entropy regularisation scheme is applied to obtain the field map compatible with the data and with the lowest information content \citep[for more details see][]{Skilling1984,DonatiBrown1997}. The magnetic field vector is modelled as the sum of a poloidal and a toroidal component, which are both expressed with spherical harmonics formalism \citep{Lehmann2022}. We used the python implementation \texttt{zdipy} with the inclusion of the Unno-Rachkovsky's solutions to polarised radiative transfer equations \citep[see][]{Folsom2018,Bellotti2024a} and the filling factor formalism of \citet{Morin2008}.

The parameters of the Unno-Rachkovsky models \citep[see e.g.][]{delToroIniesta2003,Landi2004} are the Gaussian width ($w_G$), the Lorentzian width ($w_L$), the ratio of the line to continuum absorption coefficients ($\eta_0$), and the slope of the source function in the Milne-Eddington atmosphere ($\beta$). More information on their implementation in the \texttt{zdipy} code can be found in \citet{Erba2024} and \citet{Bellotti2024a}. We performed a $\chi^2$ minimisation between synthetic and observed Stokes~$I$ profiles for a grid of parameters, in order to find the set of parameters that best represents our observations. For StKM~1-1262, we found $w_G=9.4$~km\,s$^{-1}$, $w_L=8.0$~km\,s$^{-1}$, and $\eta_0=13.0$, and for V374~Peg, we found $w_G=9.0$~km\,s$^{-1}$, $w_L=10.0$~km\,s$^{-1}$, and $\eta_0=12.2$. The value of $\beta$ is fixed to $0.25$ \citep[for more details, see][]{Erba2024, Bellotti2024a}. The filling factors f$_I$ and f$_V$ represent the fraction of the cell of the stellar surface grid covered by magnetic regions and magnetic regions producing net circular polarisation, respectively \citep{Morin2008,Kochukhov2021}. We attempted to constrain such parameters with a $\chi^2$ minimisation but the results were inconclusive, so we fixed both filling factors to 1.0 for both stars.

\subsection{Stellar input parameters}

The ZDI reconstruction requires physical and geometrical parameters of the star such as: inclination, projected equatorial velocity $v_\mathrm{eq}\sin i$, rotation period, differential rotation rate, limb darkening coefficient and maximum degree of spherical harmonics decomposition ($\ell_\mathrm{max}$). The linear limb darkening coefficient was fixed to 0.2, the value corresponding $H$ band observations \citep{Claret2011}.

For $v_\mathrm{eq}\sin i$, we set the values obtained from the spectral fitting procedure (see Sect.~\ref{sec:ZBro}), namely 25.04\,km\,s$^{-1}$ for StKM1~1-1262 and 36.73\,km\,s$^{-1}$ for V374~Peg. The value of $v_\mathrm{eq}\sin i$ determines the attainable spatial resolution encapsulated in $\ell_\mathrm{max}$ \citep[e.g.][]{Morin2008a}. We conservatively set $\ell_\mathrm{max}$ to 10 for the two stars, since most of the magnetic energy is stored in the $\ell\leq7$ modes. 

We attempted a differential rotation search for both stars as described in \citet{Petit2002} but the results were inconclusive, so we assumed solid body rotation. The rotation period of StKM~1-1262 was estimated to be $1.240\pm0.003$\,d from TESS photometry \citep{Colman2024}, and for V374~Peg we adopted a value of 0.4455\,d \citep[see][]{Morin2008a}. All these values are summarised in Table~\ref{tab:stars_properties}.

Finally, the inclination of the star is estimated comparing the stellar radius available in the literature with the value obtained geometrically from rotation period and $v_\mathrm{eq}\sin i$. Formally, $R\sin i=\mathrm{P}_\mathrm{rot}v_\mathrm{eq}\sin i/50.59$ where the denominator accounts for the unit conversion of the variables involved ($\mathrm{P}_\mathrm{rot}$ in d, $v_\mathrm{eq}\sin(i)$ in km\,s$^{-1}$, and $R$ in $R_\odot$). For StKM~1-1262, we found $R\sin i=0.633\pm0.006$~R$_\odot$ while \citet{Kervella2022} reported $R=0.630$~R$_\odot$, indicating that the viewing angle of the star is likely equator-on (i.e. $90^{\circ}$). As input for ZDI, we assumed 70$^{\circ}$ to prevent mirroring effects between the north and south hemispheres of the star. For V374~Peg we adopted a value of 70$^{\circ}$ consistently with \citet{Morin2008a}.

\subsection{Reconstruction}

\begin{figure}[t]
    \includegraphics[width=\columnwidth]{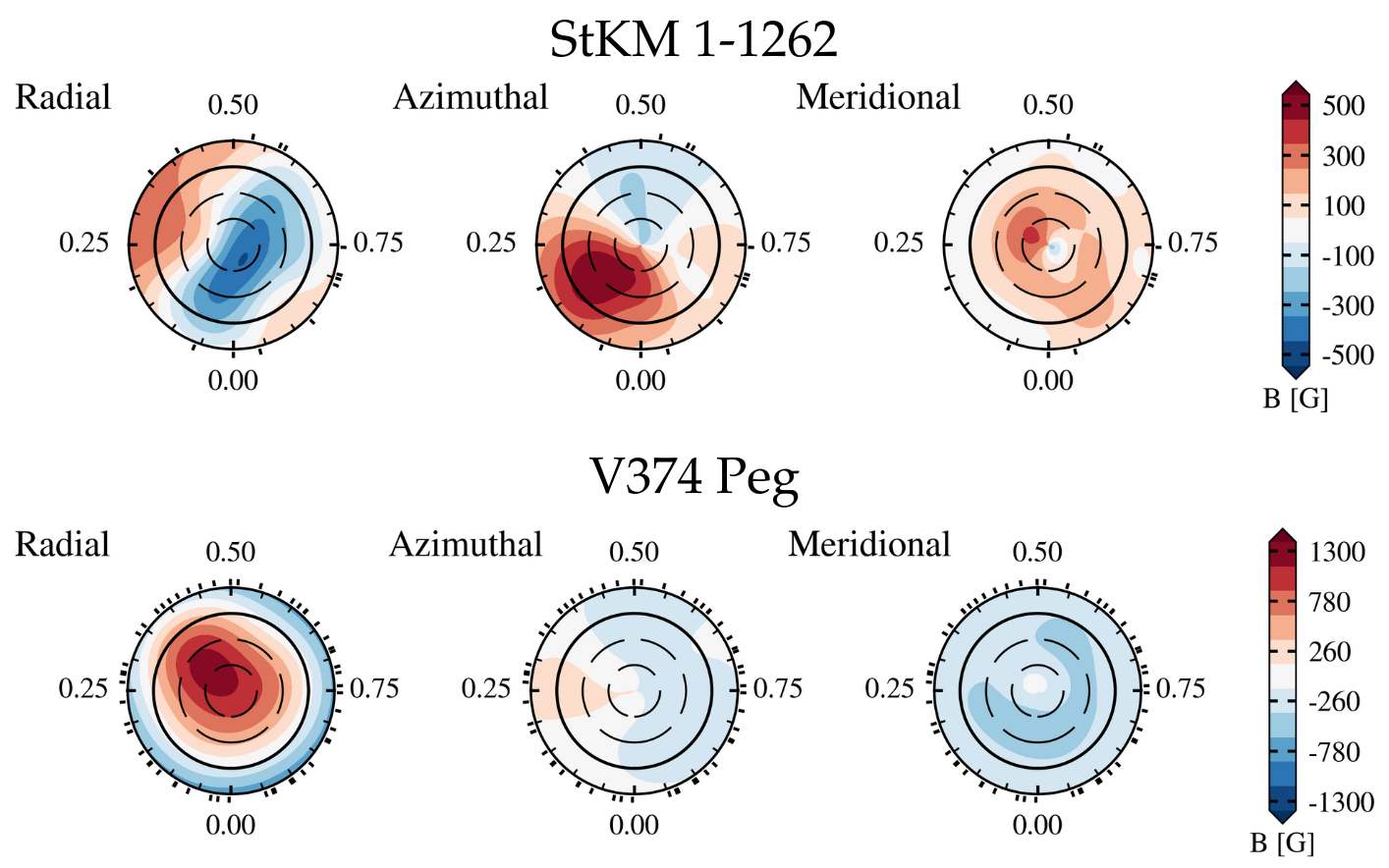}
    \caption{Reconstructed large-scale magnetic field maps in flattened polar view. StKM~1-1262 is shown in the top row and V374~Peg in the bottom row. From the left, the radial, azimuthal, and meridional components of the magnetic field vector are illustrated. Concentric circles represent different stellar latitudes: -30\,$^{\circ}$, +30\,$^{\circ}$, and +60\,$^{\circ}$ (dashed lines), as well as the equator (solid line). The radial ticks are located at the rotational phases when the observations were collected. The rotational phases are computed with Eq.~\ref{eq:ephemeris} using the first observation of each individual epoch (see Table~\ref{tab:log}). The colour bar indicates the polarity and strength (in G) of the magnetic field.}
    \label{fig:zdi_maps}
\end{figure}

The Stokes~$V$ profiles of StKM~1-1262 and their ZDI models are shown in Fig.~\ref{fig:stokesV_StKM} and the ZDI map is shown in Fig.~\ref{fig:zdi_maps}. We fitted the observed Stokes~$V$ LSD profiles down to $\chi^2_r=1.0$. As reported in Table~\ref{tab:zdi_output}, the large-scale magnetic geometry is predominantly poloidal, with a strong dipolar component accounting for 75\% of the poloidal magnetic energy and also a significant (20\%) quadrupolar component. The toroidal field is also substantial, containing 13\% of the magnetic energy. The magnetic field is tilted, with only 47\% of the total energy stored in the axisymmetric modes (that is, $m=0$). This mainly follows the fact that the poloidal component is non-axisymmetric (41\%), whereas the toroidal component is axisymmetric (82\%). The average magnetic field strength is 300\,G.

The Stokes~$V$ profiles of V374~Peg are shown in Fig.~\ref{fig:stokesV_vpeg} and the ZDI map is shown in Fig.~\ref{fig:zdi_maps}. We fitted the observed Stokes~$V$ LSD profiles to a $\chi^2_r$ of 1.0. The large-scale magnetic geometry is predominantly poloidal ($99\%$), dipolar ($97\%$), and axisymmetric ($92\%$). The average magnetic field strength is 780\,G (see Table~\ref{tab:zdi_output}).

\begin{table*}[!t]
\caption{Properties of the magnetic maps.} 
\label{tab:zdi_output}     
\centering                       
\begin{tabular}{l c c c c c c c c c c c c}      
\toprule
Star & $\chi^2_r$ & $\langle|$B$_V|\rangle$   & $|$B$_\mathrm{max}|$ & $\langle B^2\rangle$ & $f_\mathrm{pol}$ & $f_\mathrm{tor}$  & $f_\mathrm{dip}$   & $f_\mathrm{quad}$  & $f_\mathrm{oct}$   & $f_\mathrm{axi}$ & $f_\mathrm{axi,pol}$ & $f_\mathrm{axi,tor}$ \\
& & [G] & [G] & [$\times10^5$\,G$^2$] & [\%] & [\%] & [\%] & [\%] & [\%] & [\%] & [\%] & [\%] \\
\midrule
StKM~1-1262 & 1.00 & 300 & 646 & 1.08 & 86.2 & 13.8 & 74.6 & 20.0 & 4.3 & 46.8 & 41.1 & 82.1\\
V374~Peg & 1.00 & 783 & 1396 & 7.03 & 99.0 & 1.0 & 96.8 & 2.4 & 0.8 & 92.2 & 92.2 & 92.4\\ 
\bottomrule                                
\end{tabular}
\tablefoot{The following quantities are listed: star's name, target $\chi^2_r$ of the ZDI reconstruction, mean unsigned magnetic strength, maximum unsigned magnetic strength, total reconstructed magnetic energy, poloidal and toroidal magnetic energies as a fraction of the total energy, dipolar, quadrupolar, and octupolar magnetic energy as a fraction of the poloidal energy, axisymmetric magnetic energy as a fraction of the total energy, poloidal axisymmetric energy as a fraction of the poloidal energy, toroidal axisymmetric energy as a fraction of the toroidal energy.}
\end{table*}

\section{Doppler imaging}\label{sec:DI}

Using the time series of Stokes~$I$ LSD profiles, we applied Doppler imaging to reconstruct a brightness image of the surface of StKM~1-1262 and of V374~Peg. The principle of the technique is based on the distortions that surface inhomogeneities induce on the shape of spectral lines. Owing to the Doppler effect, there is a correlation between the location of the spectral distortion on the profile and the location of the inhomogeneity on the stellar surface. The occurrence of a distortion at a certain time yields information on the longitude of the surface inhomogeneity, while the extent of the distortion across the profile informs one of the latitude of the spot. Therefore, by monitoring the temporal modulation of such distortions one can retrieve a 2D brightness map \citep{Deutsch1957,Vogt1983,Rice1989}. 

The longitudinal resolution of the reconstructed map scales with $v_\mathrm{eq}\sin i$ \citep[e.g.][]{Hussain2009,Morin2008a}. Values of $v_\mathrm{eq}\sin i \geq15-20$ km\,s$^{-1}$ typically ensure sufficient rotational broadening of the line profile and ultimately the reliability of the reconstructed map. The $v_\mathrm{eq}\sin i$ of StKM~1-1262 and V374~Peg are suitable for such reconstruction (see Table~\ref{tab:stars_properties}). We used the Doppler imaging implementation provided in the \texttt{zdipy} code and the input parameters are those described in Sect.~\ref{sec:ZDI}.

\begin{figure}[t]
    \includegraphics[width=\columnwidth]{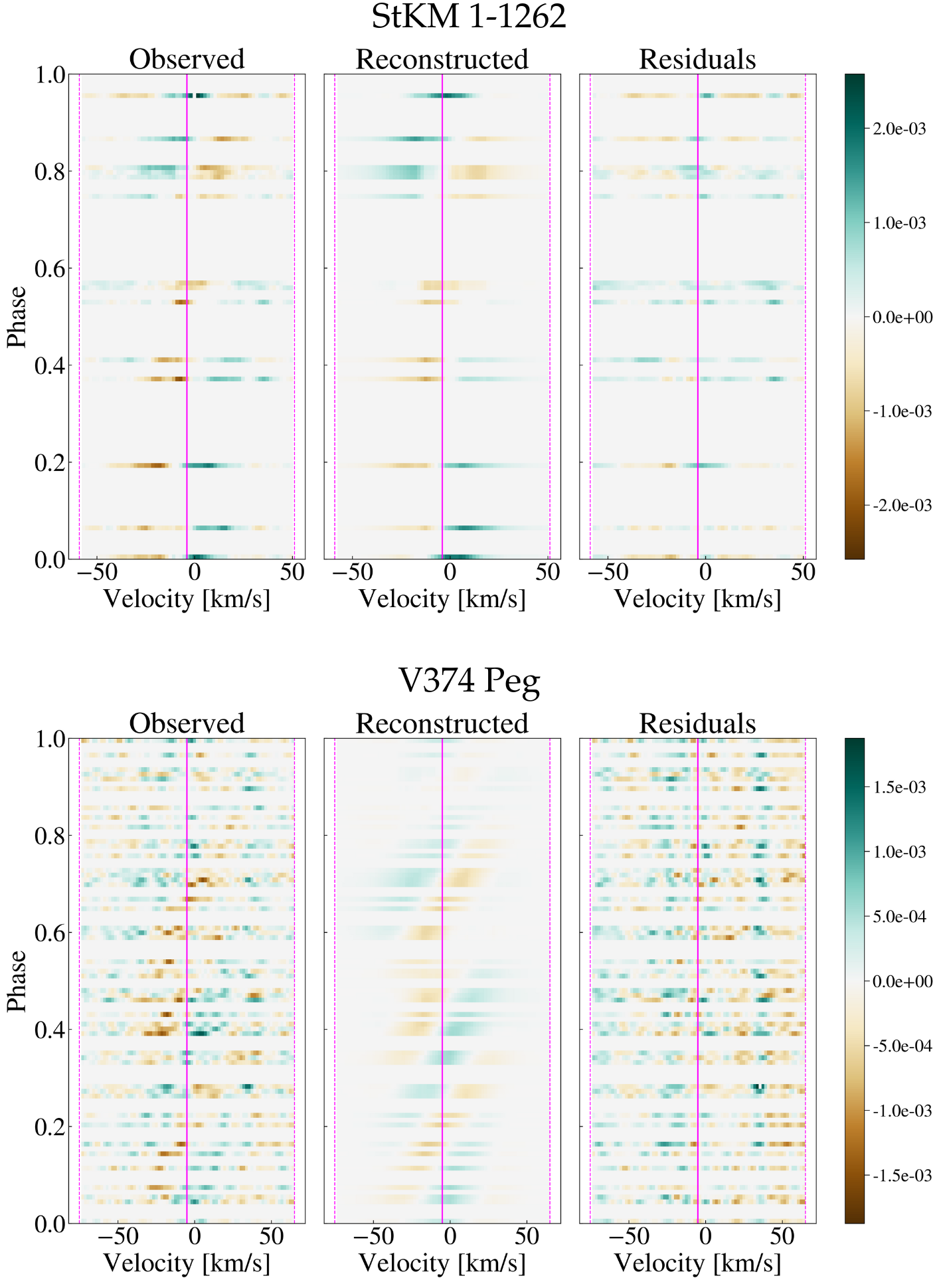}
    \caption{Dynamic spectra of StKM~1-1262 and V374~Peg. The Stokes~$I$ profiles are phase-folded according to Eq.~\ref{eq:ephemeris} and stacked vertically. From left, the panels show the observed, reconstructed and residual profiles. The colour bar indicates the flux values of Stokes~$I$ LSD profiles. In the observed and reconstructed panels, the Stokes~$I$ are median subtracted, hence positive and negative values of the colour bar indicate dark and bright features, respectively. The median subtraction removes stationary bumps, making them more evident in the residuals panel. The vertical solid line locates the radial velocity of the stars ($-3.9$~km\,s$^{-1}$ for StKM~1-1262 and $-5.0$~km\,s$^{-1}$ for V374~Peg) and the two dashed lines are located at the extremes of the intervals for the computation of B$_\ell$ and ZDI (see Sect.~\ref{sec:Blon} and Sect.~\ref{sec:ZDI}).}
    \label{fig:dynamic}
\end{figure}

For StKM~1-1262, the Stokes~$I$ data set can be fitted to a S/N level of 2800. Fig.~\ref{fig:dynamic} shows the dynamic spectrum of StKM~1-1262 corresponding to the observations (leftmost panel), model (central panel), and residuals (rightmost panel). The observations panel exhibits the signatures of a mid latitude dark spot or cluster of spots travelling over the line profiles from the blue to the red wing and crossing the line of sight around phase 0.0 (green colour), as well as a bright feature crossing at phase 0.5 (brown colour). The tilt of these signatures is reasonably the same, indicating that the brightness inhomogeneities are located at similar latitudes. The signatures are well-reproduced in the model panel, since the residuals do not feature evident systematic or rotationally modulated structure. 

\begin{figure}[t]
\centering
    \includegraphics[width=0.8\columnwidth]{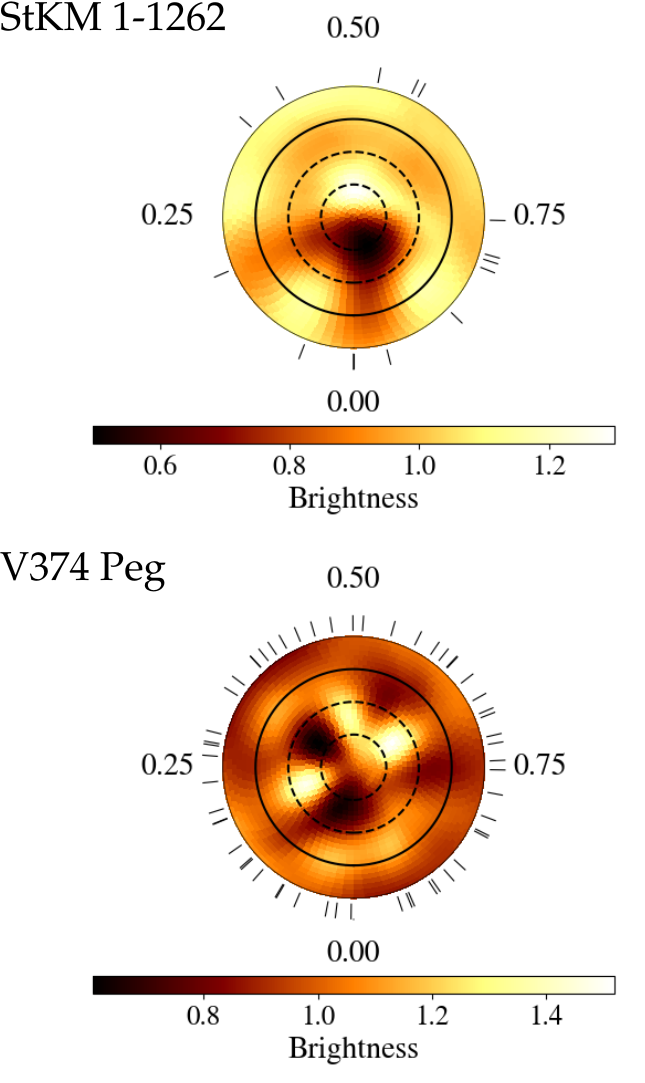}
    \caption{Reconstructed brightness maps of StKM~1-1262 and V374~Peg. The colour bar indicates the brightness relative to the quiet photosphere, which is equal to 1.0.}
    \label{fig:brimap}
\end{figure}

The brightness map of StKM~1-1262 is shown in Fig.~\ref{fig:brimap}, in which we observe consistent features as in the dynamic spectrum. There is a dark spot or a cluster of smaller ones centred around phase 0.95. This feature is located at a latitude of approximately 60$^{\circ}$ and has a surface occupancy of 9\%. The map also exhibits a bright feature at phase 0.5, which is roughly 20\% more intense than the quiet photosphere. This is slightly larger than what reported for the Sun \citep{Hirayama1979}. Overall, the observational cadence does not sample the longitudes of StKM~1-1262 densely and the attainable spatial resolution is limited by $v_\mathrm{eq}\sin i$ for Doppler imaging, therefore we caution to not over-interpret the brightness map. When compared to the results of Zeeman broadening modelling, we observe a striking correlation between the reconstructed brightness features and the rotationally modulated total magnetic flux and effective temperature derived from spectral line modelling (see Sect.~\ref{sec:ZBro} and Fig.~\ref{fig:BI}). Specifically, we observe the presence of a dark spot or a cluster of small ones at phase 0.0, which is where the total magnetic flux is highest and the effective temperature is smallest. In Appendix~\ref{app:lsd}, we describe the simultaneous modelling of Stokes~$I$ and $V$ LSD profiles.

For V374~Peg, the Stokes~$I$ data set can be fitted to a S/N level of 2700. The dynamic spectra shown in Fig.~\ref{fig:dynamic} exhibit similarities with those of \citet{Morin2008a}, with multiple brightness inhomogeneities crossing the line of sight. Signatures of dark spots are present around phases 0.0 and 0.4, and bright features at phases 0.2, 0.5, and 0.7. The rotationally modulated distortions travelling across the Stokes~$I$ profile are well-reproduced by the model, albeit a systematic discrepancy visible especially in the right part of the residuals (the red wing part of the line profile) as vertical bands. As already noted by \citet{Morin2008a}, this discrepancy does not affect the imaging process. 

Our reconstructed brightness map of V374~Peg is illustrated in Fig.~\ref{fig:brimap} and it is consistent with the dynamic spectra. We note two dark regions around latitudes 50-60$^{\circ}$ with roughly 20\% less intensity than the quiet photosphere, and three bright regions at similar latitudes with up to 40\% larger intensity. While it is not possible to compare this Doppler map with those of \citet{Morin2008a} exactly, because our model accounts for both dark and bright regions, there is an overall agreement in terms of multitude of inhomogeneities and complexity of the surface on V374~Peg. The fact that the brightness maps of V374~Peg are consistent, despite being about 14~yr apart, is also consistent with the analysis of photometric light curves of \citet{Vida2016} which suggests a spot configuration that is stable over about 16\,yr. In the case of V374~Peg, the comparison between the brightness map and the time series of total magnetic flux and effective temperature is less straightforward than StKM~1-1262, due to the lack of a clear rotational modulation of $\langle B_I \rangle$ and T$_\mathrm{eff}$.

\section{Discussion and conclusions}\label{sec:conclusions}

We have analysed SPIRou spectropolarimetric time series observations in circular polarisation mode of two stars, StKM~1-1262 and V374~Peg. Our aim is to provide robust and updated constraints on the magnetic field of the two stars, which is a fundamental ingredient to understand the mechanisms driving their detected radio emission. For StKM~1-1262, the detected radio emission has identical frequency and polarisation properties as fundamental plasma emission from a Solar Type~II burst, likely representing one of the few evidences of an extrasolar type-II burst \citep{Tasse2021,Callingham2025}. Such detection suggests that the CMEs can occur for stars whose field is substantially stronger than the Sun's \citep{AlvaradoGomez2018,Strickert2024}. The magnetic field characterisation of StKM~1-1262 performed in this work has been used in \citet{Callingham2025} to support such conclusion.

\begin{figure*}[t]
    \centering
    \includegraphics[width=\textwidth]{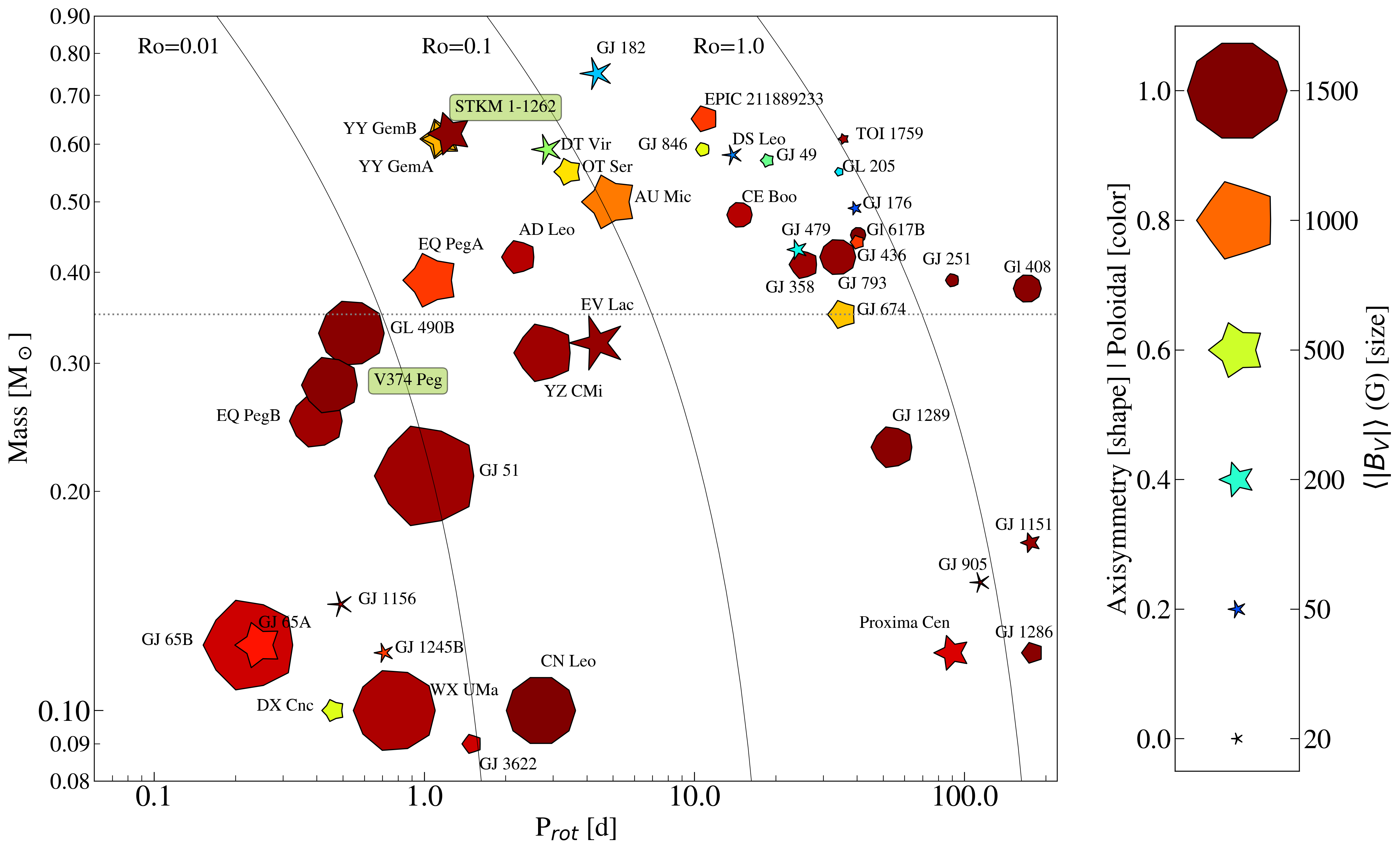}
    \caption{Properties of the magnetic topologies for cool, main-sequence stars obtained via Zeeman-Doppler Imaging. The locations of StKM~1-1262 and V374~Peg are highlighted. The $y$- and $x$-axes represent the mass and rotation period of the star. Gray curves illustrate different Rossby numbers ($Ro=0.01$, 0.1 and 1.0) and are obtained using the empirical relations of \citet{Wright2018}. The Rossby number is the ratio between the rotation period of a star and the convective turnover time. The symbol size, colour and shape encodes the ZDI average field strength, poloidal/toroidal energy fraction and axisymmetry \citep[see][for the data used]{Petit2008,Morgenthaler2012,Fares2013,BoroSaikia2015,DoNascimento2016,BoroSaikia2016,Hussain2016,Alvarado-Gomez2015,Folsom2016,Folsom2018,Kochukhov2019,Folsom2020,Klein2021,Willamo2022,Martioli2022,Bellotti2023a,Lehmann2024}. For stars with multiple reconstructions, the data represent the most recent ZDI maps.}
    \label{fig:megaconfusogram}%
\end{figure*}

From the unpolarised spectra, we modelled Zeeman broadening and intensification, constraining the total unsigned magnetic field $\langle B_I\rangle$ to be $3.53\pm0.06$\,kG for StKM~1-1262 and $5.46\pm0.09$\,kG for V374~Peg. In terms of mass and rotation period, StKM~1-1262 falls close to the binary system YY~Gem, whose magnetic field characterisation was carried out by \citet{Kochukhov2019}. The authors measured a total magnetic field of 3.2-3.4~kG for the two stars in the binary, which is compatible with our value for StKM~1-1262. Our results are also similar to the early-M~dwarf OT~Ser, with a total unsigned magnetic field of $3.2$\,kG \citep{Shulyak2019}. For V374~Peg, our measurement is compatible with the estimate of $4.4\pm1.0$~kG by \citet{Shulyak2019} within uncertainties, and it is also consistent with stars in the vicinity of V374~Peg in terms of mass and rotation period such as EQ~Peg~B \citep[$4.2\pm1.0$~kG;][]{Shulyak2017}.

We used the spectral fitting procedure to derive values of surface temperature T$_\mathrm{eff}$ and $\langle B_I\rangle$ simultaneously. We found a strong anti-correlation between total magnetic flux and effective temperature for StKM~1-1262, which is in agreement with the tight connection between the stellar surface temperature variations and the small-scale magnetic field, as already pointed out in \citet{Artigau2024} and \citet{Cristofari2025}. For V374~Peg, we only obtained a moderate anti-correlation, which is most likely due to a combination of factors such as the low S/N of the observations, the larger $v_\mathrm{eq}\sin(i)$ (hence rotational broadening), the low contrast between spots and quiet photosphere (of about 300~K), and potentially a more dense and homogeneous distribution of surface features as opposed to StKM~1-1262. The combined action of several surface spots distributed homogeneously could also explain the lack of a clear rotational modulation of T$_\mathrm{eff}$ and $\langle B_I\rangle$ for V374~Peg. This would be consistent with the fact that V374~Peg falls in the supersaturated regime of the activity-rotation relation \citep{Wright2011}, for which coronal X-ray emission decreases below the saturation level of $L_X/L_\mathrm{bol}\sim10^{-3}$ with increasing rotation \citep[e.g.][]{Prosser1996}. In fact, although V374~Peg is expected to be more active than StKM~1-1262, its X-ray emission is $L_X=2.75\times10^{28}$~erg/s \citep{Delfosse1998}, which is about one order of magnitude lower than that of StKM~1-1262 \citep[$L_X=2.20\times10^{29}$~erg/s][]{Agueros2009}. In this scenario, V374~Peg's surface would be completely covered with active regions in contrast to StKM~1-1262.

From the circularly polarised spectra, we first measured the longitudinal field. We found an average of $-67$\,G for StKM~1-1262 and 351\,G for V374~Peg, consistently showing a difference in magnetic activity level between the two stars. We then reconstructed the large-scale magnetic field by means of Zeeman-Doppler imaging, and the results are contextualised in Fig.~\ref{fig:megaconfusogram}. We obtained predominantly poloidal-dipolar configurations with moderate axisymmetry for StKM~1-1262 and high axisymmetry for V374~Peg. The average magnetic field is 300\,G and 800\,G for the two stars. The ratio $\langle|B_V|\rangle/\langle B_I\rangle$ between the magnetic field strength as recovered by circularly polarised (that is, ZDI) and unpolarised (that is, Zeeman broadening modelling) light is 8.5\% for StKM~1-1262 and 15\% for V374~Peg. Values of up to 10\% are typical for partly-convective M~dwarfs \citep{Reiners2009b,Morin2010,Kochukhov2021}, and up to 20\% for fully convective M~dwarfs depending on the large-scale field geometry \citep[see e.g.][]{Kochukhov2021}. Both our findings are therefore within expectations.

Our ZDI reconstruction of StKM~1-1262 is in overall agreement with stars of similar properties like YY~Gem \citep{Kochukhov2019}. The situation is similar for V374~Peg, since \citet{Morin2008a} also reconstructed a largely poloidal-dipolar field with high degree of axisymmetry. This suggests that the large-scale magnetic field of V374~Peg is stable over time, but the wide observational gap of approximately 10~yr observations prevent us from being conclusive. Additional observations conducted on a yearly basis will provide a definitive answer.

We also reconstructed the brightness map of both stars. For StKM~1-1262, we found the surface to be characterised by a dark spot lying approximately at rotational phase 0.0 (see Fig.~\ref{fig:brimap}). The structure of the brightness map is corroborated by the time series of $\langle B_I\rangle$ and T$_\mathrm{eff}$, since their rotational modulation correlates with the presence of such starspot. For V374~Peg, we reconstructed a more complex map, with multiple dark and bright regions, which is in overall agreement with previous reconstructions \citep{Morin2008a} and the view that this star has a more homogeneous spot distribution. In this case, we did not observe a clear rotational modulation of $\langle B_I\rangle$ or T$_\mathrm{eff}$, hence connecting them to the Doppler map is less straightforward. From our analysis of the $\langle B_I\rangle$-T$_\mathrm{eff}$ anti-correlation, we obtained spot coverages of 25\% and 48\% for StKM~1-1262 and V374~Peg, which are higher than the mean spot coverage reconstructed via Doppler imaging of 9\% and 17\%, respectively. Such discrepancy could stem from the fact that the temperature variations studied in Sect.~\ref{sec:ZBro} may be the result of large spots as well as several small ones, making the empirical relation in Eq.~\ref{eq:spot_coverage} more sensitive to different spatial scales than Doppler imaging. In fact, the spatial resolution attainable by Doppler imaging, which is further deteriorated by S/N considerations, is mostly limited to big spots or small spots clustered together.

Finally, the LoTTS survey has detected several low-mass stars with radio observations, regardless of the magnetic activity level of the star \citep{Callingham2021}. Shedding more light on the physical mechanism driving the radio emission, whether planet- or plasma-induced for instance, requires a robust constraint on the magnetic field. Therefore, spectropolarimetric observations of these low-mass stars are required to fully understand their magnetism and plasma environment.

\begin{acknowledgements}

We would like to thank Prof.~B.~Pope for the contribution to the SPIRou observing proposal of StKM~1-1262 and the comments for this manuscript.
This publication is part of the project "Exo-space weather and contemporaneous signatures of star-planet interactions" (with project number OCENW.M.22.215 of the research programme "Open Competition Domain Science- M"), which is financed by the Dutch Research Council (NWO). PC and AAV acknowledge funding from the European Research Council (ERC) under the European Union's Horizon 2020 research and innovation programme (grant agreement No 817540, ASTROFLOW). JRC acknowledges funding from the European Union via the European Research Council (ERC) grant Epaphus (project number 101166008)
AAV acknowledges funding from the Dutch Research Council (NWO), with project number VI.C.232.041 of the Talent Programme Vici. MMJ acknowledges support from STFC consolidated grant number ST/R000824/1.
Based on observations obtained at the Canada-France-Hawaii Telescope (CFHT) which is operated by the National Research Council of Canada, the Institut National des Sciences de l'Univers of the Centre National de la Recherche Scientique of France, and the University of Hawaii. This work has made use of the VALD database, operated at Uppsala University, the Institute of Astronomy RAS in Moscow, and the University of Vienna; Astropy, 12 a community-developed core Python package for Astronomy \citep{Astropy2013,Astropy2018}; NumPy \citep{VanderWalt2011}; Matplotlib: Visualization with Python \citep{Hunter2007}; SciPy \citep{Virtanen2020} and PyAstronomy \citep{Czesla2019}.

\end{acknowledgements}


\bibliographystyle{aa}
\bibliography{biblio}

@ARTICLE{Callingham2025,
       author = {{Callingham}, J.~R. and {Tasse}, C. and {Keers}, R.},
      journal = {Nature},
         year = 2025,
       volume = {accepted},
}

@ARTICLE{Tasse2021,
       author = {{Tasse}, C. and {Shimwell}, T. and {Hardcastle}, M.~J. and {O'Sullivan}, S.~P. and {van Weeren}, R. and {Best}, P.~N. and {Bester}, L. and {Hugo}, B. and {Smirnov}, O. and {Sabater}, J. and {Calistro-Rivera}, G. and {de Gasperin}, F. and {Morabito}, L.~K. and {R{\"o}ttgering}, H. and {Williams}, W.~L. and {Bonato}, M. and {Bondi}, M. and {Botteon}, A. and {Br{\"u}ggen}, M. and {Brunetti}, G. and {Chy{\.z}y}, K.~T. and {Garrett}, M.~A. and {G{\"u}rkan}, G. and {Jarvis}, M.~J. and {Kondapally}, R. and {Mandal}, S. and {Prandoni}, I. and {Repetti}, A. and {Retana-Montenegro}, E. and {Schwarz}, D.~J. and {Shulevski}, A. and {Wiaux}, Y.},
        title = "{The LOFAR Two-meter Sky Survey: Deep Fields Data Release 1. I. Direction-dependent calibration and imaging}",
      journal = {\aap},
         year = 2021,
        month = apr,
       volume = {648},
          eid = {A1},
        pages = {A1},
          doi = {10.1051/0004-6361/202038804},
archivePrefix = {arXiv},
       eprint = {2011.08328},
 primaryClass = {astro-ph.IM},
       adsurl = {https://ui.adsabs.harvard.edu/abs/2021A&A...648A...1T}
}

@article{berdyugina-2005,
	adsurl = {https://ui.adsabs.harvard.edu/abs/2005LRSP....2....8B},
	author = {{Berdyugina}, Svetlana V.},
	doi = {10.12942/lrsp-2005-8},
	eid = {8},
	journal = {Living Reviews in Solar Physics},
	month = dec,
	number = {1},
	pages = {8},
	title = {{Starspots: A Key to the Stellar Dynamo}},
	volume = {2},
	year = 2005}

@ARTICLE{Cristofari2025,
       author = {{Cristofari}, P.~I. and {Donati}, J.-F. and {Bellotti}, S. and {Artigau}, {\'E}. and {Carmona}, A. and {Moutou}, C. and {Delfosse}, X. and {Petit}, P. and {Finociety}, B. and {Dias do Nascimento}, J.},
        title = "{Rotational modulation and long-term evolution of the small-scale magnetic fields of M dwarfs observed with SPIRou}",
      journal = {\aap},
         year = 2025,
        month = oct,
       volume = {702},
          eid = {A111},
        pages = {A111},
          doi = {10.1051/0004-6361/202554902},
archivePrefix = {arXiv},
       eprint = {2508.04569},
 primaryClass = {astro-ph.SR},
       adsurl = {https://ui.adsabs.harvard.edu/abs/2025A&A...702A.111C}
}

@ARTICLE{Callingham2024,
       author = {{Callingham}, J.~R. and {Pope}, B.~J.~S. and {Kavanagh}, R.~D. and {Bellotti}, S. and {Daley-Yates}, S. and {Damasso}, M. and {Grie{\ss}meier}, J. -M. and {G{\"u}del}, M. and {G{\"u}nther}, M. and {Kao}, M.~M. and {Klein}, B. and {Mahadevan}, S. and {Morin}, J. and {Nichols}, J.~D. and {Osten}, R.~A. and {P{\'e}rez-Torres}, M. and {Pineda}, J.~S. and {Rigney}, J. and {Saur}, J. and {Stef{\'a}nsson}, G. and {Turner}, J.~D. and {Vedantham}, H. and {Vidotto}, A.~A. and {Villadsen}, J. and {Zarka}, P.},
        title = "{Radio signatures of star-planet interactions, exoplanets and space weather}",
      journal = {Nature Astronomy},
         year = 2024,
        month = nov,
       volume = {8},
        pages = {1359-1372},
          doi = {10.1038/s41550-024-02405-6},
archivePrefix = {arXiv},
       eprint = {2409.15507},
 primaryClass = {astro-ph.EP},
       adsurl = {https://ui.adsabs.harvard.edu/abs/2024NatAs...8.1359C}
}

@ARTICLE{Cuntz2015,
       author = {{Cuntz}, M.},
        title = "{S-type and P-type Habitability in Stellar Binary Systems: A Comprehensive Approach. II. Elliptical Orbits}",
      journal = {\apj},
         year = 2015,
        month = jan,
       volume = {798},
       number = {2},
          eid = {101},
        pages = {101},
          doi = {10.1088/0004-637X/798/2/101},
archivePrefix = {arXiv},
       eprint = {1409.3796},
 primaryClass = {astro-ph.SR},
       adsurl = {https://ui.adsabs.harvard.edu/abs/2015ApJ...798..101C}
}

@ARTICLE{Airapetian2020,
       author = {{Airapetian}, V.~S. and {Barnes}, R. and {Cohen}, O. and {Collinson}, G.~A. and {Danchi}, W.~C. and {Dong}, C.~F. and {Del Genio}, A.~D. and {France}, K. and {Garcia-Sage}, K. and {Glocer}, A. and {Gopalswamy}, N. and {Grenfell}, J.~L. and {Gronoff}, G. and {G{\"u}del}, M. and {Herbst}, K. and {Henning}, W.~G. and {Jackman}, C.~H. and {Jin}, M. and {Johnstone}, C.~P. and {Kaltenegger}, L. and {Kay}, C.~D. and {Kobayashi}, K. and {Kuang}, W. and {Li}, G. and {Lynch}, B.~J. and {L{\"u}ftinger}, T. and {Luhmann}, J.~G. and {Maehara}, H. and {Mlynczak}, M.~G. and {Notsu}, Y. and {Osten}, R.~A. and {Ramirez}, R.~M. and {Rugheimer}, S. and {Scheucher}, M. and {Schlieder}, J.~E. and {Shibata}, K. and {Sousa-Silva}, C. and {Stamenkovi{\'c}}, V. and {Strangeway}, R.~J. and {Usmanov}, A.~V. and {Vergados}, P. and {Verkhoglyadova}, O.~P. and {Vidotto}, A.~A. and {Voytek}, M. and {Way}, M.~J. and {Zank}, G.~P. and {Yamashiki}, Y.},
        title = "{Impact of space weather on climate and habitability of terrestrial-type exoplanets}",
      journal = {International Journal of Astrobiology},
         year = 2020,
        month = apr,
       volume = {19},
       number = {2},
        pages = {136-194},
          doi = {10.1017/S1473550419000132},
archivePrefix = {arXiv},
       eprint = {1905.05093},
 primaryClass = {astro-ph.EP},
       adsurl = {https://ui.adsabs.harvard.edu/abs/2020IJAsB..19..136A}
}

@ARTICLE{Cuntz2020,
       author = {{Cuntz}, Manfred and {Wang}, Zhaopeng},
        title = "{When more is less: The P‑type binary habitability paradox}",
      journal = {Astronomische Nachrichten},
         year = 2020,
        month = may,
       volume = {341},
       number = {4},
        pages = {402-409},
          doi = {10.1002/asna.202013758},
       adsurl = {https://ui.adsabs.harvard.edu/abs/2020AN....341..402C}
}

@ARTICLE{Seales2021,
       author = {{Seales}, Johnny and {Lenardic}, Adrian},
        title = "{The Temporal Onset of Habitability For Earth-Like Planets}",
      journal = {arXiv e-prints},
         year = 2021,
        month = jun,
          eid = {arXiv:2106.14852},
        pages = {arXiv:2106.14852},
          doi = {10.48550/arXiv.2106.14852},
archivePrefix = {arXiv},
       eprint = {2106.14852},
 primaryClass = {astro-ph.EP},
       adsurl = {https://ui.adsabs.harvard.edu/abs/2021arXiv210614852S}
}

@ARTICLE{Vida2016,
       author = {{Vida}, K. and {Kriskovics}, L. and {Ol{\'a}h}, K. and {Leitzinger}, M. and {Odert}, P. and {K{\H{o}}v{\'a}ri}, Zs. and {Korhonen}, H. and {Greimel}, R. and {Robb}, R. and {Cs{\'a}k}, B. and {Kov{\'a}cs}, J.},
        title = "{Investigating magnetic activity in very stable stellar magnetic fields. Long-term photometric and spectroscopic study of the fully convective M4 dwarf V374 Pegasi}",
      journal = {A\&A},
         year = 2016,
        month = may,
       volume = {590},
          eid = {A11},
        pages = {A11},
          doi = {10.1051/0004-6361/201527925},
archivePrefix = {arXiv},
       eprint = {1603.00867},
 primaryClass = {astro-ph.SR},
       adsurl = {https://ui.adsabs.harvard.edu/abs/2016A&A...590A..11V}
}

@ARTICLE{Ricker2014,
       author = {{Ricker}, George R. and {Winn}, Joshua N. and {Vanderspek}, Roland and {Latham}, David W. and {Bakos}, G{\'a}sp{\'a}r {\'A}. and {Bean}, Jacob L. and {Berta-Thompson}, Zachory K. and {Brown}, Timothy M. and {Buchhave}, Lars and {Butler}, Nathaniel R. and {Butler}, R. Paul and {Chaplin}, William J. and {Charbonneau}, David and {Christensen-Dalsgaard}, J{\o}rgen and {Clampin}, Mark and {Deming}, Drake and {Doty}, John and {De Lee}, Nathan and {Dressing}, Courtney and {Dunham}, Edward W. and {Endl}, Michael and {Fressin}, Francois and {Ge}, Jian and {Henning}, Thomas and {Holman}, Matthew J. and {Howard}, Andrew W. and {Ida}, Shigeru and {Jenkins}, Jon M. and {Jernigan}, Garrett and {Johnson}, John Asher and {Kaltenegger}, Lisa and {Kawai}, Nobuyuki and {Kjeldsen}, Hans and {Laughlin}, Gregory and {Levine}, Alan M. and {Lin}, Douglas and {Lissauer}, Jack J. and {MacQueen}, Phillip and {Marcy}, Geoffrey and {McCullough}, Peter R. and {Morton}, Timothy D. and {Narita}, Norio and {Paegert}, Martin and {Palle}, Enric and {Pepe}, Francesco and {Pepper}, Joshua and {Quirrenbach}, Andreas and {Rinehart}, Stephen A. and {Sasselov}, Dimitar and {Sato}, Bun'ei and {Seager}, Sara and {Sozzetti}, Alessandro and {Stassun}, Keivan G. and {Sullivan}, Peter and {Szentgyorgyi}, Andrew and {Torres}, Guillermo and {Udry}, Stephane and {Villasenor}, Joel},
        title = "{Transiting Exoplanet Survey Satellite (TESS)}",
      journal = {Journal of Astronomical Telescopes, Instruments, and Systems},
         year = 2015,
        month = jan,
       volume = {1},
          eid = {014003},
        pages = {014003},
          doi = {10.1117/1.JATIS.1.1.014003},
       adsurl = {https://ui.adsabs.harvard.edu/abs/2015JATIS...1a4003R}
}

@ARTICLE{Kopparapu2013,
       author = {{Kopparapu}, Ravi Kumar},
        title = "{A Revised Estimate of the Occurrence Rate of Terrestrial Planets in the Habitable Zones around Kepler M-dwarfs}",
      journal = {ApJl},
         year = 2013,
        month = apr,
       volume = {767},
       number = {1},
          eid = {L8},
        pages = {L8},
          doi = {10.1088/2041-8205/767/1/L8},
archivePrefix = {arXiv},
       eprint = {1303.2649},
 primaryClass = {astro-ph.EP},
       adsurl = {https://ui.adsabs.harvard.edu/abs/2013ApJ...767L...8K}
}

@ARTICLE{Rosen2012,
       author = {{Ros{\'e}n}, L. and {Kochukhov}, O.},
        title = "{How reliable is Zeeman Doppler imaging without simultaneous temperature reconstruction?}",
      journal = {A\&A},
         year = 2012,
        month = dec,
       volume = {548},
          eid = {A8},
        pages = {A8},
          doi = {10.1051/0004-6361/201219972},
archivePrefix = {arXiv},
       eprint = {1210.0789},
 primaryClass = {astro-ph.SR},
       adsurl = {https://ui.adsabs.harvard.edu/abs/2012A&A...548A...8R}
}

@ARTICLE{Cristofari2022a,
       author = {{Cristofari}, P.~I. and {Donati}, J. -F. and {Masseron}, T. and {Fouqu{\'e}}, P. and {Moutou}, C. and {Delfosse}, X. and {Artigau}, E. and {Folsom}, C.~P. and {Carmona}, A. and {Gaidos}, E. and {do Nascimento}, J. -D. and {Jahandar}, F. and {H{\'e}brard}, G.},
        title = "{Estimating fundamental parameters of nearby M dwarfs from SPIRou spectra}",
      journal = {MNRAS},
         year = 2022,
        month = apr,
       volume = {511},
       number = {2},
        pages = {1893-1912},
          doi = {10.1093/mnras/stab3679},
archivePrefix = {arXiv},
       eprint = {2112.08118},
 primaryClass = {astro-ph.SR},
       adsurl = {https://ui.adsabs.harvard.edu/abs/2022MNRAS.511.1893C}
}

@article{Cristofari2023a,
	author = {{Cristofari}, P.~I. and Donati, J-F and Folsom, C P and Masseron, T and Fouqu{\'e}, P and Moutou, C and Artigau, E and Carmona, A and Petit, P and Delfosse, X and Martioli, E and the SLS consortium},
	doi = {10.1093/mnras/stad865},
	eprint = {https://academic.oup.com/mnras/article-pdf/522/1/1342/50032228/stad865\_supplemental\_file.pdf},
	issn = {0035-8711},
	journal = {\mnras},
	month = {03},
	number = {1},
	pages = {1342-1357},
	title = {{Constraining atmospheric parameters and surface magnetic fields with ZeeTurbo: an application to SPIRou spectra}},
	url = {https://doi.org/10.1093/mnras/stad865},
	volume = {522},
	year = {2023}}

@article{Cristofari2023b,
	adsurl = {https://ui.adsabs.harvard.edu/abs/2023MNRAS.526.5648C},
	archiveprefix = {arXiv},
	author = {{Cristofari}, P.~I. and {Donati}, J. -F. and {Moutou}, C. and {Lehmann}, L.~T. and {Charpentier}, P. and {Fouqu{\'e}}, P. and {Folsom}, C.~P. and {Masseron}, T. and {Carmona}, A. and {Delfosse}, X. and {Petit}, P. and {Artigau}, E. and {Cook}, N.~J. and {SLS Consortium}},
	doi = {10.1093/mnras/stad3144},
	eprint = {2310.08386},
	journal = {\mnras},
	month = dec,
	number = {4},
	pages = {5648-5674},
	primaryclass = {astro-ph.SR},
	title = {{Measuring small-scale magnetic fields of 44 M dwarfs from SPIRou spectra with ZeeTurbo}},
	volume = {526},
	year = 2023}

@article{Landstreet1988,
	adsurl = {https://ui.adsabs.harvard.edu/abs/1988ApJ...326..967L},
	author = {{Landstreet}, J.~D.},
	bdsk-color = {3},
	doi = {10.1086/166155},
	journal = {\apj},
	month = mar,
	pages = {967},
	title = {{The Magnetic Field and Abundance Distribution Geometry of the Peculiar A Star 53 Camelopardalis}},
	volume = {326},
	year = 1988}

@article{Wade2001,
	adsurl = {https://ui.adsabs.harvard.edu/abs/2001A&A...374..265W},
	author = {{Wade}, G.~A. and {Bagnulo}, S. and {Kochukhov}, O. and {Landstreet}, J.~D. and {Piskunov}, N. and {Stift}, M.~J.},
	doi = {10.1051/0004-6361:20010735},
	journal = {\aap},
	month = jul,
	pages = {265-279},
	title = {{LTE spectrum synthesis in magnetic stellar atmospheres. The interagreement of three independent polarised radiative transfer codes}},
	volume = {374},
	year = 2001}

@misc{plez2012,
	adsurl = {https://ui.adsabs.harvard.edu/abs/2012ascl.soft05004P},
	archiveprefix = {ascl},
	author = {{Plez}, B.},
	bdsk-color = {2},
	eid = {ascl:1205.004},
	eprint = {1205.004},
	howpublished = {Astrophysics Source Code Library, record ascl:1205.004},
	month = may,
	pages = {ascl:1205.004},
	title = {{Turbospectrum: Code for spectral synthesis}},
	year = 2012}

@ARTICLE{Cook2022,
       author = {{Cook}, Neil James and {Artigau}, {\'E}tienne and {Doyon}, Ren{\'e} and {Hobson}, Melissa and {Martioli}, Eder and {Bouchy}, Fran{\c{c}}ois and {Moutou}, Claire and {Carmona}, Andres and {Usher}, Chris and {Fouqu{\'e}}, Pascal and {Arnold}, Luc and {Delfosse}, Xavier and {Boisse}, Isabelle and {Cadieux}, Charles and {Vandal}, Thomas and {Donati}, Jean-Fran{\c{c}}ois and {Desli{\`e}res}, Ariane},
        title = "{APERO: A PipelinE to Reduce Observations-Demonstration with SPIRou}",
      journal = {PASP},
         year = 2022,
        month = nov,
       volume = {134},
       number = {1041},
          eid = {114509},
        pages = {114509},
          doi = {10.1088/1538-3873/ac9e74},
archivePrefix = {arXiv},
       eprint = {2211.01358},
 primaryClass = {astro-ph.IM},
       adsurl = {https://ui.adsabs.harvard.edu/abs/2022PASP..134k4509C}
}

@ARTICLE{Donati1997,
       author = {{Donati}, J. -F. and {Semel}, M. and {Carter}, B.~D. and {Rees}, D.~E. and {Collier Cameron}, A.},
        title = "{Spectropolarimetric observations of active stars}",
      journal = {MNRAS},
         year = 1997,
        month = nov,
       volume = {291},
       number = {4},
        pages = {658-682},
          doi = {10.1093/mnras/291.4.658},
       adsurl = {https://ui.adsabs.harvard.edu/abs/1997MNRAS.291..658D}
}

@ARTICLE{Hebrard2016,
       author = {{H{\'e}brard}, {\'E}. M. and {Donati}, J. -F. and {Delfosse}, X. and {Morin}, J. and {Moutou}, C. and {Boisse}, I.},
        title = "{Modelling the RV jitter of early-M dwarfs using tomographic imaging}",
      journal = {MNRAS},
         year = 2016,
        month = sep,
       volume = {461},
       number = {2},
        pages = {1465-1497},
          doi = {10.1093/mnras/stw1346},
archivePrefix = {arXiv},
       eprint = {1606.01775},
 primaryClass = {astro-ph.SR},
       adsurl = {https://ui.adsabs.harvard.edu/abs/2016MNRAS.461.1465H}
}

@ARTICLE{Kochukhov2010a,
       author = {{Kochukhov}, O. and {Makaganiuk}, V. and {Piskunov}, N.},
        title = "{Least-squares deconvolution of the stellar intensity and polarization spectra}",
      journal = {A\&A},
         year = 2010,
        month = dec,
       volume = {524},
          eid = {A5},
        pages = {A5},
          doi = {10.1051/0004-6361/201015429},
archivePrefix = {arXiv},
       eprint = {1008.5115},
 primaryClass = {astro-ph.SR},
       adsurl = {https://ui.adsabs.harvard.edu/abs/2010A&A...524A...5K}
}

@ARTICLE{Colman2024,
       author = {{Colman}, Isabel L. and {Angus}, Ruth and {David}, Trevor and {Curtis}, Jason and {Hattori}, Soichiro and {Lu}, Yuxi (Lucy)},
        title = "{Methods for the Detection of Stellar Rotation Periods in Individual TESS Sectors and Results from the Prime Mission}",
      journal = {\aj},
         year = 2024,
        month = may,
       volume = {167},
       number = {5},
          eid = {189},
        pages = {189},
          doi = {10.3847/1538-3881/ad2c86},
archivePrefix = {arXiv},
       eprint = {2402.14954},
 primaryClass = {astro-ph.SR},
       adsurl = {https://ui.adsabs.harvard.edu/abs/2024AJ....167..189C}
}

@ARTICLE{Shimwell2022,
       author = {{Shimwell}, T.~W. and {Hardcastle}, M.~J. and {Tasse}, C. and {Best}, P.~N. and {R{\"o}ttgering}, H.~J.~A. and {Williams}, W.~L. and {Botteon}, A. and {Drabent}, A. and {Mechev}, A. and {Shulevski}, A. and {van Weeren}, R.~J. and {Bester}, L. and {Br{\"u}ggen}, M. and {Brunetti}, G. and {Callingham}, J.~R. and {Chy{\.z}y}, K.~T. and {Conway}, J.~E. and {Dijkema}, T.~J. and {Duncan}, K. and {de Gasperin}, F. and {Hale}, C.~L. and {Haverkorn}, M. and {Hugo}, B. and {Jackson}, N. and {Mevius}, M. and {Miley}, G.~K. and {Morabito}, L.~K. and {Morganti}, R. and {Offringa}, A. and {Oonk}, J.~B.~R. and {Rafferty}, D. and {Sabater}, J. and {Smith}, D.~J.~B. and {Schwarz}, D.~J. and {Smirnov}, O. and {O'Sullivan}, S.~P. and {Vedantham}, H. and {White}, G.~J. and {Albert}, J.~G. and {Alegre}, L. and {Asabere}, B. and {Bacon}, D.~J. and {Bonafede}, A. and {Bonnassieux}, E. and {Brienza}, M. and {Bilicki}, M. and {Bonato}, M. and {Calistro Rivera}, G. and {Cassano}, R. and {Cochrane}, R. and {Croston}, J.~H. and {Cuciti}, V. and {Dallacasa}, D. and {Danezi}, A. and {Dettmar}, R.~J. and {Di Gennaro}, G. and {Edler}, H.~W. and {En{\ss}lin}, T.~A. and {Emig}, K.~L. and {Franzen}, T.~M.~O. and {Garc{\'\i}a-Vergara}, C. and {Grange}, Y.~G. and {G{\"u}rkan}, G. and {Hajduk}, M. and {Heald}, G. and {Heesen}, V. and {Hoang}, D.~N. and {Hoeft}, M. and {Horellou}, C. and {Iacobelli}, M. and {Jamrozy}, M. and {Jeli{\'c}}, V. and {Kondapally}, R. and {Kukreti}, P. and {Kunert-Bajraszewska}, M. and {Magliocchetti}, M. and {Mahatma}, V. and {Ma{\l}ek}, K. and {Mandal}, S. and {Massaro}, F. and {Meyer-Zhao}, Z. and {Mingo}, B. and {Mostert}, R.~I.~J. and {Nair}, D.~G. and {Nakoneczny}, S.~J. and {Nikiel-Wroczy{\'n}ski}, B. and {Orr{\'u}}, E. and {Pajdosz-{\'S}mierciak}, U. and {Pasini}, T. and {Prandoni}, I. and {van Piggelen}, H.~E. and {Rajpurohit}, K. and {Retana-Montenegro}, E. and {Riseley}, C.~J. and {Rowlinson}, A. and {Saxena}, A. and {Schrijvers}, C. and {Sweijen}, F. and {Siewert}, T.~M. and {Timmerman}, R. and {Vaccari}, M. and {Vink}, J. and {West}, J.~L. and {Wo{\l}owska}, A. and {Zhang}, X. and {Zheng}, J.},
        title = "{The LOFAR Two-metre Sky Survey. V. Second data release}",
      journal = {\aap},
         year = 2022,
        month = mar,
       volume = {659},
          eid = {A1},
        pages = {A1},
          doi = {10.1051/0004-6361/202142484},
archivePrefix = {arXiv},
       eprint = {2202.11733},
 primaryClass = {astro-ph.GA},
       adsurl = {https://ui.adsabs.harvard.edu/abs/2022A&A...659A...1S}
}

@ARTICLE{Delfosse1998,
       author = {{Delfosse}, X. and {Forveille}, T. and {Perrier}, C. and {Mayor}, M.},
        title = "{Rotation and chromospheric activity in field M dwarfs}",
      journal = {\aap},
         year = 1998,
        month = mar,
       volume = {331},
        pages = {581-595},
       adsurl = {https://ui.adsabs.harvard.edu/abs/1998A&A...331..581D}
}

@ARTICLE{Erba2024,
       author = {{Erba}, C. and {Folsom}, C.~P. and {David-Uraz}, A. and {Wade}, G.~A. and {Seadrow}, S. and {Bellotti}, S. and {Fossati}, L. and {Petit}, V. and {Shultz}, M.~E.},
        title = "{First Observation of the Complete Rotation Period of the Ultraslowly Rotating Magnetic O Star HD 54879}",
      journal = {\apj},
         year = 2024,
        month = dec,
       volume = {977},
       number = {1},
          eid = {84},
        pages = {84},
          doi = {10.3847/1538-4357/ad865a},
archivePrefix = {arXiv},
       eprint = {2401.09722},
 primaryClass = {astro-ph.SR},
       adsurl = {https://ui.adsabs.harvard.edu/abs/2024ApJ...977...84E}
}

@ARTICLE{Ryabchikova2015,
       author = {{Ryabchikova}, T. and {Piskunov}, N. and {Kurucz}, R.~L. and {Stempels}, H.~C. and {Heiter}, U. and {Pakhomov}, Yu and {Barklem}, P.~S.},
        title = "{A major upgrade of the VALD database}",
      journal = {Phys. Scr.},
         year = 2015,
        month = may,
       volume = {90},
       number = {5},
          eid = {054005},
        pages = {054005},
          doi = {10.1088/0031-8949/90/5/054005},
       adsurl = {https://ui.adsabs.harvard.edu/abs/2015PhyS...90e4005R}
}

@ARTICLE{Wright2011,
       author = {{Wright}, Nicholas J. and {Drake}, Jeremy J. and {Mamajek}, Eric E. and {Henry}, Gregory W.},
        title = "{The Stellar-activity-Rotation Relationship and the Evolution of Stellar Dynamos}",
      journal = {ApJ},
         year = 2011,
        month = dec,
       volume = {743},
       number = {1},
          eid = {48},
        pages = {48},
          doi = {10.1088/0004-637X/743/1/48},
archivePrefix = {arXiv},
       eprint = {1109.4634},
 primaryClass = {astro-ph.SR},
       adsurl = {https://ui.adsabs.harvard.edu/abs/2011ApJ...743...48W}
}

@ARTICLE{Morin2008,
       author = {{Morin}, J. and {Donati}, J. -F. and {Petit}, P. and {Delfosse}, X. and {Forveille}, T. and {Albert}, L. and {Auri{\`e}re}, M. and {Cabanac}, R. and {Dintrans}, B. and {Fares}, R. and {Gastine}, T. and {Jardine}, M.~M. and {Ligni{\`e}res}, F. and {Paletou}, F. and {Ramirez Velez}, J.~C. and {Th{\'e}ado}, S.},
        title = "{Large-scale magnetic topologies of mid M dwarfs}",
      journal = {MNRAS},
         year = 2008,
        month = oct,
       volume = {390},
       number = {2},
        pages = {567-581},
          doi = {10.1111/j.1365-2966.2008.13809.x},
archivePrefix = {arXiv},
       eprint = {0808.1423},
 primaryClass = {astro-ph},
       adsurl = {https://ui.adsabs.harvard.edu/abs/2008MNRAS.390..567M}
}

@ARTICLE{Shulyak2019,
       author = {{Shulyak}, D. and {Reiners}, A. and {Nagel}, E. and {Tal-Or}, L. and {Caballero}, J.~A. and {Zechmeister}, M. and {B{\'e}jar}, V.~J.~S. and {Cort{\'e}s-Contreras}, M. and {Martin}, E.~L. and {Kaminski}, A. and {Ribas}, I. and {Quirrenbach}, A. and {Amado}, P.~J. and {Anglada-Escud{\'e}}, G. and {Bauer}, F.~F. and {Dreizler}, S. and {Guenther}, E.~W. and {Henning}, T. and {Jeffers}, S.~V. and {K{\"u}rster}, M. and {Lafarga}, M. and {Montes}, D. and {Morales}, J.~C. and {Pedraz}, S.},
        title = "{Magnetic fields in M dwarfs from the CARMENES survey}",
      journal = {A\&A},
         year = 2019,
        month = jun,
       volume = {626},
          eid = {A86},
        pages = {A86},
          doi = {10.1051/0004-6361/201935315},
archivePrefix = {arXiv},
       eprint = {1904.12762},
 primaryClass = {astro-ph.SR},
       adsurl = {https://ui.adsabs.harvard.edu/abs/2019A&A...626A..86S}
}

@ARTICLE{Gustafsson2008,
       author = {{Gustafsson}, B. and {Edvardsson}, B. and {Eriksson}, K. and {J{\o}rgensen}, U.~G. and {Nordlund}, {\r{A}}. and {Plez}, B.},
        title = "{A grid of MARCS model atmospheres for late-type stars. I. Methods and general properties}",
      journal = {A\&A},
         year = 2008,
        month = aug,
       volume = {486},
       number = {3},
        pages = {951-970},
          doi = {10.1051/0004-6361:200809724},
archivePrefix = {arXiv},
       eprint = {0805.0554},
 primaryClass = {astro-ph},
       adsurl = {https://ui.adsabs.harvard.edu/abs/2008A&A...486..951G}
}

@ARTICLE{Folsom2016,
       author = {{Folsom}, C.~P. and {Petit}, P. and {Bouvier}, J. and {L{\`e}bre}, A. and {Amard}, L. and {Palacios}, A. and {Morin}, J. and {Donati}, J. -F. and {Jeffers}, S.~V. and {Marsden}, S.~C. and {Vidotto}, A.~A.},
        title = "{The evolution of surface magnetic fields in young solar-type stars - I. The first 250 Myr}",
      journal = {MNRAS},
         year = 2016,
        month = mar,
       volume = {457},
       number = {1},
        pages = {580-607},
          doi = {10.1093/mnras/stv2924},
archivePrefix = {arXiv},
       eprint = {1601.00684},
 primaryClass = {astro-ph.SR},
       adsurl = {https://ui.adsabs.harvard.edu/abs/2016MNRAS.457..580F}
}

@MISC{Czesla2019,
       author = {{Czesla}, Stefan and {Schr{\"o}ter}, Sebastian and {Schneider}, Christian P. and {Huber}, Klaus F. and {Pfeifer}, Fabian and {Andreasen}, Daniel T. and {Zechmeister}, Mathias},
        title = "{PyA: Python astronomy-related packages}",
         year = 2019,
        month = jun,
          eid = {ascl:1906.010},
        pages = {ascl:1906.010},
archivePrefix = {ascl},
       eprint = {1906.010},
       adsurl = {https://ui.adsabs.harvard.edu/abs/2019ascl.soft06010C}
}

@ARTICLE{Astropy2013,
       author = {{Astropy Collaboration} and {Robitaille}, Thomas P. and {Tollerud}, Erik J. and {Greenfield}, Perry and {Droettboom}, Michael and {Bray}, Erik and {Aldcroft}, Tom and {Davis}, Matt and {Ginsburg}, Adam and {Price-Whelan}, Adrian M. and {Kerzendorf}, Wolfgang E. and {Conley}, Alexander and {Crighton}, Neil and {Barbary}, Kyle and {Muna}, Demitri and {Ferguson}, Henry and {Grollier}, Fr{\'e}d{\'e}ric and {Parikh}, Madhura M. and {Nair}, Prasanth H. and {Unther}, Hans M. and {Deil}, Christoph and {Woillez}, Julien and {Conseil}, Simon and {Kramer}, Roban and {Turner}, James E.~H. and {Singer}, Leo and {Fox}, Ryan and {Weaver}, Benjamin A. and {Zabalza}, Victor and {Edwards}, Zachary I. and {Azalee Bostroem}, K. and {Burke}, D.~J. and {Casey}, Andrew R. and {Crawford}, Steven M. and {Dencheva}, Nadia and {Ely}, Justin and {Jenness}, Tim and {Labrie}, Kathleen and {Lim}, Pey Lian and {Pierfederici}, Francesco and {Pontzen}, Andrew and {Ptak}, Andy and {Refsdal}, Brian and {Servillat}, Mathieu and {Streicher}, Ole},
        title = "{Astropy: A community Python package for astronomy}",
      journal = {A\&A},
         year = 2013,
        month = oct,
       volume = {558},
          eid = {A33},
        pages = {A33},
          doi = {10.1051/0004-6361/201322068},
archivePrefix = {arXiv},
       eprint = {1307.6212},
 primaryClass = {astro-ph.IM},
       adsurl = {https://ui.adsabs.harvard.edu/abs/2013A&A...558A..33A}
}

@ARTICLE{Astropy2018,
       author = {{Astropy Collaboration} and {Price-Whelan}, A.~M. and {Sip{\H{o}}cz}, B.~M. and {G{\"u}nther}, H.~M. and {Lim}, P.~L. and {Crawford}, S.~M. and {Conseil}, S. and {Shupe}, D.~L. and {Craig}, M.~W. and {Dencheva}, N. and {Ginsburg}, A. and {VanderPlas}, J.~T. and {Bradley}, L.~D. and {P{\'e}rez-Su{\'a}rez}, D. and {de Val-Borro}, M. and {Aldcroft}, T.~L. and {Cruz}, K.~L. and {Robitaille}, T.~P. and {Tollerud}, E.~J. and {Ardelean}, C. and {Babej}, T. and {Bach}, Y.~P. and {Bachetti}, M. and {Bakanov}, A.~V. and {Bamford}, S.~P. and {Barentsen}, G. and {Barmby}, P. and {Baumbach}, A. and {Berry}, K.~L. and {Biscani}, F. and {Boquien}, M. and {Bostroem}, K.~A. and {Bouma}, L.~G. and {Brammer}, G.~B. and {Bray}, E.~M. and {Breytenbach}, H. and {Buddelmeijer}, H. and {Burke}, D.~J. and {Calderone}, G. and {Cano Rodr{\'\i}guez}, J.~L. and {Cara}, M. and {Cardoso}, J.~V.~M. and {Cheedella}, S. and {Copin}, Y. and {Corrales}, L. and {Crichton}, D. and {D'Avella}, D. and {Deil}, C. and {Depagne}, {\'E}. and {Dietrich}, J.~P. and {Donath}, A. and {Droettboom}, M. and {Earl}, N. and {Erben}, T. and {Fabbro}, S. and {Ferreira}, L.~A. and {Finethy}, T. and {Fox}, R.~T. and {Garrison}, L.~H. and {Gibbons}, S.~L.~J. and {Goldstein}, D.~A. and {Gommers}, R. and {Greco}, J.~P. and {Greenfield}, P. and {Groener}, A.~M. and {Grollier}, F. and {Hagen}, A. and {Hirst}, P. and {Homeier}, D. and {Horton}, A.~J. and {Hosseinzadeh}, G. and {Hu}, L. and {Hunkeler}, J.~S. and {Ivezi{\'c}}, {\v{Z}}. and {Jain}, A. and {Jenness}, T. and {Kanarek}, G. and {Kendrew}, S. and {Kern}, N.~S. and {Kerzendorf}, W.~E. and {Khvalko}, A. and {King}, J. and {Kirkby}, D. and {Kulkarni}, A.~M. and {Kumar}, A. and {Lee}, A. and {Lenz}, D. and {Littlefair}, S.~P. and {Ma}, Z. and {Macleod}, D.~M. and {Mastropietro}, M. and {McCully}, C. and {Montagnac}, S. and {Morris}, B.~M. and {Mueller}, M. and {Mumford}, S.~J. and {Muna}, D. and {Murphy}, N.~A. and {Nelson}, S. and {Nguyen}, G.~H. and {Ninan}, J.~P. and {N{\"o}the}, M. and {Ogaz}, S. and {Oh}, S. and {Parejko}, J.~K. and {Parley}, N. and {Pascual}, S. and {Patil}, R. and {Patil}, A.~A. and {Plunkett}, A.~L. and {Prochaska}, J.~X. and {Rastogi}, T. and {Reddy Janga}, V. and {Sabater}, J. and {Sakurikar}, P. and {Seifert}, M. and {Sherbert}, L.~E. and {Sherwood-Taylor}, H. and {Shih}, A.~Y. and {Sick}, J. and {Silbiger}, M.~T. and {Singanamalla}, S. and {Singer}, L.~P. and {Sladen}, P.~H. and {Sooley}, K.~A. and {Sornarajah}, S. and {Streicher}, O. and {Teuben}, P. and {Thomas}, S.~W. and {Tremblay}, G.~R. and {Turner}, J.~E.~H. and {Terr{\'o}n}, V. and {van Kerkwijk}, M.~H. and {de la Vega}, A. and {Watkins}, L.~L. and {Weaver}, B.~A. and {Whitmore}, J.~B. and {Woillez}, J. and {Zabalza}, V. and {Astropy Contributors}},
        title = "{The Astropy Project: Building an Open-science Project and Status of the v2.0 Core Package}",
      journal = {AJ},
         year = 2018,
        month = sep,
       volume = {156},
       number = {3},
          eid = {123},
        pages = {123},
          doi = {10.3847/1538-3881/aabc4f},
archivePrefix = {arXiv},
       eprint = {1801.02634},
 primaryClass = {astro-ph.IM},
       adsurl = {https://ui.adsabs.harvard.edu/abs/2018AJ....156..123A}
}

@MISC{Virtanen2020,
       author = {{Virtanen}, Pauli and {Gommers}, Ralf and {Burovski}, Evgeni and {Oliphant}, Travis E. and {Weckesser}, Warren and {Cournapeau}, David and {Alexbrc} and {Peterson}, Pearu and {Reddy}, Tyler and {Wilson}, Josh and {Haberland}, Matt and {Mayorov}, Nikolay and {Endolith} and {Nelson}, Andrew and {Der Van Walt}, Stefan and {Laxalde}, Denis and {Brett}, Matthew and {Polat}, Ilhan and {Larson}, Eric and {Millman}, Jarrod and {Lars} and {Van Mulbregt}, Paul and {Eric-Jones} and {Carey}, CJ and {Moore}, Eric and {Kern}, Robert and {Leslie}, Tim and {Perktold}, Josef and {Striega}, Kai and {Feng}, Yu},
        title = "{scipy/scipy: SciPy 1.5.3}",
         year = 2020,
        month = oct,
          eid = {10.5281/zenodo.595738},
          doi = {10.5281/zenodo.595738},
      version = {v1.5.3},
    publisher = {Zenodo},
       adsurl = {https://ui.adsabs.harvard.edu/abs/2020zndo....595738V}
}

@ARTICLE{Hunter2007,
       author = {{Hunter}, John D.},
        title = "{Matplotlib: A 2D Graphics Environment}",
      journal = {Computing in Science and Engineering},
         year = 2007,
        month = may,
       volume = {9},
       number = {3},
        pages = {90-95},
          doi = {10.1109/MCSE.2007.55},
       adsurl = {https://ui.adsabs.harvard.edu/abs/2007CSE.....9...90H}
}

@ARTICLE{VanderWalt2011,
       author = {{van der Walt}, St{\'e}fan and {Colbert}, S. Chris and {Varoquaux}, Ga{\"e}l},
        title = "{The NumPy Array: A Structure for Efficient Numerical Computation}",
      journal = {Computing in Science and Engineering},
         year = 2011,
        month = mar,
       volume = {13},
       number = {2},
        pages = {22-30},
          doi = {10.1109/MCSE.2011.37},
archivePrefix = {arXiv},
       eprint = {1102.1523},
 primaryClass = {cs.MS},
       adsurl = {https://ui.adsabs.harvard.edu/abs/2011CSE....13b..22V}
}

@ARTICLE{Carolan2019,
       author = {{Carolan}, S. and {Vidotto}, A.~A. and {Loesch}, C. and {Coogan}, P.},
        title = "{The evolution of Earth's magnetosphere during the solar main sequence}",
      journal = {\mnras},
         year = 2019,
        month = nov,
       volume = {489},
       number = {4},
        pages = {5784-5801},
          doi = {10.1093/mnras/stz2422},
archivePrefix = {arXiv},
       eprint = {1908.03537},
 primaryClass = {astro-ph.EP},
       adsurl = {https://ui.adsabs.harvard.edu/abs/2019MNRAS.489.5784C}
}

@ARTICLE{Morgenthaler2012,
       author = {{Morgenthaler}, A. and {Petit}, P. and {Saar}, S. and {Solanki}, S.~K. and {Morin}, J. and {Marsden}, S.~C. and {Auri{\`e}re}, M. and {Dintrans}, B. and {Fares}, R. and {Gastine}, T. and {Lanoux}, J. and {Ligni{\`e}res}, F. and {Paletou}, F. and {Ram{\'\i}rez V{\'e}lez}, J.~C. and {Th{\'e}ado}, S. and {Van Grootel}, V.},
        title = "{Long-term magnetic field monitoring of the Sun-like star {\ensuremath{\xi}} Bootis A}",
      journal = {A\&A},
         year = 2012,
        month = apr,
       volume = {540},
          eid = {A138},
        pages = {A138},
          doi = {10.1051/0004-6361/201118139},
archivePrefix = {arXiv},
       eprint = {1109.5066},
 primaryClass = {astro-ph.SR},
       adsurl = {https://ui.adsabs.harvard.edu/abs/2012A&A...540A.138M}
}

@ARTICLE{BoroSaikia2016,
       author = {{Boro Saikia}, S. and {Jeffers}, S.~V. and {Morin}, J. and {Petit}, P. and {Folsom}, C.~P. and {Marsden}, S.~C. and {Donati}, J. -F. and {Cameron}, R. and {Hall}, J.~C. and {Perdelwitz}, V. and {Reiners}, A. and {Vidotto}, A.~A.},
        title = "{A solar-like magnetic cycle on the mature K-dwarf 61 Cygni A (HD 201091)}",
      journal = {A\&A},
         year = 2016,
        month = oct,
       volume = {594},
          eid = {A29},
        pages = {A29},
          doi = {10.1051/0004-6361/201628262},
archivePrefix = {arXiv},
       eprint = {1606.01032},
 primaryClass = {astro-ph.SR},
       adsurl = {https://ui.adsabs.harvard.edu/abs/2016A&A...594A..29B}
}

@ARTICLE{Wright2018,
       author = {{Wright}, Nicholas J. and {Newton}, Elisabeth R. and {Williams}, Peter K.~G. and {Drake}, Jeremy J. and {Yadav}, Rakesh K.},
        title = "{The stellar rotation-activity relationship in fully convective M dwarfs}",
      journal = {MNRAS},
         year = 2018,
        month = sep,
       volume = {479},
       number = {2},
        pages = {2351-2360},
          doi = {10.1093/mnras/sty1670},
archivePrefix = {arXiv},
       eprint = {1807.03304},
 primaryClass = {astro-ph.SR},
       adsurl = {https://ui.adsabs.harvard.edu/abs/2018MNRAS.479.2351W}
}

@ARTICLE{Donati2008,
       author = {{Donati}, J. -F. and {Morin}, J. and {Petit}, P. and {Delfosse}, X. and {Forveille}, T. and {Auri{\`e}re}, M. and {Cabanac}, R. and {Dintrans}, B. and {Fares}, R. and {Gastine}, T. and {Jardine}, M.~M. and {Ligni{\`e}res}, F. and {Paletou}, F. and {Ramirez Velez}, J.~C. and {Th{\'e}ado}, S.},
        title = "{Large-scale magnetic topologies of early M dwarfs}",
      journal = {MNRAS},
         year = 2008,
        month = oct,
       volume = {390},
       number = {2},
        pages = {545-560},
          doi = {10.1111/j.1365-2966.2008.13799.x},
archivePrefix = {arXiv},
       eprint = {0809.0269},
 primaryClass = {astro-ph},
       adsurl = {https://ui.adsabs.harvard.edu/abs/2008MNRAS.390..545D}
}

@ARTICLE{Vogt1983,
       author = {{Vogt}, S.~S. and {Penrod}, G.~D.},
        title = "{Doppler imaging of spotted stars : application to the RS Canum Venaticorum star HR 1099.}",
      journal = {PASP},
         year = 1983,
        month = sep,
       volume = {95},
        pages = {565-576},
          doi = {10.1086/131208},
       adsurl = {https://ui.adsabs.harvard.edu/abs/1983PASP...95..565V}
}

@ARTICLE{Deutsch1957,
       author = {{Deutsch}, Armin J.},
        title = "{A method for mapping the surfaces of some magnetic stars.}",
      journal = {AJ},
         year = 1957,
        month = jan,
       volume = {62},
        pages = {139},
          doi = {10.1086/107596},
       adsurl = {https://ui.adsabs.harvard.edu/abs/1957AJ.....62Q.139D}
}

@ARTICLE{Rice1989,
       author = {{Rice}, J.~B. and {Wehlau}, W.~H. and {Khokhlova}, V.~L.},
        title = "{Mapping stellar surfaces by Doppler imaging : technique and application.}",
      journal = {A\&A},
         year = 1989,
        month = jan,
       volume = {208},
        pages = {179-188},
       adsurl = {https://ui.adsabs.harvard.edu/abs/1989A&A...208..179R}
}

@ARTICLE{Claret2011,
       author = {{Claret}, A. and {Bloemen}, S.},
        title = "{Gravity and limb-darkening coefficients for the Kepler, CoRoT, Spitzer, uvby, UBVRIJHK, and Sloan photometric systems}",
      journal = {A\&A},
         year = 2011,
        month = may,
       volume = {529},
          eid = {A75},
        pages = {A75},
          doi = {10.1051/0004-6361/201116451},
       adsurl = {https://ui.adsabs.harvard.edu/abs/2011A&A...529A..75C}
}

@ARTICLE{Klein2021,
       author = {{Klein}, Baptiste and {Donati}, Jean-Fran{\c{c}}ois and {Moutou}, Claire and {Delfosse}, Xavier and {Bonfils}, Xavier and {Martioli}, Eder and {Fouqu{\'e}}, Pascal and {Cloutier}, Ryan and {Artigau}, {\'E}tienne and {Doyon}, Ren{\'e} and {H{\'e}brard}, Guillaume and {Morin}, Julien and {Rameau}, Julien and {Plavchan}, Peter and {Gaidos}, Eric},
        title = "{Investigating the young AU Mic system with SPIRou: large-scale stellar magnetic field and close-in planet mass}",
      journal = {MNRAS},
         year = 2021,
        month = mar,
       volume = {502},
       number = {1},
        pages = {188-205},
          doi = {10.1093/mnras/staa3702},
archivePrefix = {arXiv},
       eprint = {2011.13357},
 primaryClass = {astro-ph.EP},
       adsurl = {https://ui.adsabs.harvard.edu/abs/2021MNRAS.502..188K}
}

@ARTICLE{Martioli2022,
       author = {{Martioli}, E. and {H{\'e}brard}, G. and {Fouqu{\'e}}, P. and {Artigau}, {\'E}. and {Donati}, J. -F. and {Cadieux}, C. and {Bellotti}, S. and {Lecavelier des Etangs}, A. and {Doyon}, R. and {do Nascimento}, J. -D. and {Arnold}, L. and {Carmona}, A. and {Cook}, N.~J. and {Cortes-Zuleta}, P. and {de Almeida}, L. and {Delfosse}, X. and {Folsom}, C.~P. and {K{\"o}nig}, P. -C. and {Moutou}, C. and {Ould-Elhkim}, M. and {Petit}, P. and {Stassun}, K.~G. and {Vidotto}, A.~A. and {Vandal}, T. and {Benneke}, B. and {Boisse}, I. and {Bonfils}, X. and {Boyd}, P. and {Brasseur}, C. and {Charbonneau}, D. and {Cloutier}, R. and {Collins}, K. and {Cristofari}, P. and {Crossfield}, I. and {D{\'\i}az}, R.~F. and {Fausnaugh}, M. and {Figueira}, P. and {Forveille}, T. and {Furlan}, E. and {Girardin}, E. and {Gnilka}, C.~L. and {Gomes da Silva}, J. and {Gu}, P. -G. and {Guerra}, P. and {Howell}, S.~B. and {Hussain}, G.~A.~J. and {Jenkins}, J.~M. and {Kiefer}, F. and {Latham}, D.~W. and {Matson}, R.~A. and {Matthews}, E.~C. and {Morin}, J. and {Naves}, R. and {Ricker}, G. and {Seager}, S. and {Takami}, M. and {Twicken}, J.~D. and {Vanderburg}, A. and {Vanderspek}, R. and {Winn}, J.},
        title = "{TOI-1759 b: A transiting sub-Neptune around a low mass star characterized with SPIRou and TESS}",
      journal = {A\&A},
         year = 2022,
        month = apr,
       volume = {660},
          eid = {A86},
        pages = {A86},
          doi = {10.1051/0004-6361/202142540},
archivePrefix = {arXiv},
       eprint = {2202.01259},
 primaryClass = {astro-ph.EP},
       adsurl = {https://ui.adsabs.harvard.edu/abs/2022A&A...660A..86M}
}

@ARTICLE{Cutri2003,
       author = {{Cutri}, R.~M. and {Skrutskie}, M.~F. and {van Dyk}, S. and {Beichman}, C.~A. and {Carpenter}, J.~M. and {Chester}, T. and {Cambresy}, L. and {Evans}, T. and {Fowler}, J. and {Gizis}, J. and {Howard}, E. and {Huchra}, J. and {Jarrett}, T. and {Kopan}, E.~L. and {Kirkpatrick}, J.~D. and {Light}, R.~M. and {Marsh}, K.~A. and {McCallon}, H. and {Schneider}, S. and {Stiening}, R. and {Sykes}, M. and {Weinberg}, M. and {Wheaton}, W.~A. and {Wheelock}, S. and {Zacarias}, N.},
        title = "{VizieR Online Data Catalog: 2MASS All-Sky Catalog of Point Sources (Cutri+ 2003)}",
      journal = {VizieR Online Data Catalog},
         year = 2003,
        month = jun,
          eid = {II/246},
        pages = {II/246},
       adsurl = {https://ui.adsabs.harvard.edu/abs/2003yCat.2246....0C}
}

@ARTICLE{Callingham2021,
       author = {{Callingham}, J.~R. and {Vedantham}, H.~K. and {Shimwell}, T.~W. and {Pope}, B.~J.~S. and {Davis}, I.~E. and {Best}, P.~N. and {Hardcastle}, M.~J. and {R{\"o}ttgering}, H.~J.~A. and {Sabater}, J. and {Tasse}, C. and {van Weeren}, R.~J. and {Williams}, W.~L. and {Zarka}, P. and {de Gasperin}, F. and {Drabent}, A.},
        title = "{The population of M dwarfs observed at low radio frequencies}",
      journal = {Nature Astronomy},
         year = 2021,
        month = oct,
       volume = {5},
        pages = {1233-1239},
          doi = {10.1038/s41550-021-01483-0},
archivePrefix = {arXiv},
       eprint = {2110.03713},
 primaryClass = {astro-ph.SR},
       adsurl = {https://ui.adsabs.harvard.edu/abs/2021NatAs...5.1233C}
}

@ARTICLE{Vedantham2020,
       author = {{Vedantham}, H.~K. and {Callingham}, J.~R. and {Shimwell}, T.~W. and {Tasse}, C. and {Pope}, B.~J.~S. and {Bedell}, M. and {Snellen}, I. and {Best}, P. and {Hardcastle}, M.~J. and {Haverkorn}, M. and {Mechev}, A. and {O'Sullivan}, S.~P. and {R{\"o}ttgering}, H.~J.~A. and {White}, G.~J.},
        title = "{Coherent radio emission from a quiescent red dwarf indicative of star-planet interaction}",
      journal = {Nature Astronomy},
         year = 2020,
        month = feb,
       volume = {4},
        pages = {577-583},
          doi = {10.1038/s41550-020-1011-9},
archivePrefix = {arXiv},
       eprint = {2002.08727},
 primaryClass = {astro-ph.EP},
       adsurl = {https://ui.adsabs.harvard.edu/abs/2020NatAs...4..577V}
}

@ARTICLE{Nichols2012,
       author = {{Nichols}, J.~D. and {Burleigh}, M.~R. and {Casewell}, S.~L. and {Cowley}, S.~W.~H. and {Wynn}, G.~A. and {Clarke}, J.~T. and {West}, A.~A.},
        title = "{Origin of Electron Cyclotron Maser Induced Radio Emissions at Ultracool Dwarfs: Magnetosphere-Ionosphere Coupling Currents}",
      journal = {ApJ},
         year = 2012,
        month = nov,
       volume = {760},
       number = {1},
          eid = {59},
        pages = {59},
          doi = {10.1088/0004-637X/760/1/59},
archivePrefix = {arXiv},
       eprint = {1210.1864},
 primaryClass = {astro-ph.SR},
       adsurl = {https://ui.adsabs.harvard.edu/abs/2012ApJ...760...59N}
}

@ARTICLE{Zarka2007,
       author = {{Zarka}, Philippe},
        title = "{Plasma interactions of exoplanets with their parent star and associated radio emissions}",
      journal = {Planetary and Space Science},
         year = 2007,
        month = apr,
       volume = {55},
       number = {5},
        pages = {598-617},
          doi = {10.1016/j.pss.2006.05.045},
       adsurl = {https://ui.adsabs.harvard.edu/abs/2007P&SS...55..598Z}
}

@ARTICLE{Vidotto2019,
       author = {{Vidotto}, A.~A. and {Feeney}, N. and {Groh}, J.~H.},
        title = "{Can we detect aurora in exoplanets orbiting M dwarfs?}",
      journal = {MNRAS},
         year = 2019,
        month = sep,
       volume = {488},
       number = {1},
        pages = {633-644},
          doi = {10.1093/mnras/stz1696},
archivePrefix = {arXiv},
       eprint = {1906.07089},
 primaryClass = {astro-ph.EP},
       adsurl = {https://ui.adsabs.harvard.edu/abs/2019MNRAS.488..633V}
}

@ARTICLE{DonatiBrown1997,
       author = {{Donati}, J. -F. and {Brown}, S.~F.},
        title = "{Zeeman-Doppler imaging of active stars. V. Sensitivity of maximum entropy magnetic maps to field orientation.}",
      journal = {A\&A},
         year = 1997,
        month = oct,
       volume = {326},
        pages = {1135-1142},
       adsurl = {https://ui.adsabs.harvard.edu/abs/1997A&A...326.1135D}
}

@ARTICLE{Kasting1993,
       author = {{Kasting}, James F. and {Whitmire}, Daniel P. and {Reynolds}, Ray T.},
        title = "{Habitable Zones around Main Sequence Stars}",
      journal = {Icarus},
         year = 1993,
        month = jan,
       volume = {101},
       number = {1},
        pages = {108-128},
          doi = {10.1006/icar.1993.1010},
       adsurl = {https://ui.adsabs.harvard.edu/abs/1993Icar..101..108K}
}

@ARTICLE{Lenardic2016,
       author = {{Lenardic}, A. and {Jellinek}, A.~M. and {Foley}, B. and {O'Neill}, C. and {Moore}, W.~B.},
        title = "{Climate-tectonic coupling: Variations in the mean, variations about the mean, and variations in mode}",
      journal = {Journal of Geophysical Research (Planets)},
         year = 2016,
        month = oct,
       volume = {121},
       number = {10},
        pages = {1831-1864},
          doi = {10.1002/2016JE005089},
       adsurl = {https://ui.adsabs.harvard.edu/abs/2016JGRE..121.1831L}
}

@ARTICLE{Driscoll2013,
       author = {{Driscoll}, P. and {Bercovici}, D.},
        title = "{Divergent evolution of Earth and Venus: Influence of degassing, tectonics, and magnetic fields}",
      journal = {Icarus},
         year = 2013,
        month = nov,
       volume = {226},
       number = {2},
        pages = {1447-1464},
          doi = {10.1016/j.icarus.2013.07.025},
       adsurl = {https://ui.adsabs.harvard.edu/abs/2013Icar..226.1447D}
}

@ARTICLE{Lammer2003,
       author = {{Lammer}, H. and {Selsis}, F. and {Ribas}, I. and {Guinan}, E.~F. and {Bauer}, S.~J. and {Weiss}, W.~W.},
        title = "{Atmospheric Loss of Exoplanets Resulting from Stellar X-Ray and Extreme-Ultraviolet Heating}",
      journal = {ApJl},
         year = 2003,
        month = dec,
       volume = {598},
       number = {2},
        pages = {L121-L124},
          doi = {10.1086/380815},
       adsurl = {https://ui.adsabs.harvard.edu/abs/2003ApJ...598L.121L}
}

@ARTICLE{Owen2012,
       author = {{Owen}, James E. and {Jackson}, Alan P.},
        title = "{Planetary evaporation by UV \& X-ray radiation: basic hydrodynamics}",
      journal = {MNRAS},
         year = 2012,
        month = oct,
       volume = {425},
       number = {4},
        pages = {2931-2947},
          doi = {10.1111/j.1365-2966.2012.21481.x},
archivePrefix = {arXiv},
       eprint = {1206.2367},
 primaryClass = {astro-ph.EP},
       adsurl = {https://ui.adsabs.harvard.edu/abs/2012MNRAS.425.2931O}
}

@ARTICLE{Rugheimer2015,
       author = {{Rugheimer}, S. and {Kaltenegger}, L. and {Segura}, A. and {Linsky}, J. and {Mohanty}, S.},
        title = "{Effect of UV Radiation on the Spectral Fingerprints of Earth-like Planets Orbiting M Stars}",
      journal = {ApJ},
         year = 2015,
        month = aug,
       volume = {809},
       number = {1},
          eid = {57},
        pages = {57},
          doi = {10.1088/0004-637X/809/1/57},
archivePrefix = {arXiv},
       eprint = {1506.07202},
 primaryClass = {astro-ph.EP},
       adsurl = {https://ui.adsabs.harvard.edu/abs/2015ApJ...809...57R}
}

@ARTICLE{Petit2002,
       author = {{Petit}, P. and {Donati}, J. -F. and {Collier Cameron}, A.},
        title = "{Differential rotation of cool active stars: the case of intermediate rotators}",
      journal = {MNRAS},
         year = 2002,
        month = aug,
       volume = {334},
       number = {2},
        pages = {374-382},
          doi = {10.1046/j.1365-8711.2002.05529.x},
       adsurl = {https://ui.adsabs.harvard.edu/abs/2002MNRAS.334..374P}
}

@ARTICLE{Vidotto2013,
       author = {{Vidotto}, A.~A. and {Jardine}, M. and {Morin}, J. and {Donati}, J. -F. and {Lang}, P. and {Russell}, A.~J.~B.},
        title = "{Effects of M dwarf magnetic fields on potentially habitable planets}",
      journal = {A\&A},
         year = 2013,
        month = sep,
       volume = {557},
          eid = {A67},
        pages = {A67},
          doi = {10.1051/0004-6361/201321504},
archivePrefix = {arXiv},
       eprint = {1306.4789},
 primaryClass = {astro-ph.EP},
       adsurl = {https://ui.adsabs.harvard.edu/abs/2013A&A...557A..67V}
}

@ARTICLE{Cockell2016,
       author = {{Cockell}, C.~S. and {Bush}, T. and {Bryce}, C. and {Direito}, S. and {Fox-Powell}, M. and {Harrison}, J.~P. and {Lammer}, H. and {Landenmark}, H. and {Martin-Torres}, J. and {Nicholson}, N. and {Noack}, L. and {O'Malley-James}, J. and {Payler}, S.~J. and {Rushby}, A. and {Samuels}, T. and {Schwendner}, P. and {Wadsworth}, J. and {Zorzano}, M.~P.},
        title = "{Habitability: A Review}",
      journal = {Astrobiology},
         year = 2016,
        month = jan,
       volume = {16},
       number = {1},
        pages = {89-117},
          doi = {10.1089/ast.2015.1295},
       adsurl = {https://ui.adsabs.harvard.edu/abs/2016AsBio..16...89C}
}

@ARTICLE{PerezTorres2021,
       author = {{P{\'e}rez-Torres}, M. and {G{\'o}mez}, J.~F. and {Ortiz}, J.~L. and {Leto}, P. and {Anglada}, G. and {G{\'o}mez}, J.~L. and {Rodr{\'\i}guez}, E. and {Trigilio}, C. and {Amado}, P.~J. and {Alberdi}, A. and {Anglada-Escud{\'e}}, G. and {Osorio}, M. and {Umana}, G. and {Berdi{\~n}as}, Z. and {L{\'o}pez-Gonz{\'a}lez}, M.~J. and {Morales}, N. and {Rodr{\'\i}guez-L{\'o}pez}, C. and {Chibueze}, J.},
        title = "{Monitoring the radio emission of Proxima Centauri}",
      journal = {\aap},
         year = 2021,
        month = jan,
       volume = {645},
          eid = {A77},
        pages = {A77},
          doi = {10.1051/0004-6361/202039052},
archivePrefix = {arXiv},
       eprint = {2012.02116},
 primaryClass = {astro-ph.SR},
       adsurl = {https://ui.adsabs.harvard.edu/abs/2021A&A...645A..77P}
}

@ARTICLE{Turner2021,
       author = {{Turner}, Jake D. and {Zarka}, Philippe and {Grie{\ss}meier}, Jean-Mathias and {Lazio}, Joseph and {Cecconi}, Baptiste and {Emilio Enriquez}, J. and {Girard}, Julien N. and {Jayawardhana}, Ray and {Lamy}, Laurent and {Nichols}, Jonathan D. and {de Pater}, Imke},
        title = "{The search for radio emission from the exoplanetary systems 55 Cancri, {\ensuremath{\upsilon}} Andromedae, and {\ensuremath{\tau}} Bo{\"o}tis using LOFAR beam-formed observations}",
      journal = {\aap},
         year = 2021,
        month = jan,
       volume = {645},
          eid = {A59},
        pages = {A59},
          doi = {10.1051/0004-6361/201937201},
archivePrefix = {arXiv},
       eprint = {2012.07926},
 primaryClass = {astro-ph.EP},
       adsurl = {https://ui.adsabs.harvard.edu/abs/2021A&A...645A..59T}
}

@ARTICLE{Donati2023,
       author = {{Donati}, J. -F. and {Cristofari}, P.~I. and {Finociety}, B. and {Klein}, B. and {Moutou}, C. and {Gaidos}, E. and {Cadieux}, C. and {Artigau}, E. and {Correia}, A.~C.~M. and {Bou{\'e}}, G. and {Cook}, N.~J. and {Carmona}, A. and {Lehmann}, L.~T. and {Bouvier}, J. and {Martioli}, E. and {Morin}, J. and {Fouqu{\'e}}, P. and {Delfosse}, X. and {Doyon}, R. and {H{\'e}brard}, G. and {Alencar}, S.~H.~P. and {Laskar}, J. and {Arnold}, L. and {Petit}, P. and {K{\'o}sp{\'a}l}, {\'A}. and {Vidotto}, A. and {Folsom}, C.~P. and {collaboration}, the S L S},
        title = "{The magnetic field and multiple planets of the young dwarf AU Mic}",
      journal = {\mnras},
         year = 2023,
        month = oct,
       volume = {525},
       number = {1},
        pages = {455-475},
          doi = {10.1093/mnras/stad1193},
archivePrefix = {arXiv},
       eprint = {2304.09642},
 primaryClass = {astro-ph.SR},
       adsurl = {https://ui.adsabs.harvard.edu/abs/2023MNRAS.525..455D}
}

@ARTICLE{Donati2020,
       author = {{Donati}, J. -F. and {Kouach}, D. and {Moutou}, C. and {Doyon}, R. and {Delfosse}, X. and {Artigau}, E. and {Baratchart}, S. and {Lacombe}, M. and {Barrick}, G. and {H{\'e}brard}, G. and {Bouchy}, F. and {Saddlemyer}, L. and {Par{\`e}s}, L. and {Rabou}, P. and {Micheau}, Y. and {Dolon}, F. and {Reshetov}, V. and {Challita}, Z. and {Carmona}, A. and {Striebig}, N. and {Thibault}, S. and {Martioli}, E. and {Cook}, N. and {Fouqu{\'e}}, P. and {Vermeulen}, T. and {Wang}, S.~Y. and {Arnold}, L. and {Pepe}, F. and {Boisse}, I. and {Figueira}, P. and {Bouvier}, J. and {Ray}, T.~P. and {Feugeade}, C. and {Morin}, J. and {Alencar}, S. and {Hobson}, M. and {Castilho}, B. and {Udry}, S. and {Santos}, N.~C. and {Hernandez}, O. and {Benedict}, T. and {Vall{\'e}e}, P. and {Gallou}, G. and {Dupieux}, M. and {Larrieu}, M. and {Perruchot}, S. and {Sottile}, R. and {Moreau}, F. and {Usher}, C. and {Baril}, M. and {Wildi}, F. and {Chazelas}, B. and {Malo}, L. and {Bonfils}, X. and {Loop}, D. and {Kerley}, D. and {Wevers}, I. and {Dunn}, J. and {Pazder}, J. and {Macdonald}, S. and {Dubois}, B. and {Carri{\'e}}, E. and {Valentin}, H. and {Henault}, F. and {Yan}, C.~H. and {Steinmetz}, T.},
        title = "{SPIRou: NIR velocimetry and spectropolarimetry at the CFHT}",
      journal = {MNRAS},
         year = 2020,
        month = nov,
       volume = {498},
       number = {4},
        pages = {5684-5703},
          doi = {10.1093/mnras/staa2569},
archivePrefix = {arXiv},
       eprint = {2008.08949},
 primaryClass = {astro-ph.IM},
       adsurl = {https://ui.adsabs.harvard.edu/abs/2020MNRAS.498.5684D}
}

@ARTICLE{Kochukhov2021,
       author = {{Kochukhov}, Oleg},
        title = "{Magnetic fields of M dwarfs}",
      journal = {A\&A},
         year = 2021,
        month = dec,
       volume = {29},
       number = {1},
          eid = {1},
        pages = {1},
          doi = {10.1007/s00159-020-00130-3},
archivePrefix = {arXiv},
       eprint = {2011.01781},
 primaryClass = {astro-ph.SR},
       adsurl = {https://ui.adsabs.harvard.edu/abs/2021A&ARv..29....1K}
}

@ARTICLE{Morin2010,
       author = {{Morin}, J. and {Donati}, J. -F. and {Petit}, P. and {Delfosse}, X. and {Forveille}, T. and {Jardine}, M.~M.},
        title = "{Large-scale magnetic topologies of late M dwarfs*}",
      journal = {MNRAS},
         year = 2010,
        month = oct,
       volume = {407},
       number = {4},
        pages = {2269-2286},
          doi = {10.1111/j.1365-2966.2010.17101.x},
archivePrefix = {arXiv},
       eprint = {1005.5552},
 primaryClass = {astro-ph.SR},
       adsurl = {https://ui.adsabs.harvard.edu/abs/2010MNRAS.407.2269M}
}

@ARTICLE{Jaime2014,
       author = {{Jaime}, Luisa G. and {Aguilar}, Luis and {Pichardo}, Barbara},
        title = "{Habitable zones with stable orbits for planets around binary systems}",
      journal = {MNRAS},
         year = 2014,
        month = sep,
       volume = {443},
       number = {1},
        pages = {260-274},
          doi = {10.1093/mnras/stu1052},
archivePrefix = {arXiv},
       eprint = {1401.1006},
 primaryClass = {astro-ph.EP},
       adsurl = {https://ui.adsabs.harvard.edu/abs/2014MNRAS.443..260J}
}

@ARTICLE{Penz2008,
       author = {{Penz}, T. and {Micela}, G.},
        title = "{X-ray induced mass loss effects on exoplanets orbiting dM stars}",
      journal = {A\&A},
         year = 2008,
        month = feb,
       volume = {479},
       number = {2},
        pages = {579-584},
          doi = {10.1051/0004-6361:20078873},
       adsurl = {https://ui.adsabs.harvard.edu/abs/2008A&A...479..579P}
}

@ARTICLE{Bellotti2024a,
       author = {{Bellotti}, S. and {Morin}, J. and {Lehmann}, L.~T. and {Petit}, P. and {Hussain}, G.~A.~J. and {Donati}, J. -F. and {Folsom}, C.~P. and {Carmona}, A. and {Martioli}, E. and {Klein}, B. and {Fouqu{\'e}}, P. and {Moutou}, C. and {Alencar}, S. and {Artigau}, E. and {Boisse}, I. and {Bouchy}, F. and {Bouvier}, J. and {Cook}, N.~J. and {Delfosse}, X. and {Doyon}, R. and {H{\'e}brard}, G.},
        title = "{Long-term monitoring of large-scale magnetic fields across optical and near-infrared domains with ESPaDOnS, Narval, and SPIRou. The cases of EV Lac, DS Leo, and CN Leo}",
      journal = {\aap},
         year = 2024,
        month = jun,
       volume = {686},
          eid = {A66},
        pages = {A66},
          doi = {10.1051/0004-6361/202348043},
archivePrefix = {arXiv},
       eprint = {2403.08590},
 primaryClass = {astro-ph.SR},
       adsurl = {https://ui.adsabs.harvard.edu/abs/2024A&A...686A..66B}
}

@ARTICLE{Kervella2022,
       author = {{Kervella}, Pierre and {Arenou}, Fr{\'e}d{\'e}ric and {Th{\'e}venin}, Fr{\'e}d{\'e}ric},
        title = "{Stellar and substellar companions from Gaia EDR3. Proper-motion anomaly and resolved common proper-motion pairs}",
      journal = {\aap},
         year = 2022,
        month = jan,
       volume = {657},
          eid = {A7},
        pages = {A7},
          doi = {10.1051/0004-6361/202142146},
archivePrefix = {arXiv},
       eprint = {2109.10912},
 primaryClass = {astro-ph.SR},
       adsurl = {https://ui.adsabs.harvard.edu/abs/2022A&A...657A...7K}
}

@ARTICLE{Kavanagh2022,
       author = {{Kavanagh}, Robert D. and {Vidotto}, Aline A. and {Vedantham}, Harish K. and {Jardine}, Moira M. and {Callingham}, Joe R. and {Morin}, Julien},
        title = "{Radio masers on WX UMa: hints of a Neptune-sized planet, or magnetospheric reconnection?}",
      journal = {\mnras},
         year = 2022,
        month = jul,
       volume = {514},
       number = {1},
        pages = {675-688},
          doi = {10.1093/mnras/stac1264},
archivePrefix = {arXiv},
       eprint = {2205.01661},
 primaryClass = {astro-ph.SR},
       adsurl = {https://ui.adsabs.harvard.edu/abs/2022MNRAS.514..675K}
}

@ARTICLE{Kavanagh2021,
       author = {{Kavanagh}, Robert D. and {Vidotto}, Aline A. and {Klein}, Baptiste and {Jardine}, Moira M. and {Donati}, Jean-Fran{\c{c}}ois and {{\'O} Fionnag{\'a}in}, D{\'u}alta},
        title = "{Planet-induced radio emission from the coronae of M dwarfs: the case of Prox Cen and AU Mic}",
      journal = {\mnras},
         year = 2021,
        month = jun,
       volume = {504},
       number = {1},
        pages = {1511-1518},
          doi = {10.1093/mnras/stab929},
archivePrefix = {arXiv},
       eprint = {2103.16318},
 primaryClass = {astro-ph.SR},
       adsurl = {https://ui.adsabs.harvard.edu/abs/2021MNRAS.504.1511K}
}

@ARTICLE{PenaMonino2024,
       author = {{Pe{\~n}a-Mo{\~n}ino}, L. and {P{\'e}rez-Torres}, M. and {Varela}, J. and {Zarka}, P.},
        title = "{Magnetohydrodynamic simulations of the space weather in Proxima b: Habitability conditions and radio emission}",
      journal = {\aap},
         year = 2024,
        month = aug,
       volume = {688},
          eid = {A138},
        pages = {A138},
          doi = {10.1051/0004-6361/202349042},
archivePrefix = {arXiv},
       eprint = {2405.19116},
 primaryClass = {astro-ph.EP},
       adsurl = {https://ui.adsabs.harvard.edu/abs/2024A&A...688A.138P}
}

@ARTICLE{Kavanagh2019,
       author = {{Kavanagh}, R.~D. and {Vidotto}, A.~A. and {{\'O}. Fionnag{\'a}in}, D. and {Bourrier}, V. and {Fares}, R. and {Jardine}, M. and {Helling}, Ch and {Moutou}, C. and {Llama}, J. and {Wheatley}, P.~J.},
        title = "{MOVES - II. Tuning in to the radio environment of HD189733b}",
      journal = {\mnras},
         year = 2019,
        month = jun,
       volume = {485},
       number = {4},
        pages = {4529-4538},
          doi = {10.1093/mnras/stz655},
archivePrefix = {arXiv},
       eprint = {1903.01809},
 primaryClass = {astro-ph.SR},
       adsurl = {https://ui.adsabs.harvard.edu/abs/2019MNRAS.485.4529K}
}

@ARTICLE{Alvarado-Gomez2022,
       author = {{Alvarado-G{\'o}mez}, Juli{\'a}n D. and {Cohen}, Ofer and {Drake}, Jeremy J. and {Fraschetti}, Federico and {Poppenhaeger}, Katja and {Garraffo}, Cecilia and {Chebly}, Judy and {Ilin}, Ekaterina and {Harbach}, Laura and {Kochukhov}, Oleg},
        title = "{Simulating the Space Weather in the AU Mic System: Stellar Winds and Extreme Coronal Mass Ejections}",
      journal = {\apj},
         year = 2022,
        month = apr,
       volume = {928},
       number = {2},
          eid = {147},
        pages = {147},
          doi = {10.3847/1538-4357/ac54b8},
archivePrefix = {arXiv},
       eprint = {2202.07949},
 primaryClass = {astro-ph.SR},
       adsurl = {https://ui.adsabs.harvard.edu/abs/2022ApJ...928..147A}
}

@ARTICLE{Elekes2023,
       author = {{Elekes}, F. and {Saur}, J.},
        title = "{Space environment and magnetospheric Poynting fluxes of the exoplanet {\ensuremath{\tau}} Bo{\"o}tis b}",
      journal = {\aap},
         year = 2023,
        month = mar,
       volume = {671},
          eid = {A133},
        pages = {A133},
          doi = {10.1051/0004-6361/202244947},
archivePrefix = {arXiv},
       eprint = {2301.05015},
 primaryClass = {astro-ph.EP},
       adsurl = {https://ui.adsabs.harvard.edu/abs/2023A&A...671A.133E}
}

@ARTICLE{Vidotto2014,
       author = {{Vidotto}, A.~A. and {Jardine}, M. and {Morin}, J. and {Donati}, J.~F. and {Opher}, M. and {Gombosi}, T.~I.},
        title = "{M-dwarf stellar winds: the effects of realistic magnetic geometry on rotational evolution and planets}",
      journal = {MNRAS},
         year = 2014,
        month = feb,
       volume = {438},
       number = {2},
        pages = {1162-1175},
          doi = {10.1093/mnras/stt2265},
archivePrefix = {arXiv},
       eprint = {1311.5063},
 primaryClass = {astro-ph.SR},
       adsurl = {https://ui.adsabs.harvard.edu/abs/2014MNRAS.438.1162V}
}

@BOOK{delToroIniesta2003,
       author = {{del Toro Iniesta}, Jos{\'e} Carlos},
        title = "{Introduction to Spectropolarimetry}",
         year = 2003,
       adsurl = {https://ui.adsabs.harvard.edu/abs/2003isp..book.....D}
}

@ARTICLE{Morin2008a,
       author = {{Morin}, J. and {Donati}, J. -F. and {Forveille}, T. and {Delfosse}, X. and {Dobler}, W. and {Petit}, P. and {Jardine}, M.~M. and {Collier Cameron}, A. and {Albert}, L. and {Manset}, N. and {Dintrans}, B. and {Chabrier}, G. and {Valenti}, J.~A.},
        title = "{The stable magnetic field of the fully convective star V374 Peg}",
      journal = {MNRAS},
         year = 2008,
        month = feb,
       volume = {384},
       number = {1},
        pages = {77-86},
          doi = {10.1111/j.1365-2966.2007.12709.x},
archivePrefix = {arXiv},
       eprint = {0711.1418},
 primaryClass = {astro-ph},
       adsurl = {https://ui.adsabs.harvard.edu/abs/2008MNRAS.384...77M}
}

@ARTICLE{Shulyak2017,
       author = {{Shulyak}, D. and {Reiners}, A. and {Engeln}, A. and {Malo}, L. and {Yadav}, R. and {Morin}, J. and {Kochukhov}, O.},
        title = "{Strong dipole magnetic fields in fast rotating fully convective stars}",
      journal = {Nature Astronomy},
         year = 2017,
        month = aug,
       volume = {1},
          eid = {0184},
        pages = {0184},
          doi = {10.1038/s41550-017-0184},
archivePrefix = {arXiv},
       eprint = {1801.08571},
 primaryClass = {astro-ph.SR},
       adsurl = {https://ui.adsabs.harvard.edu/abs/2017NatAs...1E.184S}
}

@ARTICLE{Kochukhov2017,
       author = {{Kochukhov}, Oleg and {Lavail}, Alexis},
        title = "{The Global and Small-scale Magnetic Fields of Fully Convective, Rapidly Spinning M Dwarf Pair GJ65 A and B}",
      journal = {ApJl},
         year = 2017,
        month = jan,
       volume = {835},
       number = {1},
          eid = {L4},
        pages = {L4},
          doi = {10.3847/2041-8213/835/1/L4},
archivePrefix = {arXiv},
       eprint = {1702.02946},
 primaryClass = {astro-ph.SR},
       adsurl = {https://ui.adsabs.harvard.edu/abs/2017ApJ...835L...4K}
}

@ARTICLE{Lavail2018,
       author = {{Lavail}, A. and {Kochukhov}, O. and {Wade}, G.~A.},
        title = "{A sudden change of the global magnetic field of the active M dwarf AD Leo revealed by full Stokes spectropolarimetric observations}",
      journal = {MNRAS},
         year = 2018,
        month = oct,
       volume = {479},
       number = {4},
        pages = {4836-4843},
          doi = {10.1093/mnras/sty1825},
archivePrefix = {arXiv},
       eprint = {1807.02346},
 primaryClass = {astro-ph.SR},
       adsurl = {https://ui.adsabs.harvard.edu/abs/2018MNRAS.479.4836L}
}

@ARTICLE{Semel1989,
       author = {{Semel}, M.},
        title = "{Zeeman-Doppler imaging of active stars. I - Basic principles.}",
      journal = {A\&A},
         year = 1989,
        month = nov,
       volume = {225},
        pages = {456-466},
       adsurl = {https://ui.adsabs.harvard.edu/abs/1989A&A...225..456S}
}

@ARTICLE{Lehmann2022,
       author = {{Lehmann}, L.~T. and {Donati}, J. -F.},
        title = "{Diagnosing large-scale stellar magnetic fields using PCA on spectropolarimetric data}",
      journal = {MNRAS},
         year = 2022,
        month = aug,
       volume = {514},
       number = {2},
        pages = {2333-2345},
          doi = {10.1093/mnras/stac1519},
archivePrefix = {arXiv},
       eprint = {2205.15647},
 primaryClass = {astro-ph.SR},
       adsurl = {https://ui.adsabs.harvard.edu/abs/2022MNRAS.514.2333L}
}

@ARTICLE{Hussain2009,
       author = {{Hussain}, G.~A.~J. and {Collier Cameron}, A. and {Jardine}, M.~M. and {Dunstone}, N. and {Ramirez Velez}, J. and {Stempels}, H.~C. and {Donati}, J. -F. and {Semel}, M. and {Aulanier}, G. and {Harries}, T. and {Bouvier}, J. and {Dougados}, C. and {Ferreira}, J. and {Carter}, B.~D. and {Lawson}, W.~A.},
        title = "{Surface magnetic fields on two accreting TTauri stars: CVCha and CRCha}",
      journal = {MNRAS},
         year = 2009,
        month = sep,
       volume = {398},
       number = {1},
        pages = {189-200},
          doi = {10.1111/j.1365-2966.2009.14881.x},
archivePrefix = {arXiv},
       eprint = {0905.0914},
 primaryClass = {astro-ph.SR},
       adsurl = {https://ui.adsabs.harvard.edu/abs/2009MNRAS.398..189H}
}

@ARTICLE{Folsom2018,
       author = {{Folsom}, C.~P. and {Bouvier}, J. and {Petit}, P. and {L{\`e}bre}, A. and {Amard}, L. and {Palacios}, A. and {Morin}, J. and {Donati}, J. -F. and {Vidotto}, A.~A.},
        title = "{The evolution of surface magnetic fields in young solar-type stars II: the early main sequence (250-650 Myr)}",
      journal = {MNRAS},
         year = 2018,
        month = mar,
       volume = {474},
       number = {4},
        pages = {4956-4987},
          doi = {10.1093/mnras/stx3021},
archivePrefix = {arXiv},
       eprint = {1711.08636},
 primaryClass = {astro-ph.SR},
       adsurl = {https://ui.adsabs.harvard.edu/abs/2018MNRAS.474.4956F}
}

@ARTICLE{Skilling1984,
       author = {{Skilling}, J. and {Bryan}, R.~K.},
        title = "{Maximum Entropy Image Reconstruction - General Algorithm}",
      journal = {MNRAS},
         year = 1984,
        month = nov,
       volume = {211},
        pages = {111},
          doi = {10.1093/mnras/211.1.111},
       adsurl = {https://ui.adsabs.harvard.edu/abs/1984MNRAS.211..111S}
}

@ARTICLE{Lehmann2024,
       author = {{Lehmann}, L.~T. and {Donati}, J. -F. and {Fouqu{\'e}}, P. and {Moutou}, C. and {Bellotti}, S. and {Delfosse}, X. and {Petit}, P. and {Carmona}, A. and {Morin}, J. and {Vidotto}, A.~A. and {the SLS consortium}},
        title = "{SPIRou reveals unusually strong magnetic fields of slowly rotating M dwarfs}",
      journal = {\mnras},
         year = 2024,
        month = jan,
       volume = {527},
       number = {2},
        pages = {4330-4352},
          doi = {10.1093/mnras/stad3472},
archivePrefix = {arXiv},
       eprint = {2311.05039},
 primaryClass = {astro-ph.SR},
       adsurl = {https://ui.adsabs.harvard.edu/abs/2024MNRAS.527.4330L}
}

@ARTICLE{Strickert2024,
       author = {{Strickert}, K.~M. and {Evensberget}, D. and {Vidotto}, A.~A.},
        title = "{High-latitude coronal mass ejections on the young solar-like star AB Dor}",
      journal = {\mnras},
         year = 2024,
        month = sep,
       volume = {533},
       number = {1},
        pages = {1156-1165},
          doi = {10.1093/mnras/stae1884},
archivePrefix = {arXiv},
       eprint = {2408.00637},
 primaryClass = {astro-ph.SR},
       adsurl = {https://ui.adsabs.harvard.edu/abs/2024MNRAS.533.1156S}
}

@ARTICLE{Reiners2009b,
       author = {{Reiners}, A. and {Basri}, G.},
        title = "{On the magnetic topology of partially and fully convective stars}",
      journal = {\aap},
         year = 2009,
        month = mar,
       volume = {496},
       number = {3},
        pages = {787-790},
          doi = {10.1051/0004-6361:200811450},
archivePrefix = {arXiv},
       eprint = {0901.1659},
 primaryClass = {astro-ph.SR},
       adsurl = {https://ui.adsabs.harvard.edu/abs/2009A&A...496..787R}
}

@ARTICLE{Kochukhov2019,
       author = {{Kochukhov}, Oleg and {Shulyak}, Denis},
        title = "{Magnetic Field of the Eclipsing M-dwarf Binary YY Gem}",
      journal = {ApJ},
         year = 2019,
        month = mar,
       volume = {873},
       number = {1},
          eid = {69},
        pages = {69},
          doi = {10.3847/1538-4357/ab06c5},
archivePrefix = {arXiv},
       eprint = {1902.04157},
 primaryClass = {astro-ph.SR},
       adsurl = {https://ui.adsabs.harvard.edu/abs/2019ApJ...873...69K}
}

@ARTICLE{Yiu2024,
       author = {{Yiu}, T.~W.~H. and {Vedantham}, H.~K. and {Callingham}, J.~R. and {G{\"u}nther}, M.~N.},
        title = "{Radio emission as a stellar activity indicator}",
      journal = {\aap},
         year = 2024,
        month = apr,
       volume = {684},
          eid = {A3},
        pages = {A3},
          doi = {10.1051/0004-6361/202347657},
archivePrefix = {arXiv},
       eprint = {2312.07162},
 primaryClass = {astro-ph.SR},
       adsurl = {https://ui.adsabs.harvard.edu/abs/2024A&A...684A...3Y}
}

@ARTICLE{Korhonen2010,
       author = {{Korhonen}, H. and {Vida}, K. and {Husarik}, M. and {Mahajan}, S. and {Szczygie{\l}}, D. and {Ol{\'a}h}, K.},
        title = "{Photometric and spectroscopic observations of three rapidly rotating late-type stars: EY Dra, V374 Peg, and GSC 02038-00293}",
      journal = {Astronomische Nachrichten},
         year = 2010,
        month = aug,
       volume = {331},
       number = {8},
        pages = {772},
          doi = {10.1002/asna.201011407},
archivePrefix = {arXiv},
       eprint = {1007.0242},
 primaryClass = {astro-ph.SR},
       adsurl = {https://ui.adsabs.harvard.edu/abs/2010AN....331..772K}
}

@article{Folsom2025, doi = {10.21105/joss.07891}, url = {https://doi.org/10.21105/joss.07891}, year = {2025}, publisher = {The Open Journal}, volume = {10}, number = {111}, pages = {7891}, author = {Folsom, Colin P. and Erba, Christiana and Petit, Veronique and Seadrow, Shaquann and Stanley, Patrick and Natan, Tali and Zaire, Bonnie and Oksala, Mary E. and Villadiego Forero, Federico and Moore, Robin and Catalan Olais, Marisol}, title = {SpecpolFlow: a new software package for spectropolarimetry using Python}, journal = {Journal of Open Source Software} }

@dataset{GaiaCollaboration2020,
       author = {{Gaia Collaboration}},
        title = "{VizieR Online Data Catalog: Gaia EDR3 (Gaia Collaboration, 2020)}",
 howpublished = {VizieR On-line Data Catalog: I/350.  Originally published in: 2021A\&A...649A...1G; doi:10.5270/esa-1ug},
         year = 2020,
        month = nov,
          eid = {I/350},
          doi = {10.26093/cds/vizier.1350},
       adsurl = {https://ui.adsabs.harvard.edu/abs/2020yCat.1350....0G}
}

@ARTICLE{Artigau2024,
       author = {{Artigau}, {\'E}tienne and {Cadieux}, Charles and {Cook}, Neil J. and {Doyon}, Ren{\'e} and {Dauplaise}, Laurie and {Arnold}, Luc and {Cadieux}, Maya and {Donati}, Jean-Fran{\c{c}}ois and {Cristofari}, Paul and {Delfosse}, Xavier and {Fouqu{\'e}}, Pascal and {Moutou}, Claire and {Larue}, Pierre and {Allart}, Romain},
        title = "{Measuring Sub-Kelvin Variations in Stellar Temperature with High-resolution Spectroscopy}",
      journal = {\aj},
         year = 2024,
        month = dec,
       volume = {168},
       number = {6},
          eid = {252},
        pages = {252},
          doi = {10.3847/1538-3881/ad7b30},
archivePrefix = {arXiv},
       eprint = {2409.07260},
 primaryClass = {astro-ph.SR},
       adsurl = {https://ui.adsabs.harvard.edu/abs/2024AJ....168..252A}
}

@ARTICLE{Llama2018,
       author = {{Llama}, Joe and {Jardine}, Moira M. and {Wood}, Kenneth and {Hallinan}, Gregg and {Morin}, Julien},
        title = "{Simulating Radio Emission from Low-mass Stars}",
      journal = {\apj},
         year = 2018,
        month = feb,
       volume = {854},
       number = {1},
          eid = {7},
        pages = {7},
          doi = {10.3847/1538-4357/aaa59f},
archivePrefix = {arXiv},
       eprint = {1801.01507},
 primaryClass = {astro-ph.SR},
       adsurl = {https://ui.adsabs.harvard.edu/abs/2018ApJ...854....7L}
}

\begin{appendix}

\onecolumn

\section{Journal of observations}\label{app:log}

In this appendix, we report the log of the observations for StKM~1-1262 and V374~Peg. The table includes the $\langle B_I \rangle$, T$_\mathrm{eff}$, and longitudinal field measurements for the two stars.

\begin{table*}[h]
\caption{\label{tab:log} List of observations for StKM~1-1262.}    
\centering                       
\begin{tabular}{l c c c c c c r c c}      
\toprule
Date & UT & HJD & $n_\mathrm{cyc}$ & $t_{exp}$ & S/N & $\sigma_\mathrm{LSD}$ & B$_\ell$ & $\langle B_I \rangle$ & T$_\mathrm{eff}$\\
 & [hh:mm:ss] & [$-2460330.1563$] & [s] & & & [$10^{-5}I_c$] & [G] & [kG] & [K]\\
\midrule
January 20 & 15:45:04.50 & 0.0000 & 0.00 & 4x523 & 254 & 8.7 & $-80.0\pm12.9$ & $4.38\pm0.07$ & $3906\pm4$\\
January 21 & 15:50:58.02 & 1.0041 & 0.81 & 4x523 & 271 & 7.2 & $-56.8\pm10.7$ & $4.04\pm0.08$ & $3921\pm4$\\
January 23 & 15:36:47.78 & 2.9943 & 2.41 & 4x523 & 244 & 9.3 & $-44.8\pm12.4$ & $3.48\pm0.07$ & $3945\pm3$\\
January 30 & 15:34:46.05 & 9.9928 & 8.06 & 4x523 & 271 & 7.4 & $-127.0\pm10.9$ & $4.24\pm0.06$ & $3912\pm3$\\
February 15 & 12:55:56.27 & 25.8825 & 20.87 & 4x523 & 254 & 7.6 & $-7.9\pm11.5$ & $4.19\pm0.07$ & $3918\pm3$\\
February 17 & 14:12:10.61 & 27.9355 & 22.53 & 4x523 & 229 & 9.7 & $-98.9\pm13.4$ & $3.51\pm0.07$ & $3961\pm3$\\
February 18 & 15:05:37.19 & 28.9726 & 23.37 & 4x523 & 216 & 11.3 & $-13.3\pm14.6$ & $3.48\pm0.07$ & $3948\pm4$\\
February 19 & 15:30:16.00 & 29.9897 & 24.19 & 4x523 & 235 & 8.8 & $-77.9\pm11.7$ & $3.94\pm0.07$ & $3915\pm3$\\
February 20 & 15:45:20.56 & 31.0002 & 25.00 & 4x523 & 210 & 9.1 & $-73.6\pm13.4$ & $4.36\pm0.06$ & $3908\pm3$\\
February 21 & 14:10:33.49 & 31.9344 & 25.75 & 4x523 & 268 & 8.1 & $-87.5\pm11.0$ & $3.77\pm0.07$ & $3934\pm3$\\
February 22 & 14:43:39.96 & 32.9574 & 26.58 & 4x523 & 275 & 7.5 & $-86.4\pm10.5$ & $3.43\pm0.07$ & $3955\pm3$\\
February 25 & 13:37:16.21 & 35.9112 & 28.96 & 4x523 & 251 & 8.7 & $-40.4\pm12.8$ & $4.39\pm0.08$ & $3911\pm3$\\
February 26 & 14:24:21.07 & 36.9439 & 29.79 & 4x523 & 227 & 8.8 & $-71.9\pm12.0$ & $3.79\pm0.06$ & $3930\pm3$\\
February 27 & 13:31:08.50 & 37.9070 & 30.57 & 4x523 & 213 & 8.7 & $-75.2\pm12.6$ & $3.40\pm0.07$ & $3953\pm3$\\
March 02 & 13:37:56.10 & 41.9117 & 33.80 & 4x523 & 278 & 7.7 & $-64.5\pm10.8$ & $3.84\pm0.07$ & $3932\pm3$\\
\bottomrule                                
\end{tabular}
\tablefoot{The columns are: (1 and 2) date and universal time of the observations, (3) heliocentric Julian date, (4) rotational cycle of the observations found using Eq.~\ref{eq:ephemeris}, (5) exposure time of a polarimetric sequence, (6) signal-to-noise ratio at 1650 nm per polarimetric sequence, (7) RMS noise level of Stokes $V$ signal in units of unpolarised continuum, (8) longitudinal magnetic field (see Eq~\ref{eq:Bl}) with error bar estimated from formal propagation, (9) average unsigned magnetic field, and (10) effective surface temperature.}
\end{table*}

\begin{longtable}{l c c c c c c r c c}   
\caption{\label{tab:log2} List of observations for V374~Peg with the same format as Table~\ref{tab:log}.}\\
\hline\hline
Date & UT & HJD & $n_\mathrm{cyc}$ & $t_{exp}$ & S/N & $\sigma_\mathrm{LSD}$ & B$_\ell$ & $\langle B_I \rangle$ & T$_\mathrm{eff}$\\
 & [hh:mm:ss] & [$-2459440.7902$]  & [s] & & & [$10^{-5}I_c$] & [G] & [kG] & [K]\\
\hline
August 14 & 06:57:49.49 & 0.0000 & 0.00 & 4x100 & 178 & 17.5 & $335.7\pm49.2$ & $6.25\pm0.15$ & $3243\pm5$\\
August 14 & 07:30:58.80 & 0.0230 & 0.05 & 4x100 & 186 & 18.0 & $383.5\pm47.6$ & $6.25\pm0.15$ & $3222\pm4$\\
August 14 & 09:16:44.06 & 0.0965 & 0.22 & 4x100 & 187 & 18.2 & $425.0\pm46.7$ & $6.03\pm0.13$ & $3236\pm5$\\
August 14 & 10:35:33.47 & 0.1512 & 0.34 & 4x100 & 178 & 18.3 & $680.3\pm48.9$ & $6.27\pm0.09$ & $3226\pm4$\\
August 14 & 11:11:49.73 & 0.1764 & 0.40 & 4x100 & 157 & 18.0 & $534.3\pm50.4$ & $6.07\pm0.13$ & $3248\pm4$\\
August 14 & 12:46:18.85 & 0.2420 & 0.54 & 4x100 & 163 & 16.3 & $178.5\pm51.0$ & $6.19\pm0.07$ & $3229\pm4$\\
August 14 & 13:22:01.20 & 0.2668 & 0.60 & 4x100 & 150 & 18.4 & $280.0\pm56.3$ & $5.97\pm0.14$ & $3233\pm5$\\
August 14 & 14:28:20.57 & 0.3129 & 0.70 & 4x100 & 146 & 20.9 & $264.0\pm60.9$ & $6.19\pm0.17$ & $3230\pm5$\\
August 16 & 06:19:28.34 & 1.9734 & 4.43 & 4x100 & 189 & 16.0 & $467.9\pm45.3$ & $6.03\pm0.14$ & $3235\pm5$\\
August 16 & 07:14:20.93 & 2.0115 & 4.52 & 4x100 & 184 & 17.0 & $309.5\pm47.2$ & $5.33\pm0.23$ & $3236\pm4$\\
August 16 & 08:03:33.62 & 2.0457 & 4.60 & 4x100 & 179 & 15.7 & $223.7\pm45.9$ & $5.25\pm0.10$ & $3241\pm4$\\
August 16 & 08:42:23.54 & 2.0726 & 4.66 & 4x100 & 187 & 16.2 & $313.8\pm45.4$ & $6.12\pm0.07$ & $3250\pm4$\\
August 16 & 09:25:00.04 & 2.1022 & 4.72 & 4x100 & 169 & 17.1 & $270.6\pm48.8$ & $5.89\pm0.13$ & $3244\pm4$\\
August 16 & 10:10:27.33 & 2.1338 & 4.79 & 4x100 & 175 & 16.1 & $237.6\pm46.0$ & $6.23\pm0.10$ & $3229\pm4$\\
August 16 & 10:54:42.21 & 2.1645 & 4.86 & 4x100 & 173 & 15.3 & $265.6\pm47.8$ & $6.13\pm0.10$ & $3254\pm4$\\
August 16 & 11:43:39.33 & 2.1985 & 4.94 & 4x100 & 162 & 17.3 & $255.8\pm53.8$ & $6.25\pm0.10$ & $3232\pm4$\\
August 17 & 06:16:08.47 & 2.9711 & 6.68 & 4x100 & 172 & 20.7 & $178.7\pm51.2$ & $6.17\pm0.15$ & $3244\pm4$\\
August 17 & 06:58:20.23 & 3.0004 & 6.74 & 4x100 & 167 & 18.0 & $298.8\pm49.3$ & $5.86\pm0.09$ & $3222\pm5$\\
August 17 & 07:50:23.44 & 3.0365 & 6.82 & 4x100 & 184 & 18.1 & $364.5\pm49.4$ & $6.11\pm0.14$ & $3229\pm5$\\
August 17 & 09:07:30.40 & 3.0901 & 6.94 & 4x100 & 177 & 17.5 & $136.7\pm48.6$ & $6.40\pm0.16$ & $3230\pm4$\\
August 17 & 09:50:16.61 & 3.1198 & 7.01 & 4x100 & 178 & 16.4 & $281.0\pm54.8$ & $6.28\pm0.08$ & $3227\pm5$\\
August 17 & 10:33:16.01 & 3.1496 & 7.08 & 4x100 & 169 & 16.7 & $455.6\pm52.8$ & $6.07\pm0.14$ & $3234\pm4$\\
August 17 & 11:20:59.43 & 3.1828 & 7.15 & 4x100 & 146 & 21.8 & $444.6\pm54.4$ & $6.25\pm0.09$ & $3249\pm4$\\
August 17 & 12:09:46.87 & 3.2166 & 7.23 & 4x100 & 150 & 18.9 & $435.5\pm57.2$ & $6.15\pm0.09$ & $3242\pm5$\\
August 20 & 06:06:45.26 & 5.9645 & 13.40 & 4x100 & 168 & 20.5 & $544.2\pm53.3$ & $6.13\pm0.14$ & $3249\pm4$\\
August 20 & 06:50:38.79 & 5.9950 & 13.47 & 4x100 & 155 & 18.0 & $395.9\pm58.8$ & $6.05\pm0.14$ & $3242\pm5$\\
August 20 & 07:26:51.78 & 6.0202 & 13.53 & 4x100 & 172 & 18.5 & $271.9\pm52.2$ & $5.89\pm0.12$ & $3242\pm4$\\
August 20 & 08:14:10.99 & 6.0530 & 13.60 & 4x100 & 161 & 18.7 & $174.8\pm58.3$ & $5.62\pm0.22$ & $3242\pm5$\\
August 20 & 09:29:52.77 & 6.1056 & 13.72 & 4x100 & 106 & 26.9 & $308.0\pm77.8$ & $6.33\pm0.21$ & $3244\pm6$\\
August 20 & 10:05:36.35 & 6.1304 & 13.78 & 4x100 & 118 & 31.8 & $380.9\pm98.8$ & $6.30\pm0.21$ & $3226\pm6$\\
August 20 & 11:33:59.06 & 6.1918 & 13.91 & 4x100 & 128 & 30.2 & $128.4\pm87.4$ & $6.39\pm0.19$ & $3221\pm5$\\
August 20 & 12:19:20.61 & 6.2233 & 13.98 & 4x100 & 124 & 26.6 & $294.2\pm77.4$ & $6.55\pm0.16$ & $3223\pm5$\\
August 21 & 06:54:34.48 & 6.9977 & 15.73 & 4x100 & 150 & 19.4 & $381.7\pm57.5$ & $6.02\pm0.15$ & $3229\pm5$\\
August 21 & 07:41:44.13 & 7.0305 & 15.80 & 4x100 & 173 & 19.1 & $327.1\pm52.5$ & $5.97\pm0.16$ & $3227\pm5$\\
August 21 & 08:17:44.29 & 7.0555 & 15.86 & 4x100 & 178 & 18.6 & $150.1\pm51.6$ & $6.29\pm0.15$ & $3237\pm4$\\
August 21 & 09:12:20.80 & 7.0934 & 15.94 & 4x100 & 186 & 16.6 & $117.0\pm48.2$ & $6.12\pm0.15$ & $3243\pm4$\\
August 21 & 10:28:37.68 & 7.1464 & 16.06 & 4x100 & 86 & 30.9 & $276.2\pm98.5$ & $6.80\pm0.20$ & $3216\pm6$\\
August 21 & 11:10:31.26 & 7.1755 & 16.12 & 4x100 & 154 & 23.3 & $481.9\pm63.2$ & $6.21\pm0.14$ & $3242\pm5$\\
August 21 & 12:07:31.67 & 7.2151 & 16.21 & 4x100 & 145 & 23.8 & $360.6\pm59.4$ & $6.32\pm0.11$ & $3225\pm4$\\
August 21 & 12:58:38.35 & 7.2506 & 16.29 & 4x100 & 167 & 19.1 & $489.2\pm57.0$ & $6.30\pm0.11$ & $3225\pm4$\\
August 22 & 06:44:30.64 & 7.9908 & 17.96 & 4x100 & 121 & 27.0 & $138.5\pm77.6$ & $6.43\pm0.16$ & $3235\pm6$\\
August 22 & 07:24:01.95 & 8.0182 & 18.02 & 4x100 & 123 & 30.7 & $378.4\pm88.2$ & $6.55\pm0.17$ & $3211\pm6$\\
August 22 & 07:58:51.70 & 8.0424 & 18.07 & 4x100 & 132 & 22.2 & $337.0\pm69.6$ & $6.22\pm0.14$ & $3224\pm5$\\
August 22 & 08:52:52.15 & 8.0799 & 18.16 & 4x100 & 156 & 20.8 & $531.6\pm62.6$ & $6.24\pm0.13$ & $3226\pm5$\\
August 22 & 10:14:18.03 & 8.1364 & 18.28 & 4x100 & 159 & 19.1 & $434.7\pm56.9$ & $5.95\pm0.13$ & $3248\pm4$\\
August 22 & 10:58:24.09 & 8.1671 & 18.35 & 4x100 & 162 & 16.7 & $613.0\pm52.6$ & $6.29\pm0.13$ & $3227\pm4$\\
August 22 & 11:51:18.61 & 8.2038 & 18.44 & 4x100 & 173 & 17.2 & $486.0\pm50.6$ & $6.17\pm0.14$ & $3231\pm4$\\
August 22 & 12:35:16.19 & 8.2343 & 18.50 & 4x100 & 173 & 17.6 & $311.7\pm49.7$ & $5.87\pm0.13$ & $3245\pm4$\\
August 23 & 06:28:23.06 & 8.9796 & 20.18 & 4x100 & 176 & 18.6 & $380.5\pm49.8$ & $6.06\pm0.12$ & $3223\pm4$\\
August 23 & 07:07:08.09 & 9.0065 & 20.24 & 4x100 & 182 & 17.9 & $442.9\pm49.0$ & $5.91\pm0.10$ & $3233\pm4$\\
August 23 & 07:40:43.06 & 9.0298 & 20.29 & 4x100 & 182 & 17.8 & $545.0\pm49.3$ & $6.14\pm0.13$ & $3226\pm4$\\
August 23 & 08:33:30.83 & 9.0665 & 20.37 & 4x100 & 180 & 17.8 & $543.8\pm52.0$ & $6.25\pm0.11$ & $3224\pm3$\\
August 23 & 09:47:38.32 & 9.1179 & 20.49 & 4x100 & 184 & 18.8 & $528.6\pm49.7$ & $5.90\pm0.11$ & $3234\pm4$\\
August 23 & 10:22:11.97 & 9.1419 & 20.54 & 4x100 & 166 & 21.6 & $326.3\pm51.8$ & $5.52\pm0.30$ & $3246\pm4$\\
August 23 & 11:17:33.79 & 9.1804 & 20.63 & 4x100 & 155 & 17.3 & $206.2\pm53.4$ & $5.76\pm0.10$ & $3233\pm4$\\
\hline
\end{longtable}

\section{Additional figures}

In this appendix, we provide additional figures related to the atmospheric characterisation of the two stars and the correlation analysis between the total unsigned magnetic field, the absolute value of the longitudinal magnetic field and the stellar surface temperature. 

\begin{figure*}[h]
    \centering
    \includegraphics[width=0.49\textwidth]{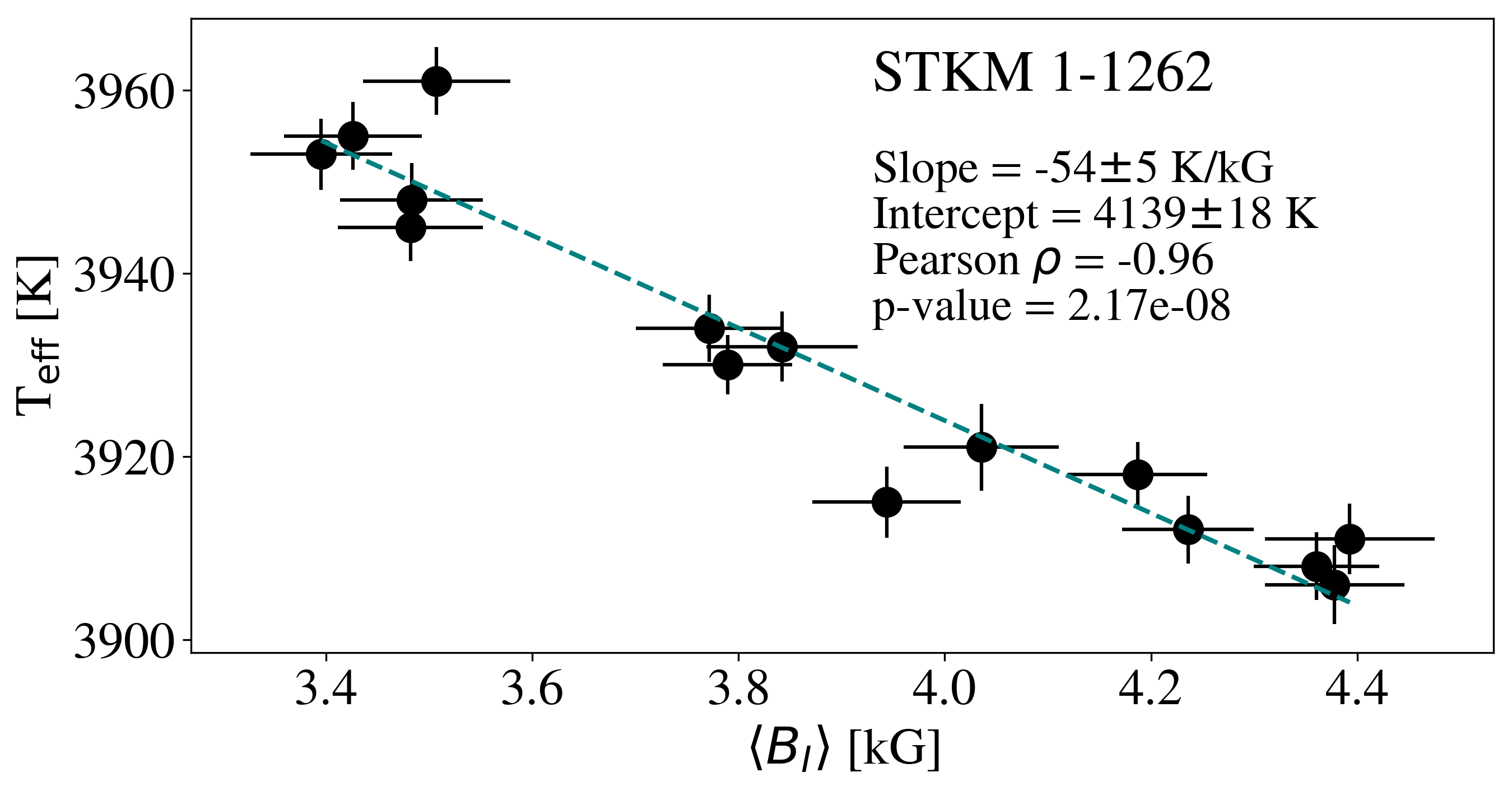}
    \includegraphics[width=0.49\textwidth]{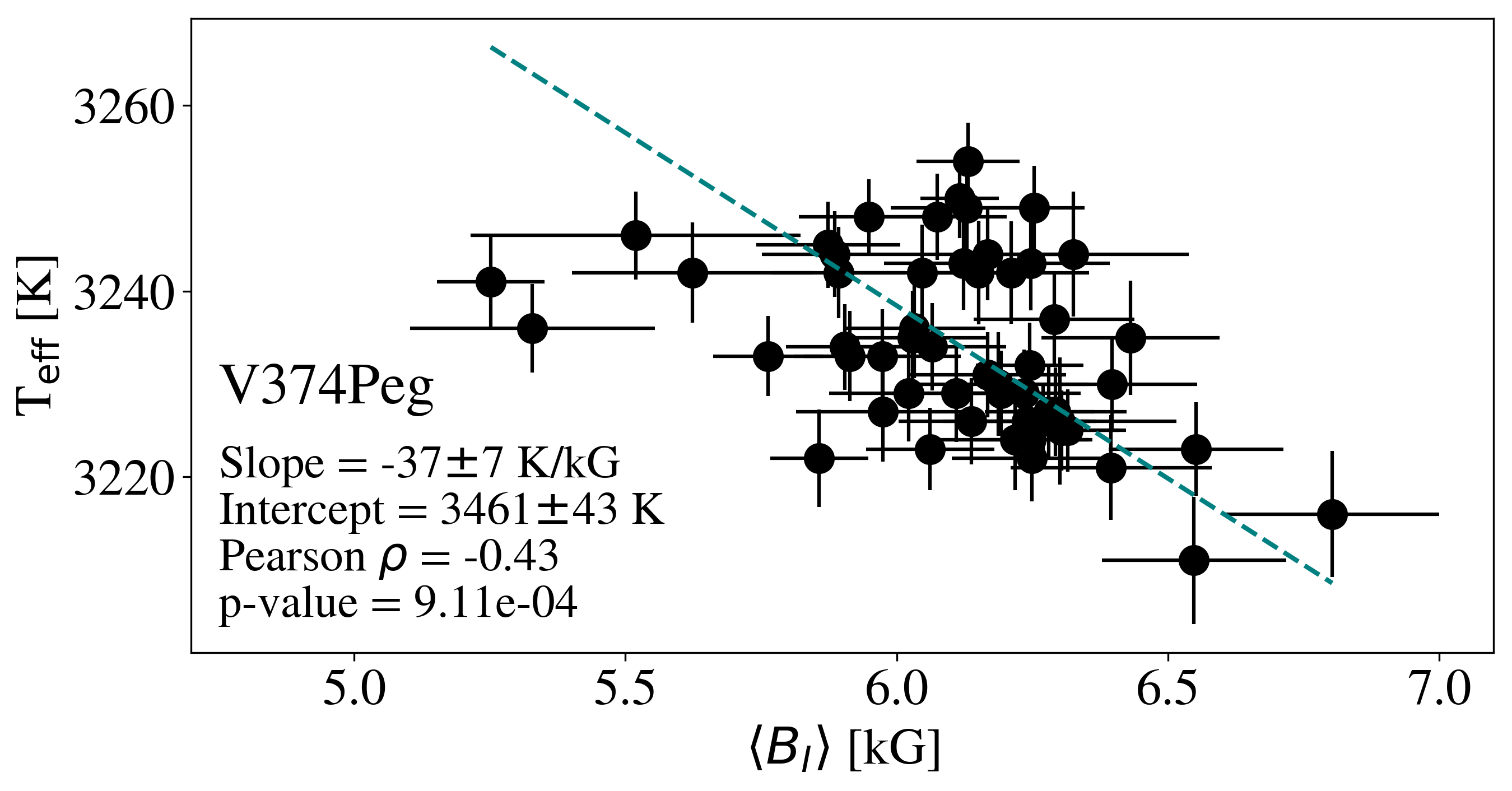}
    \includegraphics[width=0.49\textwidth]{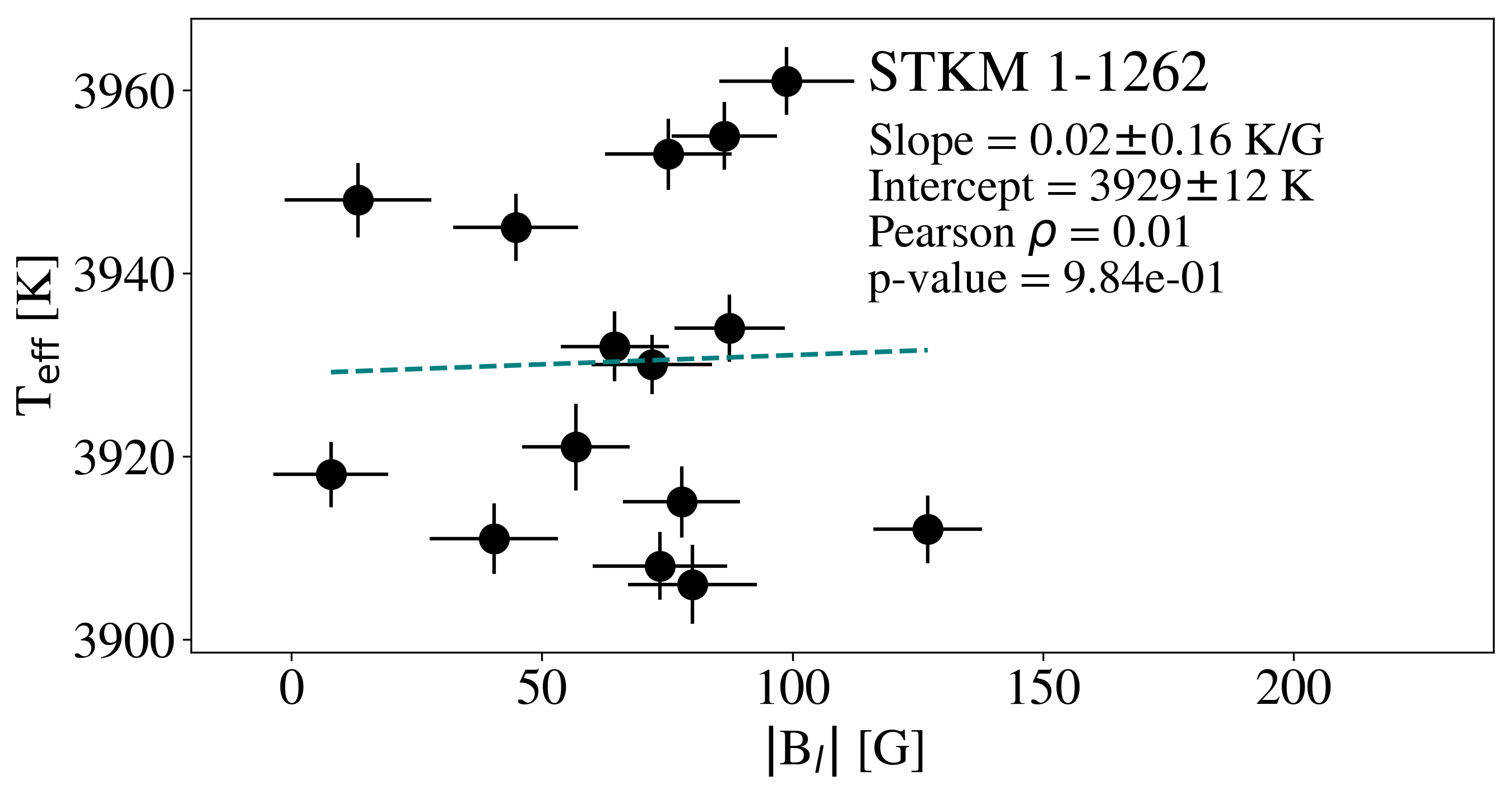}
    \includegraphics[width=0.49\textwidth]{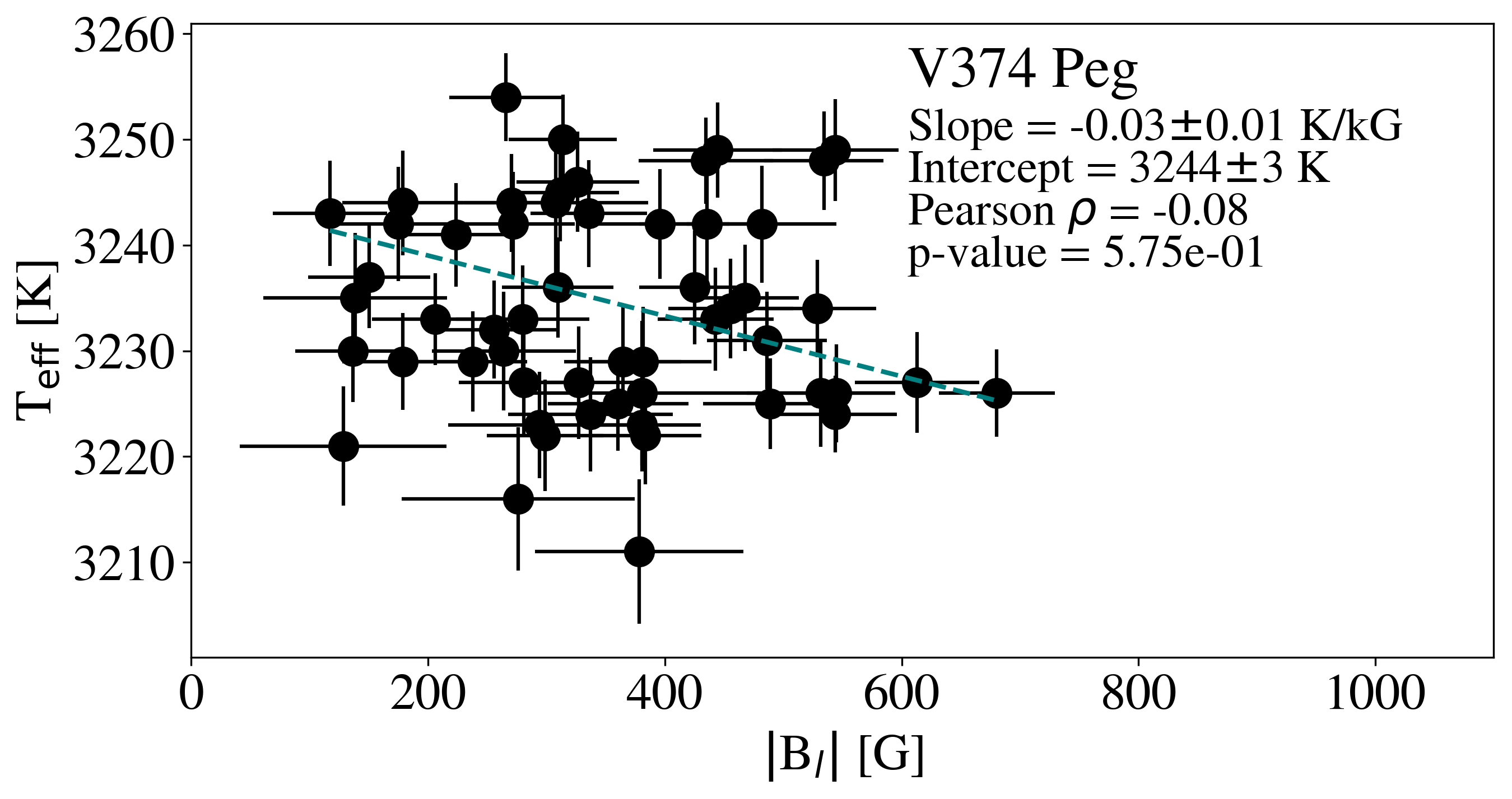}
    \caption{Correlation of the total unsigned field and longitudinal field with the stellar surface temperature. The left panels refer to StKM~1-1262 and the right ones to V374~Peg. The blue line indicates a linear fit, whose best-fit values are given in the each legend. We also include the Pearson correlation coefficient and the corresponding p-value.}
    \label{fig:correlations}
\end{figure*}

\begin{figure*}
    \centering
    \includegraphics[width=1.0\linewidth]{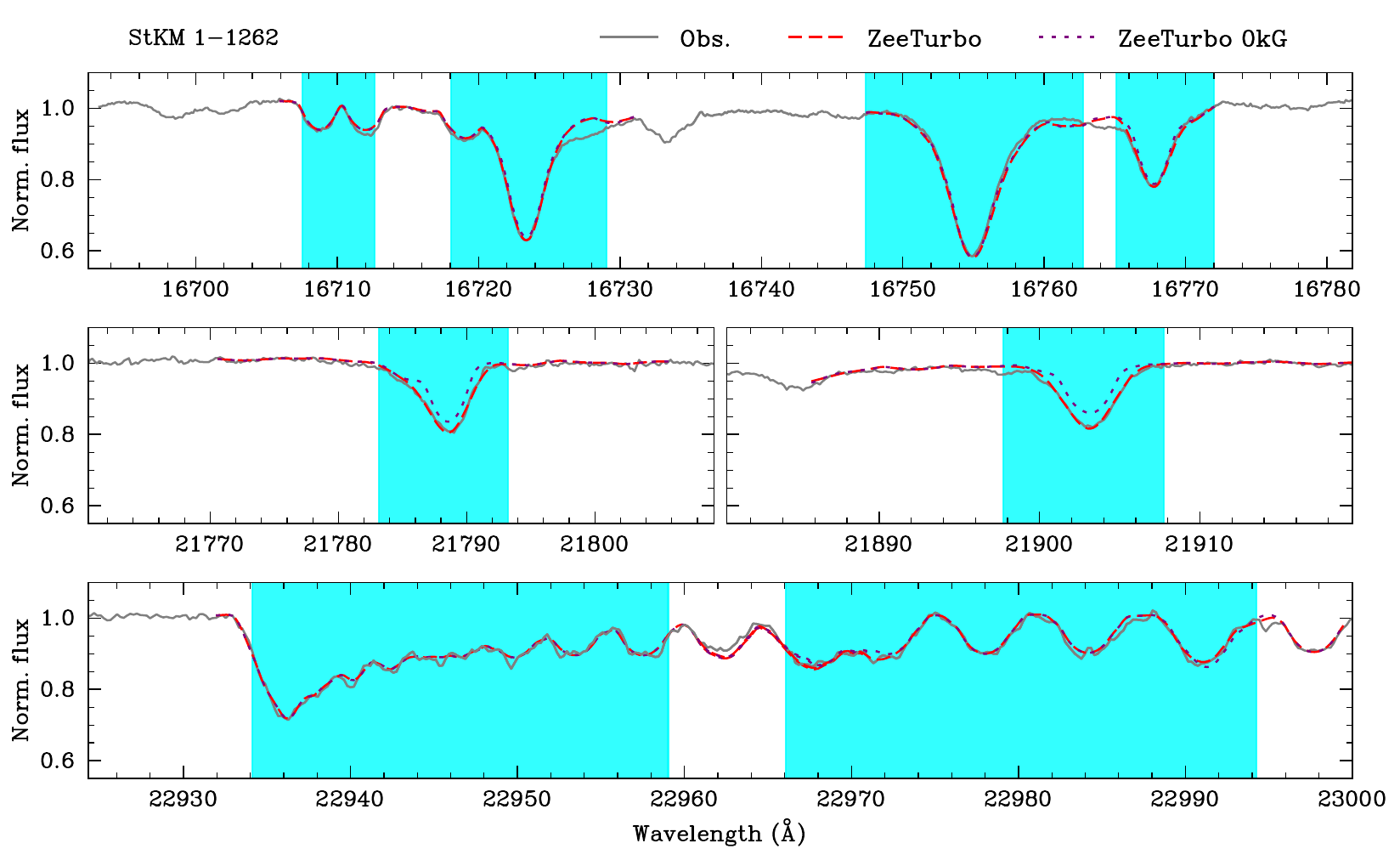}
    \caption{Example fit obtained on the template spectrum of StKM~1-1262. The cyan region shows regions on which the fit is performed.}
    \label{fig:example_spec_stkm}
\end{figure*}

\begin{figure*}
    \centering
    \includegraphics[width=1.0\linewidth]{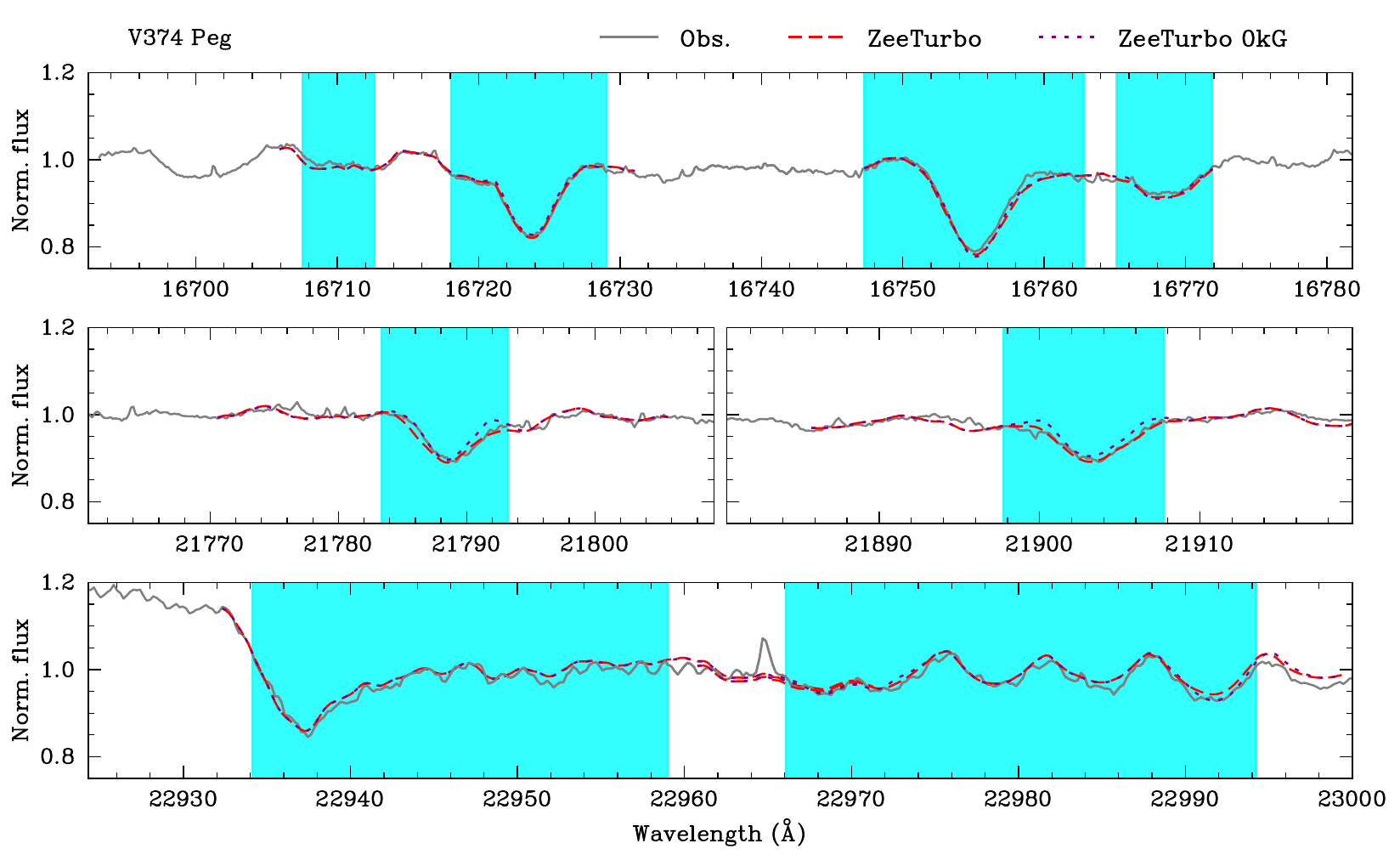}
    \caption{Same as Fig.~\ref{fig:example_spec_stkm} for V374~Peg.}
    \label{fig:example_spec_v374peg}
\end{figure*}

\section{Simultaneous Stokes~$I$ and $V$ modelling}\label{app:lsd}

\begin{figure}[h]
\centering
    \includegraphics[width=0.8\textwidth]{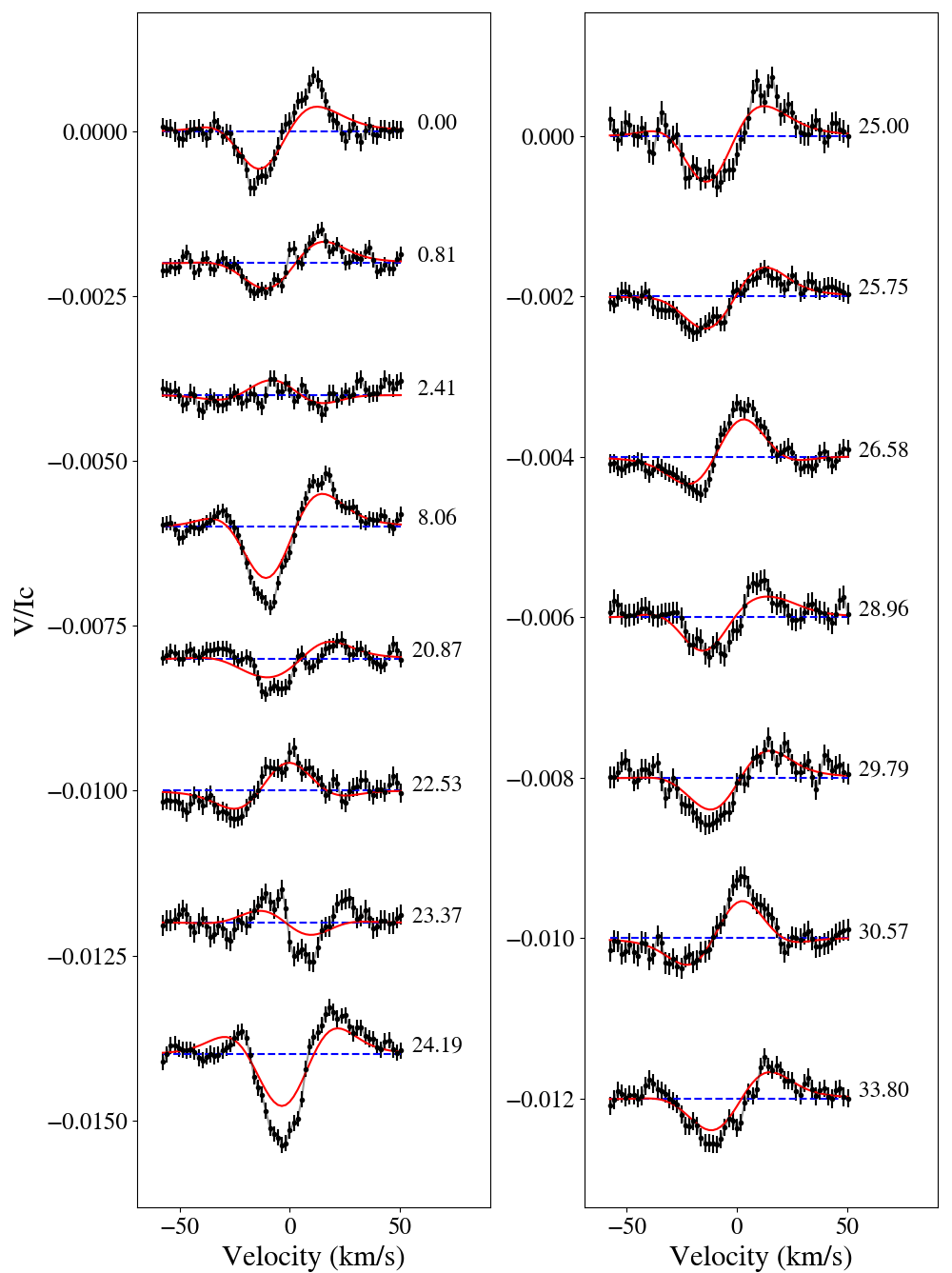}
    \caption{Time series of Stokes~$V$ LSD profiles and the ZDI models for StKM~1-1262. The observations are shown in black and the models in red. The numbers on the right indicate the rotational cycle computed from Eq.~\ref{eq:ephemeris} using the first observation of an epoch as reference date. The horizontal line represents the zero point of the profiles, which are shifted vertically based on their rotational phase for visualisation purposes. }
    \label{fig:stokesV_StKM}
\end{figure}

\begin{figure}[h]
    \centering
    \includegraphics[width=0.8\textwidth]{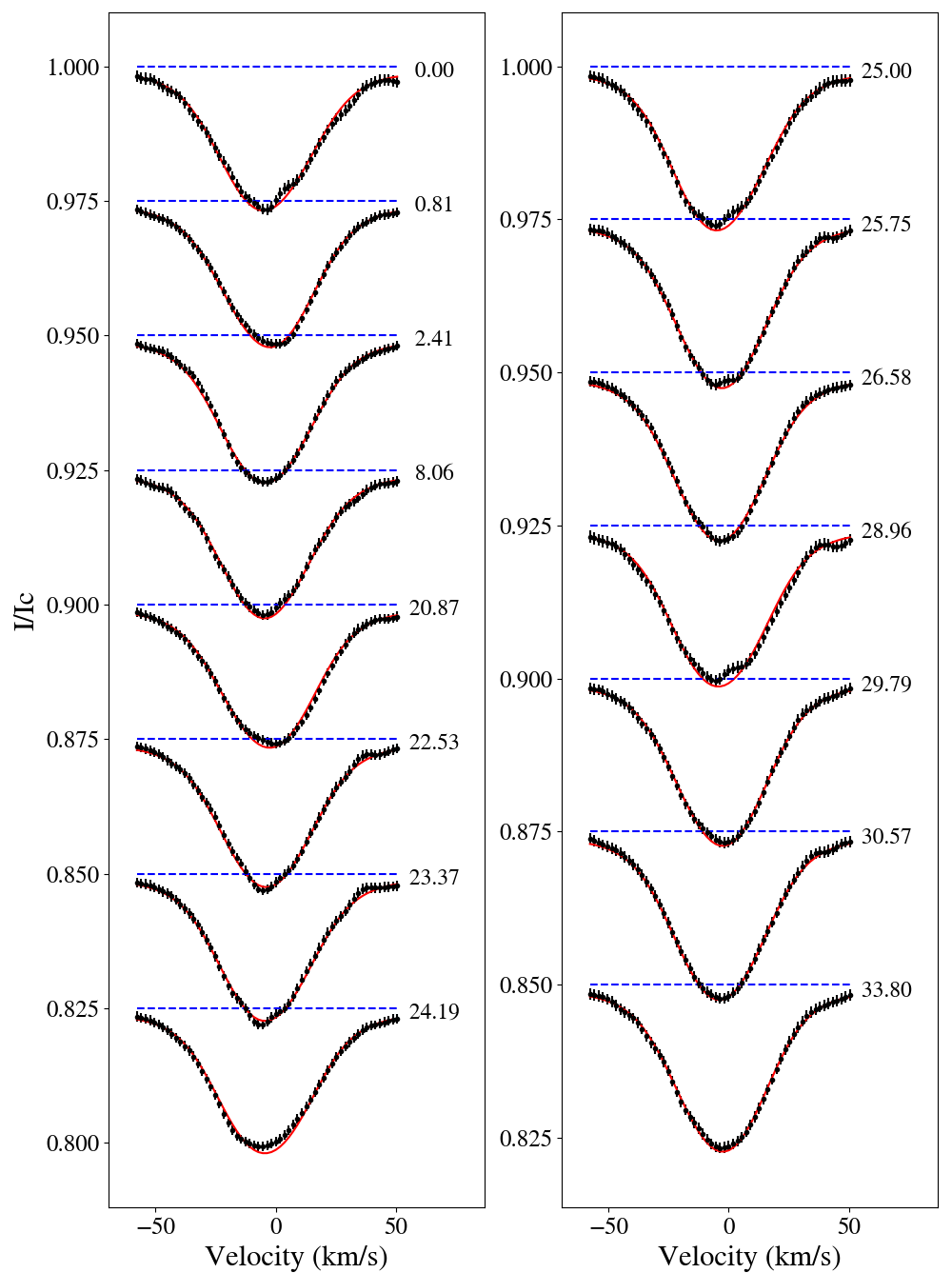}
    \caption{Time series of Stokes~$I$ LSD profiles and the DI models for StKM~1-1262. The format is the same as Fig~\ref{fig:stokesV_StKM}. The horizontal blue dashed line indicates the continuum level at 1.0, which is shifted vertically for visualisation purposes.}
    \label{fig:stokesI_StKM}
\end{figure}

\begin{figure*}[h]
\centering
    \includegraphics[width=0.9\textwidth]{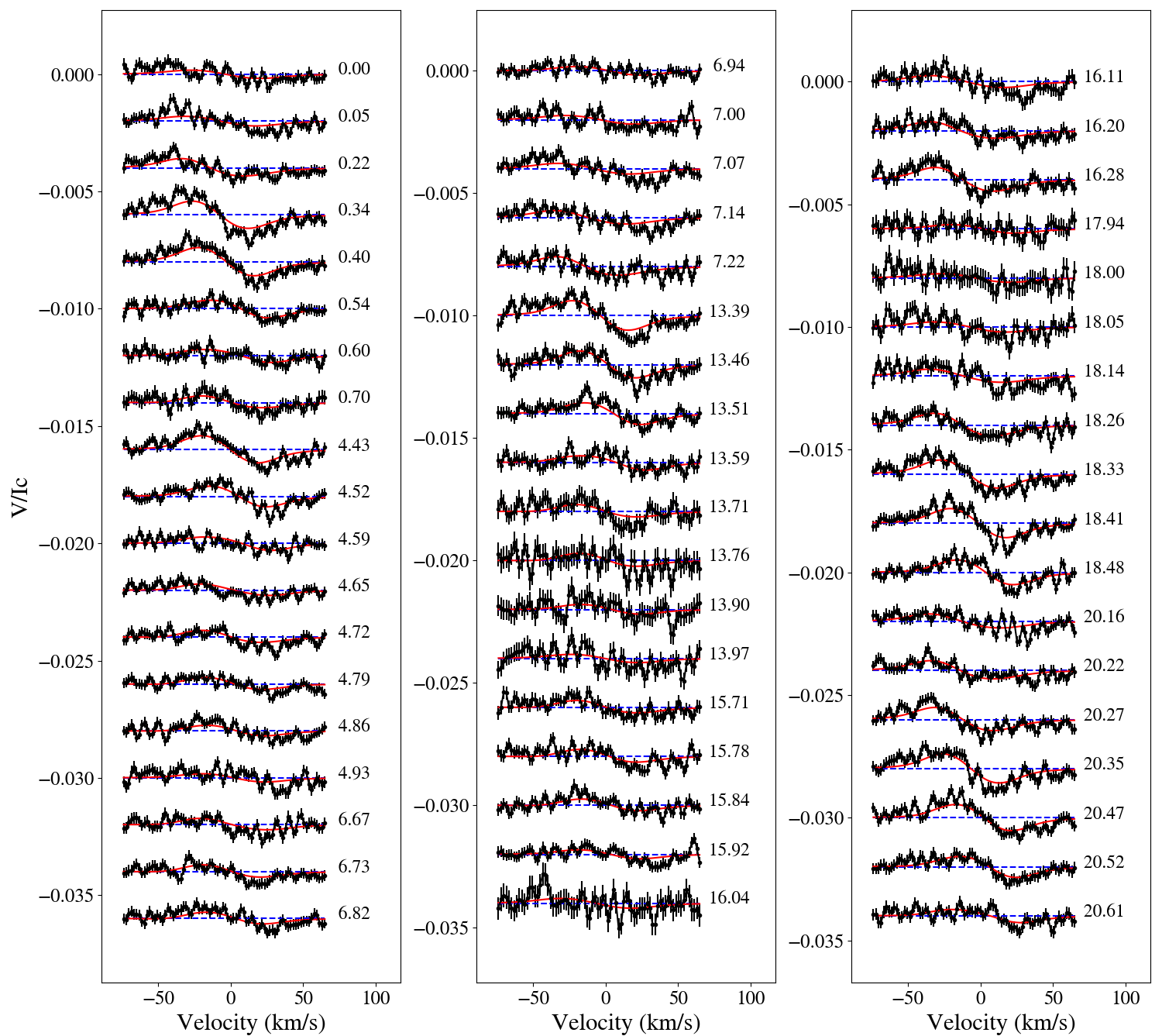}
    \caption{Time series of Stokes~$V$ LSD profiles and the ZDI models for V374~Peg. The format is the same as Fig.~\ref{fig:stokesV_StKM}. }
    \label{fig:stokesV_vpeg}
\end{figure*}

\begin{figure*}[h]
    \centering
    \includegraphics[width=0.9\textwidth]{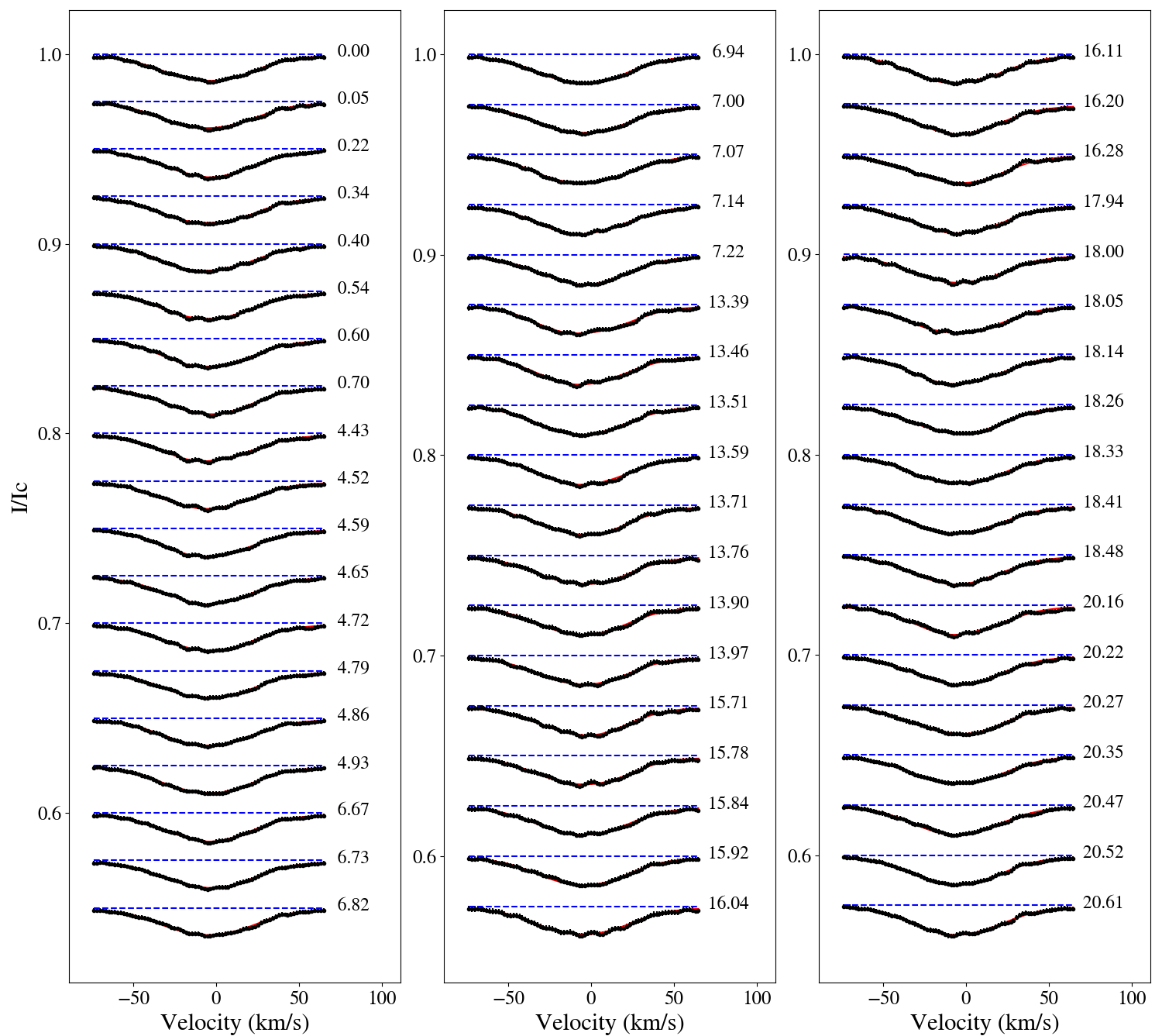}
    \caption{Time series of Stokes~$I$ LSD profiles and the DI models for V374~Peg. The format is the same as Fig.~\ref{fig:stokesI_StKM}.}
    \label{fig:stokesI_vpeg}
\end{figure*}

In this appendix, we present the Stokes~$I$ and $V$ models for both StKM~1-1262 and V374~Peg. In Fig.~\ref{fig:stokesV_StKM} and Fig.~\ref{fig:stokesV_vpeg}, we show the Stokes~$V$ LSD profiles associated to the magnetic maps presented in Sect.~\ref{sec:ZDI}. In Fig.~\ref{fig:stokesI_StKM} and Fig.~\ref{fig:stokesI_vpeg}, we show the Stokes~$I$ LSD profiles associated to the brightness maps presented in Sect.~\ref{sec:DI}.

When performing ZDI, one typical assumption is that the brightness of the photosphere is homogeneous. \citet{Rosen2012} tested the reliability of ZDI reconstructions when high-contrast temperature spots are modelled simultaneously to the magnetic geometry in ideal conditions (i.e. high S/N, dense phase coverage, and large $v_\mathrm{eq}\sin i$). They found that the magnetic field strength is underestimated by 10-15\% when temperature and magnetic spots are included, and by 30-60\% when only magnetic spots are included. We decided to reconstruct the large-scale magnetic field of STKM~1-1262 and V374~Peg together with their brightness distribution, to double check the consistency of our results. 

\begin{figure*}[h]
    \centering
    \includegraphics[width=0.8\textwidth]{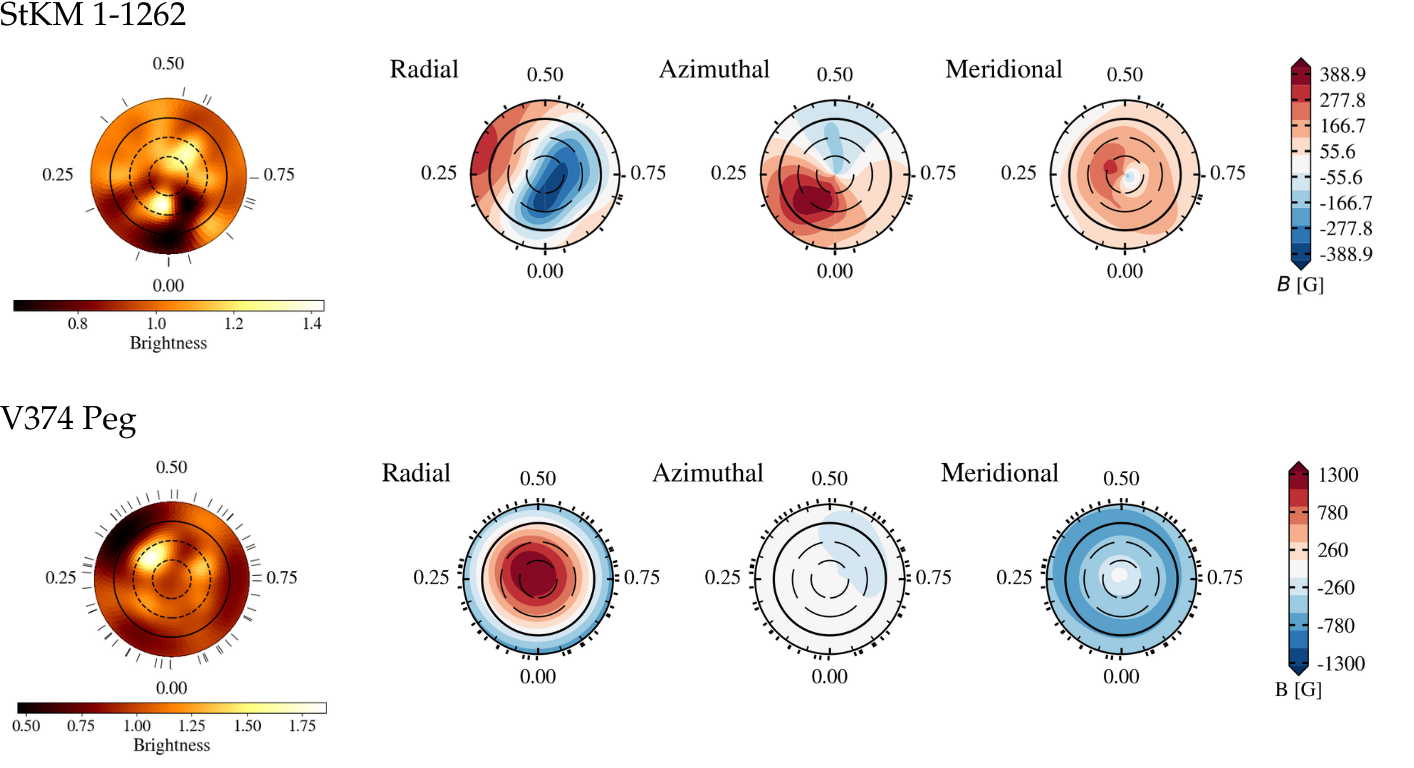}
    \caption{Simultaneous brightness and magnetic reconstruction. The format is the same as Fig.~\ref{fig:brimap} for the brightness part and Fig.~\ref{fig:zdi_maps} for the magnetic part.}
    \label{fig:zdi_di}
\end{figure*}

The maps are shown in Fig.~\ref{fig:zdi_di}. The magnetic field topology of both stars does not change substantially. For StKM~1-1262, the poloidal energy fraction is increased by 3\%, the dipolar by 6\% (with consequent decrease in quadrupolar and octupolar fractions) and the axisymmetric component by 6\%. The average field strength obtained when reconstructing Stokes~$I$ and $V$ simultaneously is 15\% lower. For V374~Peg, the configuration is more dipolar by 3\% and axisymmetric by 6\%, and the average field strength is 18\% larger.

For the brightness map of StKM~1-1262, we observe a dark spot at a similar location than when reconstructing Stokes~$I$ alone, with a coverage slightly decreased to 8.5\% and a lower contrast of 20\% with respect to the quiet photosphere. In this case, the algorithm puts a bright feature at around phase 0.10 at a higher latitude with respect to the previous run with Stokes~$I$ alone, and shifts the phase of the already-present bright feature from 0.5 to 0.6. In the new reconstruction, the position of the bright features correlates with the location of the large-scale field component with negative polarity. For V374~Peg, the contrast of the brightness map is mostly dominated by a bright feature at around phase 0.4, which is the same pointing phase of the positive polarity of the large-scale magnetic field configuration. The code still reconstructs the same bright features at phases 0.2 and 0.6 and at a similar latitude, albeit with a smaller contrast with the quiet photosphere. 

We note that magnetic effects can dominate the width of Stokes~$I$ LSD profiles rather than brightness inhomogeneities. Recent examples are the M~dwarfs AU~Mic and EV~Lac described by \citet{Donati2023} and \citet{Donati2025}, for which the authors find substantial differences in the width of Stokes~$I$ LSD profiles when computed with magnetically sensitive and insensitive lines separately. They assumed that the small-scale field locally scales up with the large-scale field, which allowed them to fit the Stokes~$I$ profiles width simultaneously with the Stokes~$V$ profiles, producing a large-scale magnetic field reconstruction consistent with the small-scale field. In our case, the change in width of the Sokes~$I$ LSD profiles is not large, likely due to rotational broadening being a dominant component, and its rotational modulation is not more or less evident when using magnetically insensitive and sensitive lines, respectively. Thus, we cannot apply the same methodology.

\end{appendix}

\end{document}